\documentclass{iopart}
\usepackage{graphicx} 
\usepackage{bm,mleftright,url,cite}
\usepackage{iopams}  
\usepackage[utf8]{inputenc}
\usepackage{xcolor}
\begin{document}

\title{Cuprates, Pnictides and Sulfosalts: Lessons in Functional Materials}
\author{N.~Bari\v si\'c$^{1,2}$ \& D. K. Sunko$^2$}

\address{$^1$ Institute of Solid State Physics, TU Wien, 1040 Vienna, Austria.}
\address{$^2$ Department of Physics, Faculty of Science, University of Zagreb, Zagreb, Croatia.}
\ead{nbarisic@phy.hr}
\ead{dks@phy.hr}

\date{January 2025}


\begin{abstract}
Murunskite K$_2$Cu$_3$FeS$_4$ is a representative sulfosalt, isostructural to the pnictides, but with electronic properties more similar to the insulating parent compounds of the cuprates. We use it as a bridge to compare the chemical and physical roles of metal and ligand orbitals in cuprates and pnictides.

In cuprates, ionicity, covalency, and metallicity are tightly interwoven to give rise to high-temperature superconductivity (SC). Their most remarkable property is the interaction of an ionically localized hole on the copper (Cu) with a Fermi liquid (FL) on the oxygens (O), which is critically important for understanding all key properties of these materials. The localization is due to strong correlations on the Cu $3d$ orbital. We describe a scenario in which the localized hole gives rise both to SC by Cooper scattering of O holes, and to Fermi arcs, as observed in cuprate spectroscopy, the latter by a purely kinematic projection of the static local disorder, without invoking any residual interactions between the mobile O FL carriers.

In the pnictides, the orbitals responsible for binding and metallic conduction appear to be separate. The Fe $3d$ $e_{g}$ orbitals hybridized with the ligands set the lattice spacing. The $3d$ $t_{2g}$ orbitals overlap directly between the Fe atoms, resulting in several electronic bands appearing at the Fermi level. The ensuing Fermi liquid exhibits both charge and magnetic correlations. We argue that a similar SC scenario as in the cuprates is plausible in the pnictides, except that a light FL scatters on a slow nearly-antiferromagnetic (AF) one, rather than on localized holes as in the cuprates.

In murunskite, the greater covalency of sulfur orbitals compensates for total disorder in the Fe atom positions, giving rise to an agglomeration of fractal magnetic clusters coexisting with perfect crystallographic order and exhibiting surprisingly robust quarter-zone AF. Because the open ligand orbitals play an important role in it, the essentially local quarter-zone AF of murunskite is more similar to the SC of the cuprates than to the half-zone AF of the pnictides, in which the ligands are passive.
\end{abstract}

\section{Introduction.}

Cuprate high-temperature superconductors (high-T$_c$ SC) have revolutionized the field of solid-state physics since they were discovered almost $40$ years ago~\cite{Bednorz86}. They stimulated unprecedented improvement in sample synthesis, characterization, and measurement. They also drew attention to the effect of strong Coulomb interactions on functional electrons, underlining the need to go beyond textbook descriptions of ionic, metallic, and covalent bonding~\cite{Anderson87}. In our view, both of these developments mean that the chemical and physical aspects of cuprates cannot be separated, thus opening a new research field at the interface of physics and chemistry.

This realization matured over a long period. Cuprate research branched at the very beginning between a top-down (abstract) and bottom-up (concrete) approach. The abstract approach begins with electrons in a single band on a square lattice, and ascribes all unusual properties of cuprates to strong on-site interactions between them. Despite most of the theoretical efforts having been concentrated on it over nearly four decades, all key aspects of the cuprates---normal-state, SC, pseudogap, strange metal, etc.---have remained mysterious from the one-band point of view, in both its dispersive and diffusive ``Planckian limit'' variants~\cite{Keimer15,Phillips22}.

Over the same period, an essential understanding of the physical regime of the cuprates has developed in the concrete approach, based on the actual chemical orbitals available to the functional electrons, and the time and energy scales through which they interact~\cite{Zaanen85,Barisic87,Emery87,Mazumdar89,Barlingay90,Barisic90,Eskes93,Barriquand94,Barisic22}. The language of orbitals is capable of describing a large class of functional materials in addition to the cuprates. That should not be surprising, because it is the standard language of chemistry. The specific challenge which, we argue, has been largely met in the cuprates, is to have it describe the physical functionality as well.

In this review, we explore the landscape shaped by this approach. We compare three classes of functional materials, the high-T$_c$ cuprates, ferropnictides, and murunskite, a sulfosalt structurally identical to the latter but exhibiting intriguing electronic similarities with the former. We shall occasionally make qualitative comparisons with the abstract approach, though without attempting to provide a comprehensive account of it. Our first step is to draw a distinction between chemical bonding and physical functionality.

In metallic bonding, the two are the same. The metallic bond provides the condensation energy, while the functionality is conduction. The success of the tight-binding model of solids in this case obscures the fact that it is chemically rather exceptional. Most compounds are ionically or covalently bound without conductivity. In such cases, the physical functionality is traditionally magnetism, although that has also been changing with an increased interest in ferroelectrics and topological materials.

In the textbook ionic case, e.g., rock salt, an electron is transferred from one atom to another, where it stays. In the insulating covalent case, a pair of electrons is shared between neighboring atoms, but goes no further. We shall not consider covalently bound materials here, although the widely investigated organic metals show that they, too, can develop interesting functionality upon doping.

The evolution of cuprate functionality with doping is accompanied by an evolution of the Cu $3d$--O $2p$ chemical bond. It crosses over from ionic to covalent between under- and over-doped materials, resulting in an increase in the concentration of mobile carriers, from $p$ to $1+p$, with temperature and doping $p$~\cite{Barisic15,Barisic19,Pelc19,Barisic22,Badoux16,Putzke21,Nicholls25}. Doped cuprates are \emph{ionic metals}, that is, dominantly ionically bound materials with conducting functionality. We shall describe the coexistence of high-energy (chemical) and low-energy (physical) degrees of freedom in them in some detail. These insights will be compared with ferropnictides and murunskite. The common thread is that one must always keep in mind the atomic orbitals responsible for various aspects of material behavior. Abstract (one- or non-band) models are not sufficient because the chemical balance of electrons in bonds keeps changing with doping and temperature, in a remarkably universal fashion in all cuprates. The language of atomic orbitals is flexible enough to keep track of this evolution, while the language of scattering rates is not. The physical and chemical points of view come together in the dual role of the chemical potential, which controls both the concentration of functional electrons and the balance of chemical reactions between constituents in the solid state.

Thermodynamically, the distinction to keep in mind is between first- and second-order phase transitions. It reflects the time scales of processes which stabilize the material. If a local degree of freedom is free to fluctuate during a thermalization interval, the molecular-field approach is a good starting point to describe the equilibrium, which is typically destabilized by a second-order transition. If the local degree of freedom flips from one long-lived discrete state to another without such fluctuations, the molecular-field approach is inapplicable, because no individual site is representative of the average. If conditions that provoked the local change of state pertain throughout the material, it will undergo a first-order transition with hysteresis, or a crossover, without a diverging correlation length. In cuprates, there is a variation of this scenario in which a local (first-order) orbital transition affects a transition on neighboring sites, so a global transition proceeds by a percolation network~\cite{Li16,Pelc18,Popcevic18,Pelc19,Phillips03,Kresin06,Fratini10}. On the level of structure, this is observed as random local tilts, possibly locally ordered into stripes or nematicity, depending on the compound.

In this review, we first establish a general framework for the description of functional materials. We describe cuprates in this framework next, with emphasis on local effects. In particular, we show how the well-known though controversial Fermi arcs~\cite{Norman98,Lee07-1,Kunisada20} can appear as an effect of localization due to strong correlations, without invoking any interactions between remaining mobile carriers. The physical regime of the pnictides is compared to the cuprates, emphasizing both the parallels and the illuminating distinctions. Finally, murunskite provides a unifying perspective for the relation between local coherence and disorder. We end with a discussion of the parallels between local magnetism in murunskite and local superconductivity in cuprates.

\section{Quantum mechanics of functional materials.}

In the condensed state, quantum mechanics and thermodynamics coexist in a balance between dispersion and diffusion. Dispersion reflects coherent processes among indistinguishable particles. Diffusion involves dissipative (irreversible) processes that destroy coherence. The ionic, covalent, and metallic bonds are the extreme limits of this competition. Ionic bonding can be understood as the attraction of classical charged balls. The ionic conductivity of solid salt at high temperature is a textbook example of diffusion. The two electrons in a covalent bond are indistinguishably entangled. The metallic bond is a super-covalent bond, in which the whole sample is a single molecule. Its conductivity at zero temperature is purely dispersive.

Functional materials are characterized by being in an intermediate position between these limits. In ionic metals, the local orbitals are partially ionic, which provides the bulk of the condensation energy, and partially metallic, which provides the functionality. The key characteristic to distinguish among the various materials is the particular separation of roles between the available orbitals. A sensible classification must respect a natural hierarchy among them.

Zeroth-order effects reflect the orbitals and symmetries available to the electrons for a given chemical potential. These determine what can happen at all, so they are a non-negotiable framework for the dynamics. First-order effects are responsible for binding. These are high-energy, usually called chemical. Second-order effects are low-energy, usually called physical. When the two scales are separated, physical models can neglect the underlying chemistry, the paradigm for which are elemental metals. Functional materials are the opposite case. Because the physical scales are directly affected by the much larger chemical ones, the sensitivity of the physical properties to nuances of preparation, doping in particular, is at the same time a challenge and an opportunity.

The difference between elemental and ionic metals is best illustrated by their different doping mechanisms. In the former, doping proceeds by alloying. The dopant electrons delocalize on equal footing with those of the host atoms, which is why alloying does not appear to be a chemical reaction, although chemical bonds between different atom species are formed in substitutional alloys. Another well-known doping  mechanism is overdoping (degeneration) of semiconductors, which leads to their metallization. In that case, localized impurity orbitals overlap when the dopant concentration becomes too high, creating a new impurity band in a band gap of the host. This mechanism is not pertinent to the materials we consider here~\cite{Bozin05}.

In cuprates, doping appears to be essentially \emph{ionic}. In the original proposal of this mechanism~\cite{Mazumdar89} and in further studies~\cite{Barlingay90,Pelc15,Mazumdar18}, the dopant electrons are assumed to remain localized in the vicinity of their host, while their Coulomb field in the dielectric layers induces an orbital transition in the copper-oxide planes, leading to the transfer of a hole from copper to oxygen:
\begin{equation}
\mathrm{Cu}^{2+} + \mathrm{O}^{2-} \rightleftharpoons \mathrm{Cu}^{+} + \mathrm{O}^{-}.
\label{orbtrans}
\end{equation}
The ensuing hole on the oxygen then delocalizes via (as argued below, indirect) O $2p_x$--$2p_y$ hopping between the two ($x$ and $y$) O orbitals in the planar CuO$_2$ unit cell, providing the carriers which metallize the planes. The chemical potential, shifted by long-range Coulomb fields in the dielectric layer, affects both the concentration of mobile carriers and the balance of the reaction~(\ref{orbtrans}). In this scenario, localized and delocalized holes in the copper-oxygen planes are in thermodynamic equilibrium.

It cannot be said in general how much the net charge of the conducting planes changes upon doping, because particular studies indicate that the answer depends on the material. In the original ionic-doping picture outlined above, it does not change at all~\cite{Mazumdar89}. This picture may be relevant in cuprates when doped through interstitial oxygen. In a realistic calculation~\cite{Lee06}, the charge of the interstitial O remained localized in its vicinity, creating a peroxide O$_2^{2-}$ with the apical oxygen that was initially in the O$^{2-}$ closed-shell configuration, thus not changing the overall charge at all, so the ionic doping effect is triggered by the same localized charge $-2$ being smeared out over a larger region. On the other hand, a calculation in the pnictides uncovered a more subtle cumulative phase-shift effect of substitutional dopants on the conducting sector, in which net charge was transferred~\cite{Berlijn12}. Finally, a direct measurement of the effect of electrostatic gating on NdBa$_2$Cu$_3$O$_7$ found a transfer of holes from CuO chains to CuO$_2$ planes~\cite{Salluzzo08}, clearly proving that doping was due to the dielectric response of the insulating sector to the gating field, without net injection of carriers into the sample, thus validating the main idea of the ionic doping scenario even in this case. In brief, individual bonds cannot distinguish between external and internal fields that shift all their chemical equilibria, not only the one in Eq.~(\ref{orbtrans}), simultaneously and with varying effect on the net charge of the CuO$_2$ planes. As we shall see in Eq.~(\ref{chargebalance}) below, their influence on the effective \emph{mobile} charge is even more elaborate.

Self-doping, often described in chemistry by non-integer oxidation numbers,  may thus provide a broadly applicable paradigm of metalization in ionic functional materials, given that such oxidation states are common in ionically bound compounds. Self-doping means that there is no clear separation between closed orbitals in the valence band and open orbitals in the conduction band. Instead, there is some charge transfer between partially open orbitals in the stoichiometric material. For example, in insulating murunskite the chemical potential is found not to lie in the middle of the insulating gap, which has been attributed to the breaking of particle-hole symmetry by charge transfer to the ligands~\cite{Tolj21}. That could also be the case in cuprates, where the charge-transfer gap is asymmetric between particles and holes because the Cu$^{2+}$--Cu$^{3+}$ ($3d^9$--$3d^8$) fluctuation around the equilibrium $3d^9$ configuration is suppressed, due to the high energy of the triply-ionized Cu$^{3+}$, while the closed-shell $3d^{10}$ (Cu$^+$) is energetically close to the chemical potential~\cite{Zaanen85,Mazumdar89}. (Spectroscopists sometimes call the low-energy $3d^{10}$ signal ``upper Hubbard band''~\cite{Montanaro24}, which may be confusing to theoreticians, who invariably apply that term to $3d^8$, whose high energy is modelled by the on-site \emph{hole} repulsion $U_d$ in the tight-binding approach~\cite{Emery87}.) An extreme case of this asymmetry is self-doping to metallicity, where the chemical potential moves into a band. That situation is exceptional in the cuprates, though it is reported in the stoichiometric superconductor  Y$_2$Ba$_4$Cu$_8$O$_{16}$ (YBCO-248, T$_c=73.2$~K~\cite{Martinez91}). Self-doping encompasses any mixture of in-plane and out-of-plane charge transfers, with or without dopants.

In the pnictides, the self-doping paradigm has a twist. Their metallicity is due to the direct overlap of Fe $3d$ $t_{2g}$ orbitals~\cite{Fink19}, without involving the ligands, so one may well ask, how is that different from the metallic bond in elemental Fe. The answer is that elemental Fe contributes the $4s$ electrons to the metallic binding, while they are involved in the ionic binding in the pnictides. Hence, self-doping in the latter has the effect that the orbitals involved in the metallicity are not the same as in elemental Fe. The phenomenological difference is that the electron bands in the pnictides show effects of intra-orbital repulsion that can be modelled by the multi-band Hubbard model, but without ferromagnetism. A similar situation occurs in the SC nickelates~\cite{Wang24}, which we do not consider here.

The balance between binding and functionality reflects an underlying quantum-mechanical competition, visible already in an elementary diagonalization of a $2\times 2$ Hamiltonian. In the two solutions for the energy, the square root appears with opposing signs, usually called ``bonding'' and ``antibonding.'' Compounds are most stable when orbitals are closed, because opening them requires promoting some electrons into less bound states. On the other hand, functionality requires some to be open, because electrons need to be manipulated easily. Thus, the physically functional ones are chemically the ``odd ones out.'' To reveal where they are, and what they are doing, is to understand the material from both points of view.

The competition between binding and functionality is dynamic and controlled by the chemical potential. In crystals, there is an even more basic competition between binding and geometry, which is kinematic. It is due simply to the fact that atoms are too large or too small to fill space neatly according to the chemically optimal crystal structure. The consequence is an instability towards deformation. Because of the inherent Coulomb instability of point-charge configurations, such geometric mismatches make ionically bound materials prone to short-range disorder. Geometry feeds back into chemistry, because any departure of a bond angle from its ideal (stoichiometric) value impacts the hybridization of orbitals, which impacts the physical functionality in turn.

Finally, one needs to distinguish crystallographic and interstitial positions of atoms in the crystal. Interstitials are in principle neutral impurities which migrate to ``weak,'' i.e., non-reactive positions (interstices) between atoms in the ``strong'' crystallographic positions that created the lattice by a network of chemical bonds. This distinction is complicated in the context of ionic doping, because the presence of interstitials affects the long-range Coulomb fields responsible for the self-doping of the material, thus changing the chemical balance between the constituent ions, as in Eq.~(\ref{orbtrans}) above. In cuprates, the apical oxygen appears in a wide range from strongly reactive to non-reactive, which can be inferred from how easy it is to displace. The non-reactive limit is achieved in some electron-doped compounds, where it can be completely removed by annealing at high temperature without other change in structure, but with AF replaced by SC all the way to stoichiometry, which was attributed to the ensuing reduced crystal-field splitting between the Cu $3d$ and O $2p$ orbitals~\cite{Tsukada05,Adachi13}.

\section{Cuprates in the underdoped regime.}

\subsection{Relevant orbitals.}

Cuprates encompass a wide variation of crystal structure and complexity, all built around copper-oxide planes as the universal functional element. The universality of their functional properties in the face of such chemical complexity is impressive~\cite{NBarisic13,Barisic15,Barisic19,Zhou03}: The only key macroscopic material-dependent property is related to superconductivity (SC), specifically, it is the value of T$_c$ itself. We proceed to describe them along the lines sketched above.

There are generally six orbitals in the vicinity of the Fermi level, which participate in the charge transfers responsible for binding, and also in the conductivity emergent upon doping~\cite{Andersen01,Pavarini01}. Three are on copper: $3d_{x^2-y^2}$, $3d_{z^2}$, and $4s$. Copper is nominally twice-ionized, Cu$^{2+}$, relative to the atomic configuration $3d^{10}4s^1$, so the $4s$ orbital is empty, while the $3d$ orbital has a single hole ($3d^{9}$ configuration) which resides in the $3d_{x^2-y^2}$ suborbital. Triple ionization (Cu$^{3+}$) is energetically prohibitive, which is encoded in tight-binding models by a large (Hubbard) repulsion $U_d\sim 7$--$8$~eV, which dynamically prevents two holes from occupying the same d orbital ($3d^{8}$ configuration). The next two orbitals are O $2p_{x,y}$ orbitals belonging to the two oxygen atoms per unit cell in the Cu--O plane, while the sixth is the O $2p_z$ orbital belonging to the apical oxygen, if present. All three are closed (full, O$^{2-}$) in the stoichiometric compounds.

The symmetries, overlaps, and energies of these orbitals work together uniquely to open a path for SC in the cuprates. First, the Cu $4s$, $3d_{z^2}$, and O $2p_z$ orbitals present to the Cu--O plane as a single axially symmetric effective orbital that we denote by $s$ to distinguish it from the real Cu $4s$, so the six-band model can be reduced to four bands without loss of chemical realism. In the ensuing planar four-band model, the $s$ orbital is the only one which couples (implicitly) to the third dimension.

The effective $s$ orbital plays a surprisingly active role both in the antiferromagnetic (AF) parent cuprates and in the doped conducting materials, despite the large $4s$--$2p$ crystal-field splitting, $6<\Delta_{ps}< 16$~eV. The reason is a very large Cu $4s$--O $2p_{x,y}$ overlap $t_{ps}\sim 2$~eV. It gives rise to an effective O $2p_x$--$2p_y$ hopping via the second-order virtual O--Cu--O hopping, which can be estimated as $t_{ps}^2/\Delta_{ps}\sim 0.3$--$0.7$~eV, leading to a significant upward renormalization of the native (chemical) $2p_x$--$2p_y$ overlap $t_{pp}\sim 0.1$~eV to an effective value $t'_{pp}$ mostly around $0.7$~eV, judging by model fits to observed Fermi surfaces~\cite{Emery87,Qimiao90,Qimiao93,Mrkonjic03,Sunko07,Hashimoto08}.

Finally, the four-band model has a remarkable orbital symmetry, implying that carriers at the diagonal of the Brillouin zone (BZ) in cuprates (nodal carriers) are strictly 2D. Namely, when one inserts $k_x=k_y$ into the (fourth-order) secular equation of the model, it factorizes into two second-order equations, where the one that contains the open band does not depend on the $s$-orbital parameters~\cite{Andersen01,Lazic15}. Recalling that the $4s$ orbital couples the planar wave functions to the perpendicular direction, it follows that the wave functions at the BZ diagonal are orthogonal to the out-of-plane wave functions in Hilbert space. The remarkable feature of this two-dimensionality is that it evolves continuously towards three-dimensionality as one moves along the Fermi surface away from the diagonal, i.e., as the anti-nodal (van Hove, vH) points are approached, provided that the Cu $4s$ orbital is hybridized with the two out-of-plane orbitals. However, if it is not, then the whole of the Fermi surface (FS) is 2D. It has been noted that the less hybridized it is, i.e., the more the $s$ orbital is pure Cu $4s$, the higher the T$_c$~\cite{Pavarini01}. Thus, we conclude that SC in the cuprates is natively two-dimensional. This conclusion has historically been avoided because the well-known Hohenberg-Mermin-Wagner (HMW) theorem states that SC is thermodynamically unstable in 2D~\cite{Hohenberg67,Mermin66}. However, it has recently become clear that the HMW argument is irrelevant for 2D SC in samples smaller than the visible universe~\cite{Palle21}.

\subsection{Antiferromagnetism.}

Stoichiometric cuprates are AF insulators. While there is a long-standing consensus that the localization of the $3d^9$ hole in the half-filled Cu--O band is due to the large excitation energy (Hubbard $U_d$) of the $3d^8$ configuration, determining the mechanism behind the AF ordering has not been so straightforward. The physics of the AF interaction $J$ between the Cu $3d^9$ orbitals can be inferred from perturbation theory. The simplest idea is a direct exchange between coppers, as if the intervening oxygen were absent, yielding $J\sim t_{pd}^2/U_d$, where $t_{pd}\sim 1$~eV is the Cu~3d--O~2p overlap. But of course the oxygen is there, leading to a fork in the road. One can continue to look for dynamic mechanisms, i.e., involving $U_d$ in the denominator, such as $J\sim t_{pd}^2/(U_d+\Delta_{pd})$, where $\Delta_{pd}\sim 3$~eV is the Cu~3d--O~2p energy splitting. These are usually called ``superexchange'' in the literature~\cite{Anderson59,Zaanen85}. However, a purely kinematic superexchange was found to be responsible~\cite{Eskes93}, namely $J\sim t_{pd}^4/\Delta_{\mathrm{eff}}^3$, where $\Delta_{\mathrm{eff}}=\Delta_{pd}-4t'_{pp}$ and $t'_{pp}\sim 0.65$~eV is the effective O $2p_x$--$2p_y$ overlap, mentioned above. The factor $4$ in the renormalized value of the bare crystal-field splitting is an effect of coherence in degenerate perturbation theory, because the two planar oxygens have the same site energy.

This quantum coherence effect reduces the denominator considerably, so the fourth-order kinematic superexchange dominates lower-order dynamic superexchange in the charge-transfer limit $\Delta_{pd}\ll U_d$, pertinent for the cuprates, while dynamic mechanisms are suppressed by the large bare $U_d$ in the denominator. Thus, cuprates are charge-transfer, as opposed to Mott-Hubbard insulators, which are in the opposite limit $\Delta_{pd}\gg U_d$, where dynamic superexchange dominates~\cite{Zaanen85}.

Kinematic superexchange explains the significantly greater sensitivity of cuprate AF to hole than electron doping. In hole doping, the added holes open the oxygen orbitals. The kinematic mechanism requires the oxygen orbitals to be closed, because the two $3d^9$ holes on neighboring coppers need to hop simultaneously on the bridging oxygen in order to correlate antiferromagnetically. Because doped holes are delocalized and paramagnetic, they undermine AF correlation efficiently over the whole lattice. Thus, AF long-range order disappears already by $3$\% doping, but short-range AF correlations can persist as long as there are localized holes. In electron doping, the extra electrons close the Cu $3d$ orbital, but the ensuing still-localized $3d^{10}$ configuration only removes some spins from the lattice, allowing for conductivity presumably via Cu but leaving the long-range AF in place~\cite{Onose01,Onose04,Li16}. The electron-doped cuprates remain AF long-range ordered until about $13$\% doping, after which both hole-like Fermi-surface and SC appear simultaneously~\cite{Hirsch89,Barlingay90,Li19}. It is appealing to associate the appearance of the hole-like conductivity with the same metallization of O~$2p_{x,y}$ orbitals as in hole doping. Instabilities towards phase separation when synthesizing electron-doped cuprates near the AF/SC border~\cite{Fournier98} indicate that the underlying AF-to-SC metal transition, which populates the O $2p$ orbitals with mobile holes, is first order, as corroborated by neutron scattering~\cite{Li17}.

\subsection{Normal-state conductivity.}

It has long been noted that Fermi surfaces measured by ARPES could be fitted to the three-band model (without the $s$ orbital) only with values of the in-plane O $2p_x$--$2p_y$ hopping $t_{pp}$ that were unrealistically large from the chemical point of view~\cite{Qimiao93}. In a one-band approach, good fits were possible only by a similarly unrealistic second-neighbor coupling $t'$ in the so-called $t$--$t'$--$J$ model, which in addition had to change sign between hole- and electron-doped materials~\cite{Jiang21}. The simple explanation of this phenomenology by the invariable presence of second-order hopping via the $s$ orbital~\cite{Pavarini01} is direct evidence that the latter is a universal feature~\cite{NBarisic13,Barisic15,Barisic19,Barisic22} of conductivity in the cuprates.

This insight helps understand the remarkable universality of the normal-state conductivity, which is summarized and discussed in greater detail in~\cite{Barisic22}. Briefly, it was shown that the sheet resistance is universal~\cite{NBarisic13} and, in the pseudogap regime, exhibits $T^2$ behavior, while the Hall effect correctly measures the carrier density~\cite{Ando04}. This indicates that $p$ mobile carriers display Fermi-liquid behavior, whereas exactly one hole per CuO$_2$ unit remains localized~\cite{NBarisic13,Barisic15,Barisic19,Pelc19,Barisic22}. The Fermi-liquid nature of these carriers was unambiguously confirmed by demonstrations of textbook scaling, including Kohler's rule~\cite{Chan14} and the scaling of the optical scattering rate~\cite{Mirzaei12,Kumar23}. On the overdoped side of the phase diagram, where $1+p$ carriers form a Fermi liquid, the quadratic slope of the sheet resistance appears to follow precisely the evolution of the carrier density~\cite{NBarisic13}. This led to the conclusion that the gradual delocalization of the previously localized hole begins around optimal doping and continues until the overdoped regime. However, the most remarkable normal state property of cuprates is the Hall mobility, which remains unchanged across the normal state of cuprates, being essentially independent of doping level or compound~\cite{Barisic15,Barisic19,Li16}. It cannot be emphasized enough that the measured value of the coefficient $C_2$ in its quadratic temperature dependence $1/\mu_H=C_2T^2$~\cite{Chien91} is essentially the same in electron and hole, underdoped, optimally doped, and overdoped cuprates, thus it is universal~\cite{Barisic15,Barisic19,Li16}. These experimental facts imply that the scattering rate (as well as the effective mass) is indeed universal, which in turn enables precise determination of carrier evolution, as well as the number of localized holes, across the phase diagram of cuprates~\cite{Pelc19,NBarisic15,Barisic19,Barisic22}.

Similar conclusions regarding the doping evolution of the carrier density were later drawn from Hall effect measurements. Initial YBCO studies~\cite{Badoux16} suggested a sharp transition from $p$ to $1+p$, but subsequent investigations on other compounds, including corrections for YBCO chain contributions, confirmed that the Hall number~\cite{Putzke21}, later associated with the carrier evolution~\cite{Nicholls25}, agrees precisely with the evolution determined from resistivity~\cite{Pelc19,Barisic22}. 

More recently, it was shown that optical conductivity measurements are fully consistent with the conclusions drawn from transport, directly confirming the Fermi-liquid scaling of the scattering rate, not only in the underdoped Fermi-liquid regime but also in the strange metal regime~\cite{Kumar23}. Moreover, as a spectroscopic probe, optical conductivity directly measures the spectral weight shifts associated with the localized hole as a function of doping and temperature, unambiguously revealing its presence and evolution. 

A particularly neat demonstration of the validity of the above results and conclusions comes from the analysis of the Hall coefficient in LSCO~\cite{Klebel23}, which exhibits an atypical evolution of the Fermi surface, passing through a Lifschitz transition well documented by ARPES~\cite{Zhong22}. Taking into account the specific Fermi surface in the presence of Fermi arcs, both the Hall effect and resistivity were calculated directly, showing excellent agreement with the measured values. These results clarify the previously observed large deviations of the Hall effect from the expected universal values~\cite{Ando04,Gorkov06}, demonstrating that they can be fully understood simply by taking into account the actually observed shape of the Fermi surface from case to case. An important methodological implication of this finding is that, before invoking exotic concepts to explain deviations from universal behavior, one should carefully analyze Fermi-surface effects within the standard FL paradigm in the particular system under consideration~\cite{Barisic15,Barisic19,Klebel23,Zhong22}. 

Another remarkable property of cuprates is that, in clean  Hg1201~\cite{NBarisic13} and YBCO~\cite{Rullier-Albenque08} within the underdoped regime, the residual resistivity is negligible. Moreover, quantum oscillations have been observed~\cite{NBarisic13a,Doiron-Leyraud07}. Both findings are surprising, since cuprates are highly disordered, leading to a heterogeneous and percolative localization of exactly one charge per CuO$_2$ unit~\cite{Barisic15,Barisic19,Pelc19,Barisic22}. Superconductivity and the insulating state, in which this single charge---presumably localized on Cu---plays a key role, also emerge through a percolative process~\cite{Pelc18,Popcevic18,Li16,Vuckovic25}. Finally, electron-phonon scattering, which should appear at a fraction of the Debye temperature, is likewise absent in cuprates, implying that the conducting channel is decoupled from local disorder. The simplest explanation of these observations is that the $3d$ orbital does not participate in conductivity (as expected due to the large on-site Coulomb repulsion $U_d$), leaving only the indirect hopping via the $s$ orbital, and direct via the $t_{pp}$ overlap, the latter too small to account for the data. Conversely, the $4s$ orbital is always empty, and the $t_{ps}$ overlap always large, across the cuprates, independently of doping. The only material-dependent property affecting hopping via the $4s$ orbital is the crystal-field splitting $\Delta_{ps}$, ranging from $6$--$16$~eV, but even on the high end, the second-order process is at least twice the value of the direct overlap. Furthermore, the bandwidth for O--O hopping benefits from a large coherence factor related to the degeneracy of the two oxygens in the unit cell~\cite{Mrkonjic03}, as already remarked above in the context of AF. The physical regime for normal Fermi-liquid (FL) conductivity only requires a bandwidth sufficient for a sharp Fermi edge deep ($E_F/kT\gg 1$) inside the band, a condition satisfied for all realistic parametrizations of both four- and three-band models, making the low-energy sector insensitive to the precise value of $\Delta_{ps}$. While that is not proof by itself that the value of $C_2$ should be the same everywhere, an issue currently under investigation, it suffices to establish a major separation of roles of the Cu orbitals in the cuprates: The $4s$ orbital is responsible for universal FL conductivity, the $3d$ orbital for the material-dependent properties influenced by strong correlations, including tuning the SC T$_c$.

\begin{figure}[t]
\begin{center}
\includegraphics[width=0.9\linewidth]{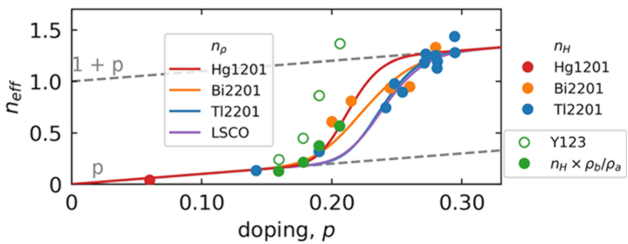}
\end{center}
\caption{The $T=0$~K evolution of $n_{\mathrm{eff}}$ in various cuprates, extrapolated from resistivity: $n_{\rho}$ (full lines~\cite{Barisic15,Barisic19,Pelc19}), and from Hall coefcient: $n_H$ (circles~\cite{Putzke21}), clearly transitioning from $p$ to $1 + p$ from under- to overdoping, with $n_{\mathrm{eff}}\approx p$ up to optimal doping. The apparent discontinuity of the open green circles in Y123~\cite{Badoux16} disappears when chain anisotropy is taken into account (full green
circles~\cite{Putzke21,Nicholls25}) and a full agreement with evolution determined from resistivity is achieved. From Ref.~\cite{Barisic22}.
\label{figptooneplusp}}
\end{figure}

The evolution of the normal-state conductivity with doping in the cuprates is summarized in the simple charge-balance equation
\begin{equation}
1+p=n_{\mathrm{eff}} + n_{\mathrm{loc}},
\label{chargebalance}
\end{equation}
where $p$ is the concentration of holes introduced by dopants, $n_{\mathrm{eff}}$ is the mobile charge, and $n_{\mathrm{loc}}$ is the localized charge. Because the doping is known and the mobile charge can be observed directly in transport (DC or Hall) measurements~\cite{Barisic15,Barisic19,Pelc19,Nicholls25,Ando04}, the concentration of localized charge is also known~\cite{Pelc19}. The $T=0$~K evolution of the mobile charge in doping is shown in Fig.~\ref{figptooneplusp}. The strongly nonlinear dependence of carrier concentration on doping, but linear-like in temperature, resolves the ``strange metal'' issues in the cuprates within the Fermi-liquid paradigm while retaining a constant (universal) scattering rate~\cite{NBarisic13,Mirzaei13,Barisic15,Barisic19,Klebel23}.

\subsection{Fermi arcs.\label{arcs}}

A long-standing challenge for the simple band (FL) picture of cuprates has been the observation of Fermi arcs, parts of the FS around the diagonal of the BZ that neither close on themselves nor touch the edge of the zone~\cite{Norman95,Lee07-1,Kunisada20}. Because such a FS is impossible in a one-body theory, numerous explanations have been proposed which invoke interactions between the mobile carriers, or that there is a Fermi-surface reconstruction (FSR) driven by a collective mode, either AF~\cite{Fang22}, charge-density wave (CDW)~\cite{Tabis21}, or structural~\cite{Beck25}. Here, we show how a Fermi arc appears already in a simple one-body model of a charge-transfer gap.

A DFT calculation in LSCO obtained Fermi arcs growing with doping~\cite{Lazic15}, in apparent contradiction with the one-body nature of DFT, which should presumably always give rise to closed Fermi surfaces, either around the $(0,0)$ point (electron-like) or around the $(\pi,\pi)$ point (hole-like). The reason was identified at the time: The calculation treated Sr-doping realistically, with the Sr atoms placed physically in a large unit cell, which corresponded to a small (supercell) Brillouin zone. In this small zone, the Fermi surfaces were closed, as required. The arcs appeared upon unfolding the small zone onto the large zone of the (undoped) primitive CuO$_2$ unit cell, using a projection procedure, well established for disordered alloys~\cite{Popescu12}.

\begin{figure}[t]
\begin{tabular}{lclcl}
(a) & \phantom{xx} & (b) & \phantom{xx} & (c) \\
\raisebox{3mm}{\includegraphics[width=0.24\linewidth]{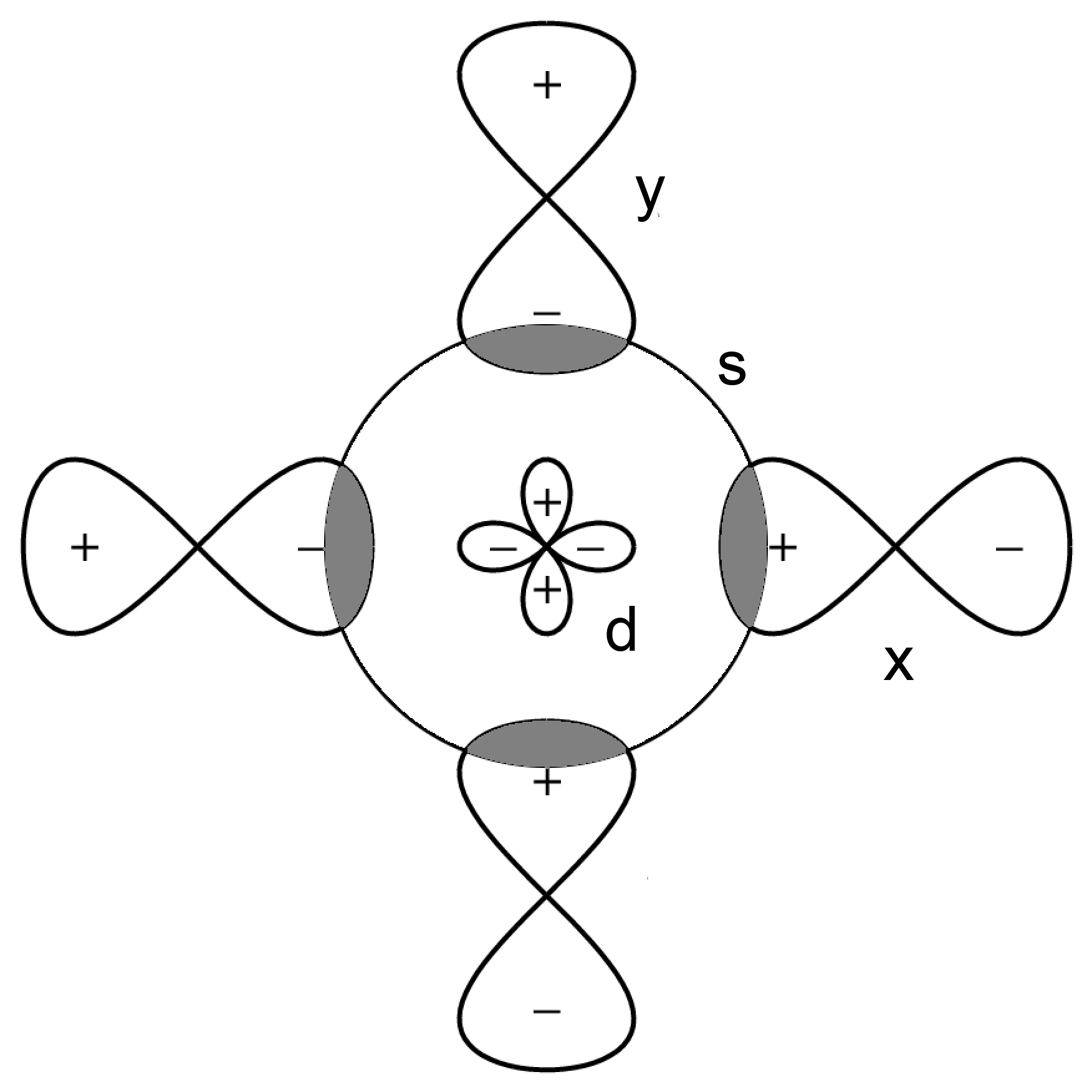}} & \phantom{x} &
\includegraphics[width=0.24\linewidth]{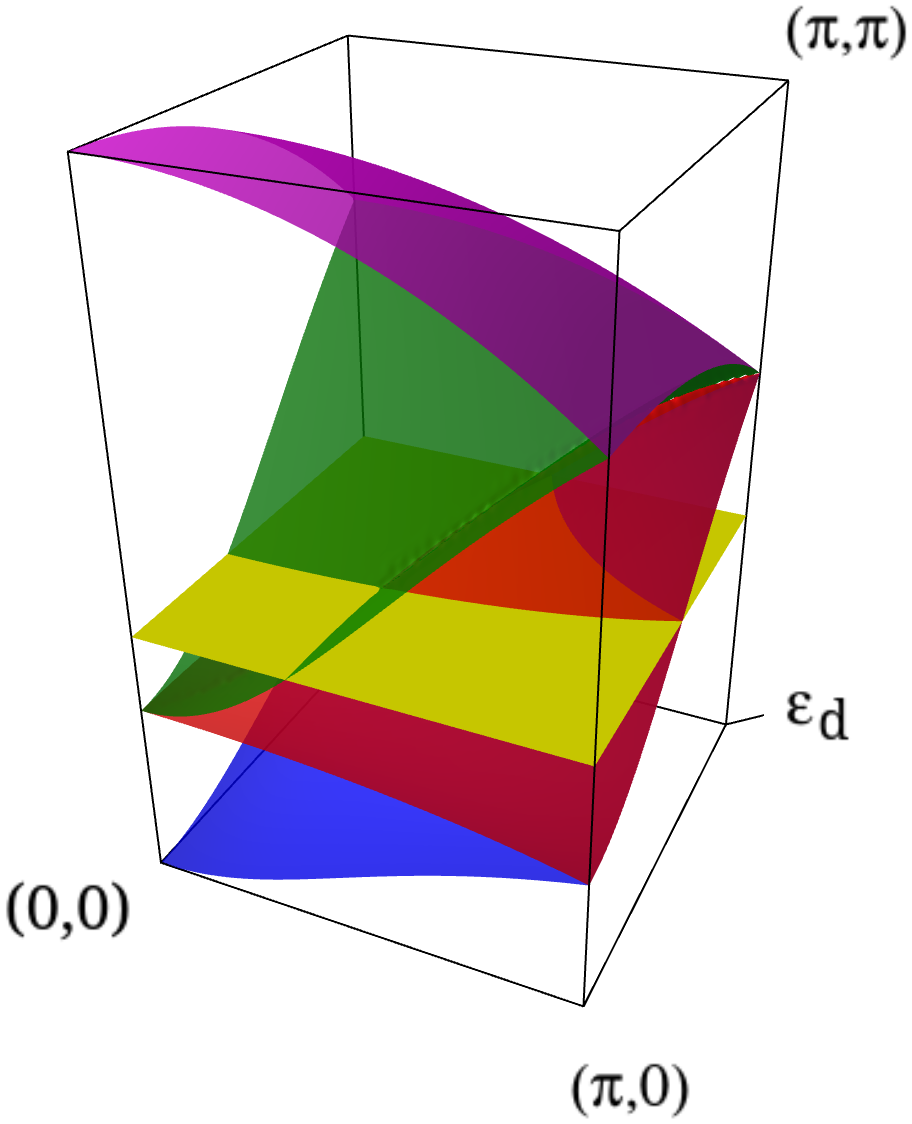} & \phantom{x} &
\includegraphics[width=0.24\linewidth]{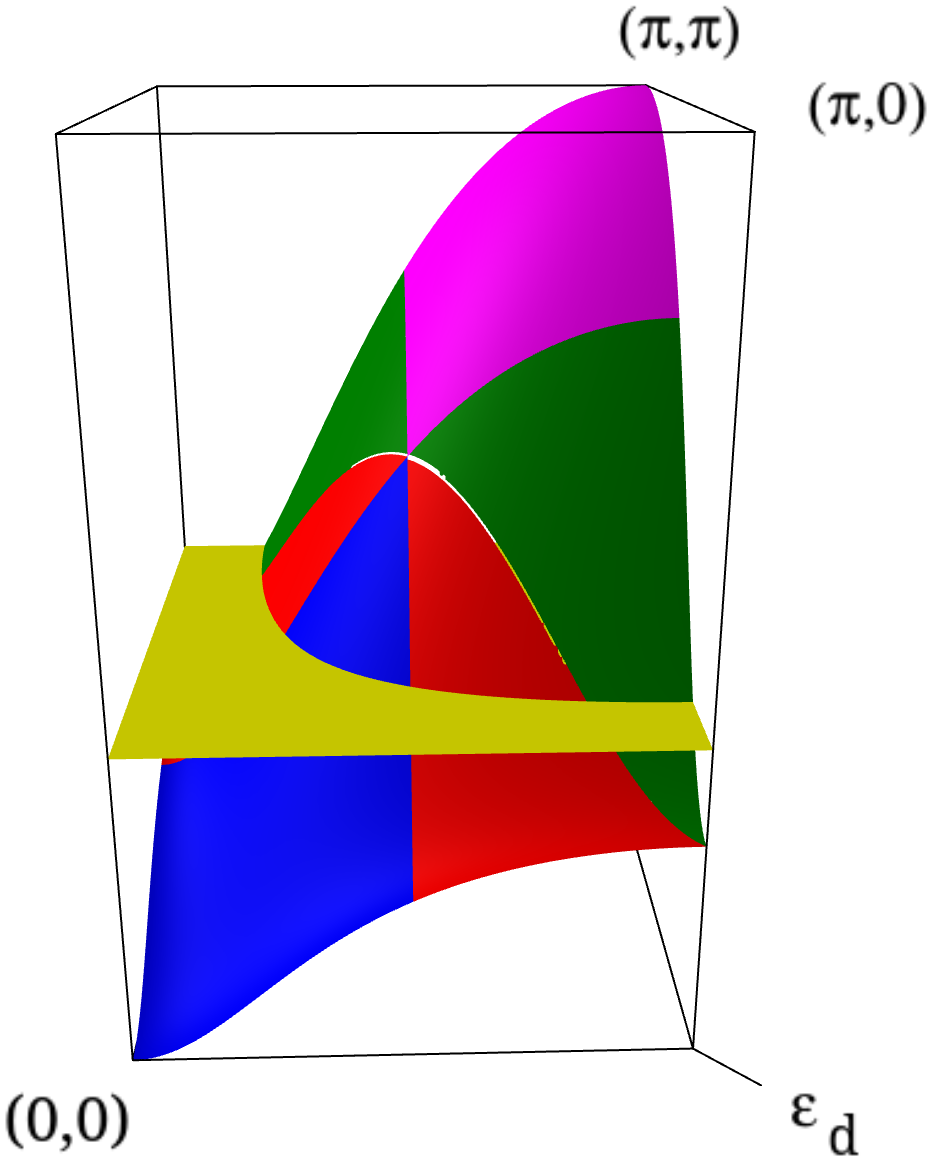} \\
(d) & \phantom{x} & (e) & \phantom{x} & (f) \\
\includegraphics[width=0.29\linewidth]{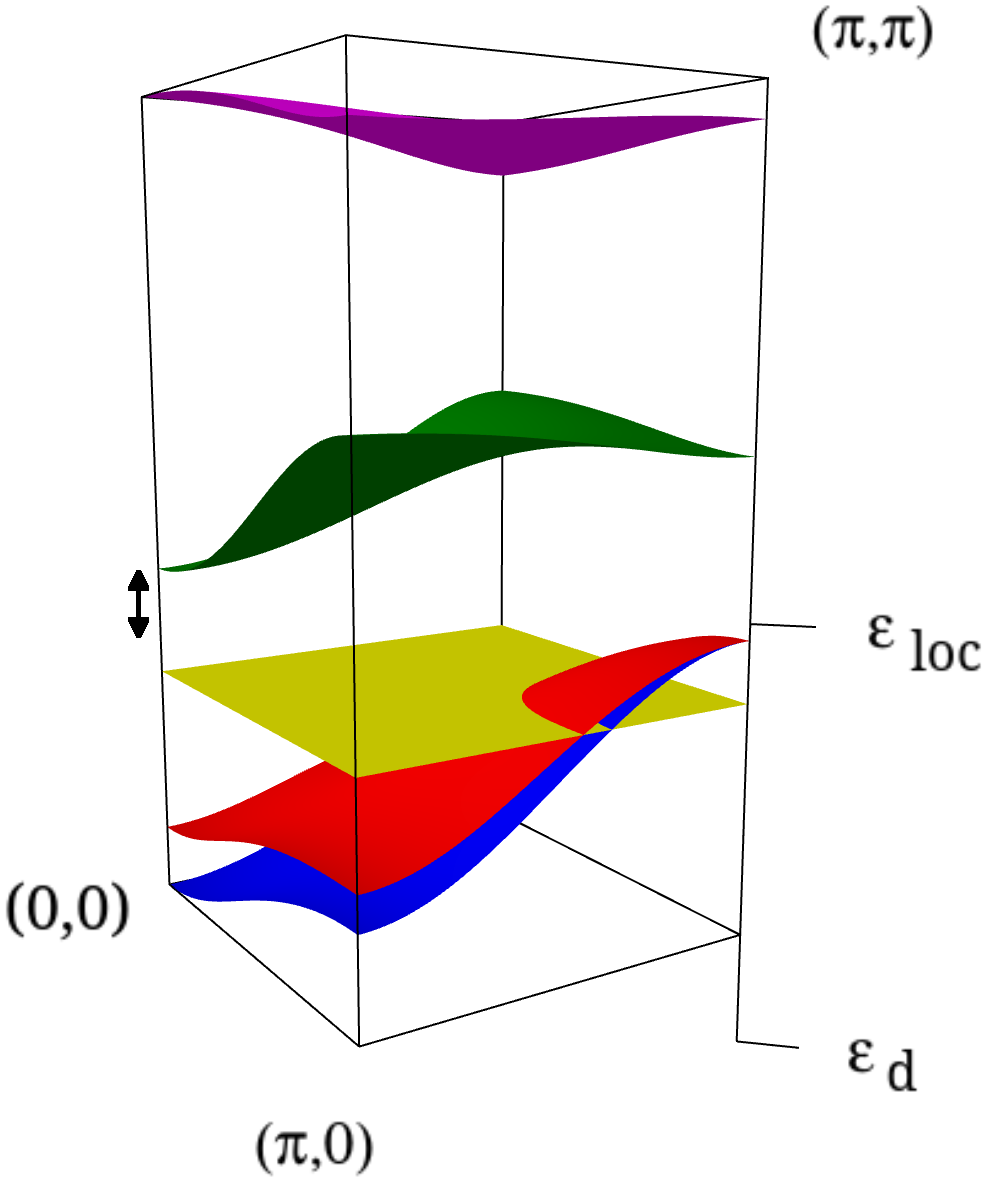} & \phantom{x} &
\includegraphics[width=0.24\linewidth]{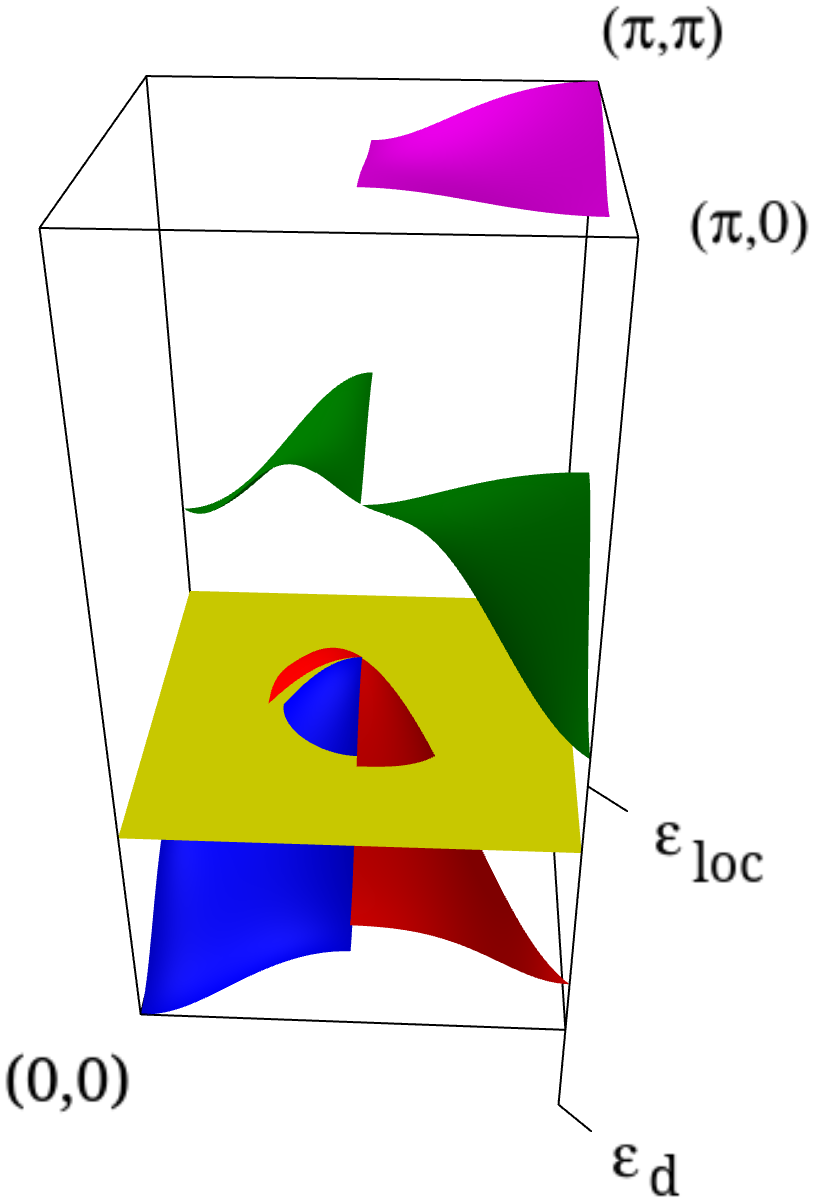} & \phantom{x} &
\includegraphics[width=0.24\linewidth]{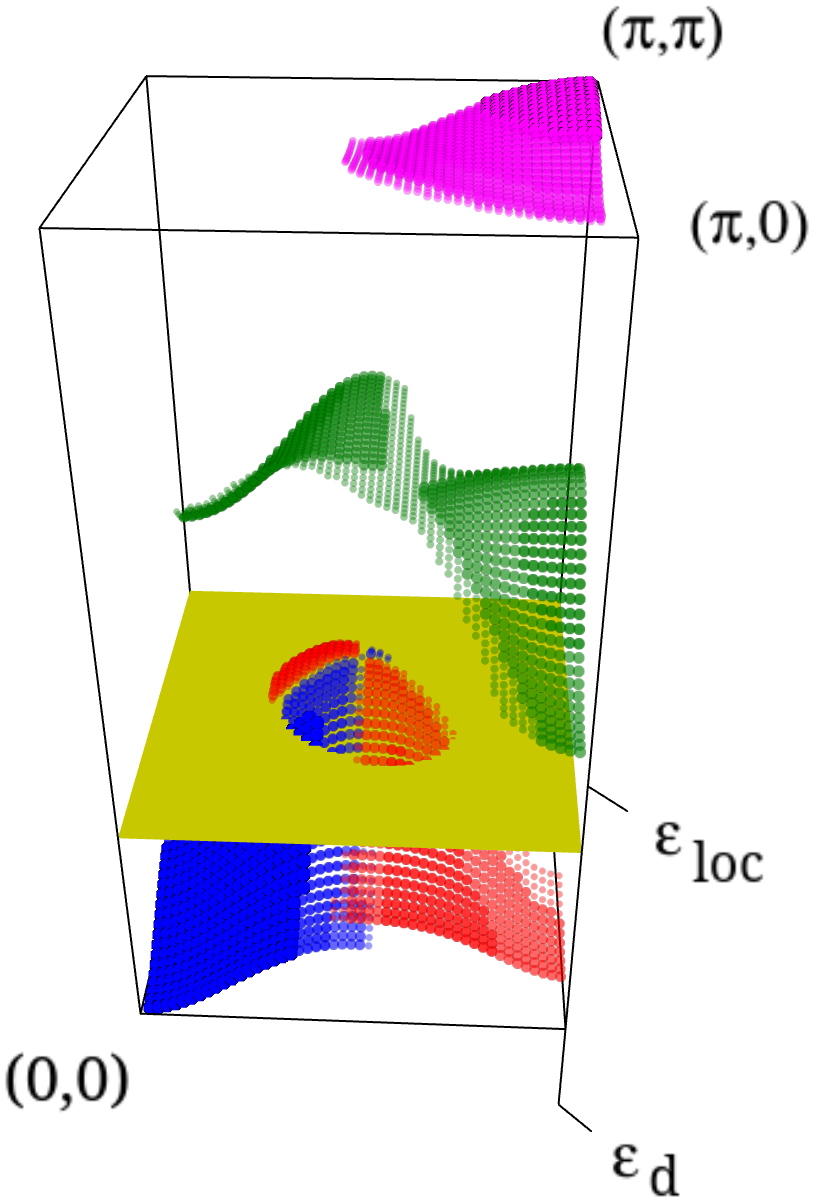}
\end{tabular}
\caption{(a) The four-band model of Eq.~(\ref{ham}). The overlaps $t_{ps}$ are marked by gray patches. (b) The four $dp$ bands of the $2\times 2$ supercell model~(\ref{ham}) in the scBZ, with the Fermi level corresponding to $15$\% hole doping marked in yellow. (c) The same bands unfolded into the pcBZ by geometric translations. The ridge separating the red and green segments connects the van Hove points of the pcBZ. (d) The four $dp$ bands of the $2\times 2$ supercell model~(\ref{ham}) in the scBZ, with a hole localized in one of the unit cells. The green band is above the Fermi level now. The double arrow marks the charge-transfer gap at half-filling. (e) The same bands unfolded into the pcBZ by geometric translations, as in (c). While the green band is gapped, the surviving Fermi surfaces of the blue and red segments are arc-like, stopping abruptly at the antidiagonal. (f) The correct projection onto the pcBZ of the schematic situation in (e). Points with less than $30$\% weight are not shown. The energy $\varepsilon_{\mathrm{loc}}$ is the energy of the localized hole in the $d$ orbital, here higher than $\varepsilon_d$ because the figure is in the electron picture.
\label{fignonint}}
\end{figure}

We construct a minimal model of that effect. The goal is to illustrate the underlying mechanism, not to reproduce the previous realistic results. A Fermi arc with the correct filling fraction $p$ out of $1+p$ emerges as soon as a hole is localized in a purely one-body tight-binding model, without explicit interactions. All local Coulomb effects are modeled through the site energies. We invoke the four-band model:
\begin{eqnarray}
H = 
\sum_{\alpha, r}\varepsilon_\alpha a_{\alpha, r}^\dagger 
a_{\alpha, r}^{\vphantom{\dagger}} + &
t_{ps}
\sum_{\left<rr'\right>\atop p=x, y} \phi_{ps}
a_{p, r}^\dagger a_{s, r'}^{\vphantom{\dagger}}
+t_{pd}
\sum_{\left<rr'\right>\atop p=x, y} \phi_{pd}
a_{p, r}^\dagger a_{d, r'}^{\vphantom{\dagger}}
\nonumber\\&
+ t_{pp}
\sum_{\left<rr'\right>} \phi_{xy}
a_{x, r}^\dagger a_{y, r'}^{\vphantom{\dagger}}
+ \mathrm{c.c.}
,
\label{ham}
\end{eqnarray}
where $\alpha=s,d,x,y$ are the $s$, Cu $3d$, O $2p_x$, and O $2p_y$ orbitals in the standard topology of the cuprate CuO$_2$ plane (Fig.~\ref{fignonint}a), with $\phi_{\alpha\beta}=\pm 1$ the phases of the respective orbital overlaps. Only nearest-neighbor overlaps appear, of which the largest is $t_{ps}= 2$~eV, while $t_{pd}= 1$~eV and $t_{pp} = 0.1$~eV. The site energies are in the electron picture: $\varepsilon_s=3$~eV, $\varepsilon_d=0$, and $\varepsilon_p=-3$~eV.

In accord with this local chemistry, the two oxygen bands are bonding (valence), the Cu~$3d$--O~$2p$ ($dp$) band is open at the Fermi level (conduction), while the $s$ band is antibonding (empty). To model a hole localized in one unit cell, we create a $2\times 2$ supercell (sc), with the corresponding small BZ (scBZ) containing $16$ bands. As long as all supercell parameters are the same, four of them are just the folded $dp$ band of the initial model, so they can be unfolded by purely geometric translations to recover it in the large BZ of the primitive cell (pcBZ), as shown in Fig.~\ref{fignonint}b--c.

Even such a simple remapping carries information which was not available in the original pcBZ. Namely, the variously colored four bands in the scBZ become colored segments of the single $dp$ band in the pcBZ, meaning that any reparametrization of the supercell model, affecting these bands, will affect the various segments of the dispersion surface in the pcBZ differently. In other words, one can manipulate these unfolded segments individually by the parameters of the supercell tight-binding model. The borders between the variously colored leaves are lines where gaps can appear. Notably, the demarcation line between the red and green sections in Fig.~\ref{fignonint}b is precisely the antidiagonal of the pcBZ.

We turn to the simplest model of one localized hole in the supercell. We reparametrize one of the four unit cells (plaquettes) by $\varepsilon_d=0\to \varepsilon_{\mathrm{loc}}=1$~eV and $t_{pd}=1\to t_{\mathrm{loc}}=1.2$~eV, and the one diagonally across from it by $\varepsilon_d\to \varepsilon_{\mathrm{loc}}/2$. The idea is for a single plaquette to become a CuO$_4$ ``molecule'' housing a hole, with some effect on a neighboring plaquette, like in a local tilt. Fig.~\ref{fignonint}d shows the four $dp$ bands corresponding to Fig.~\ref{fignonint}a. They are affected differently. The key observation is that the green band is above the Fermi level now, so a charge-transfer gap has appeared in the scBZ at half-filling in the pcBZ, because the lower two bands in the former correspond to half a band in the latter. To see the effect of the projection qualitatively in the pcBZ, we perform the same translations as in Fig.~\ref{fignonint}b, shown in Fig.~\ref{fignonint}e. The surviving segments of the Fermi surface, belonging to the blue and red bands in the scBZ, now appear as arcs, which is the main result of this section.

Of course, the geometric translations used in Fig.~\ref{fignonint}e are not the correct way to unfold the scBZ into the pcBZ when the supercell plaquettes are not all the same. The projection onto the pcBZ is a change of basis, such that all $16$ bands in the scBZ now appear at each $k$-point in the pcBZ, but with different weights~\cite{Popescu12}. The exact unfolding is shown in Fig.~\ref{fignonint}f, with the weight of each point suggested by size and opaqueness.

\begin{figure}[t]
\begin{tabular}{lclclc}
(a) & \phantom{xx} & (b) & \phantom{xx} & (c) \\
\includegraphics[width=0.25\linewidth]{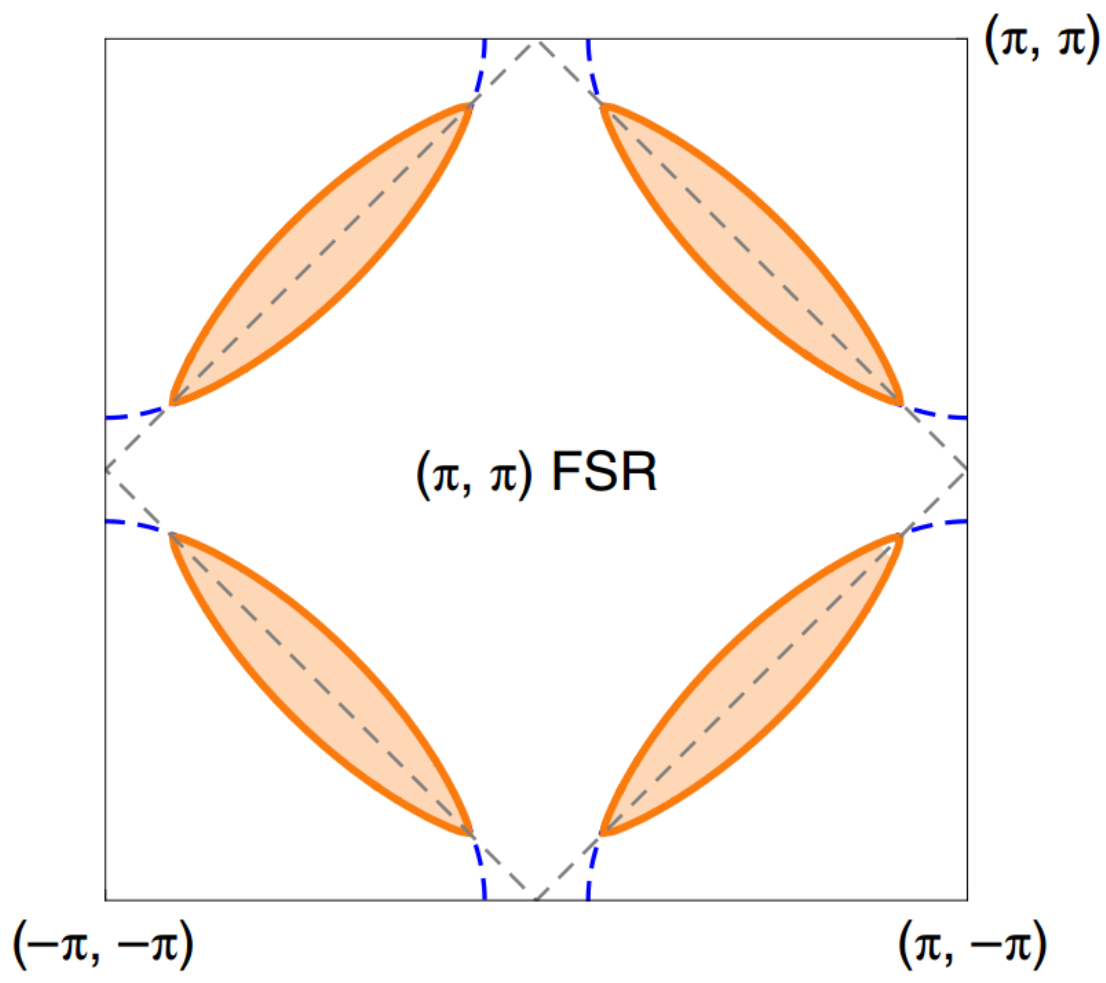} & \phantom{x} &
\includegraphics[width=0.25\linewidth]{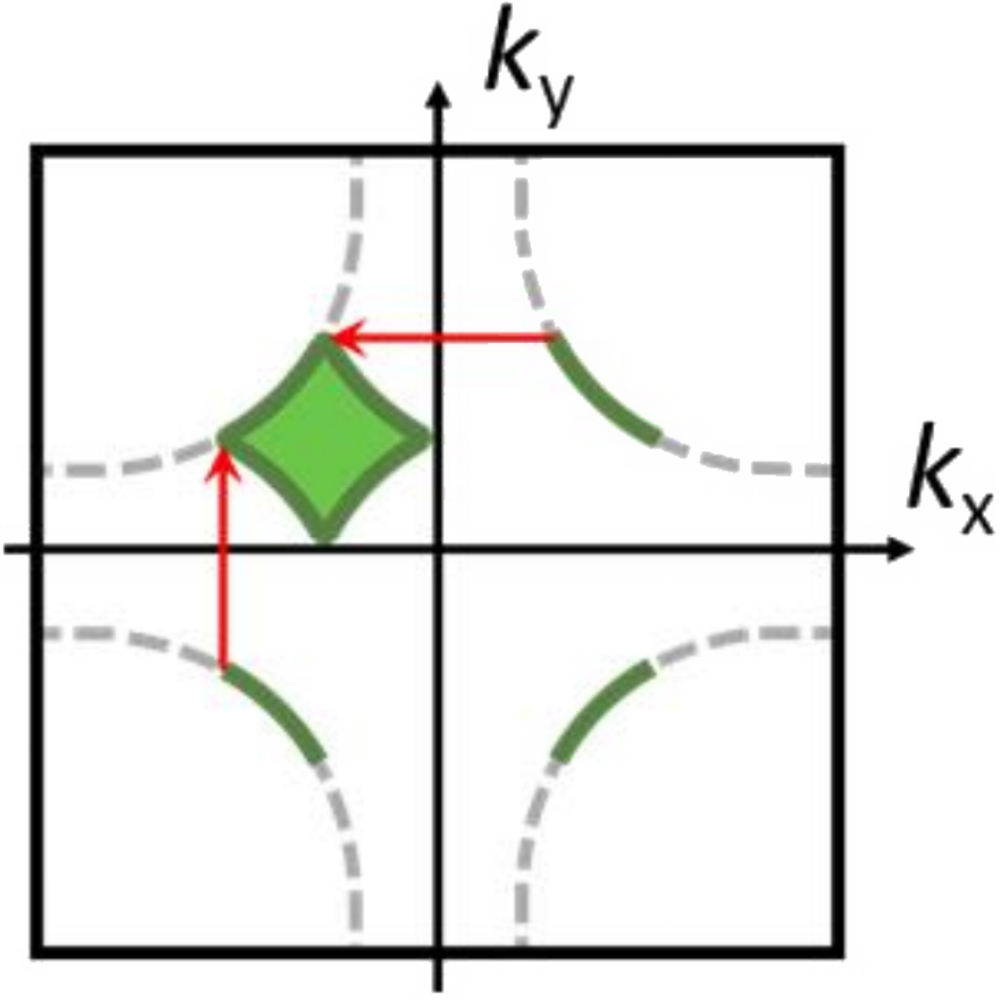} & \phantom{x} &
\raisebox{-3mm}{\includegraphics[width=0.24\linewidth]{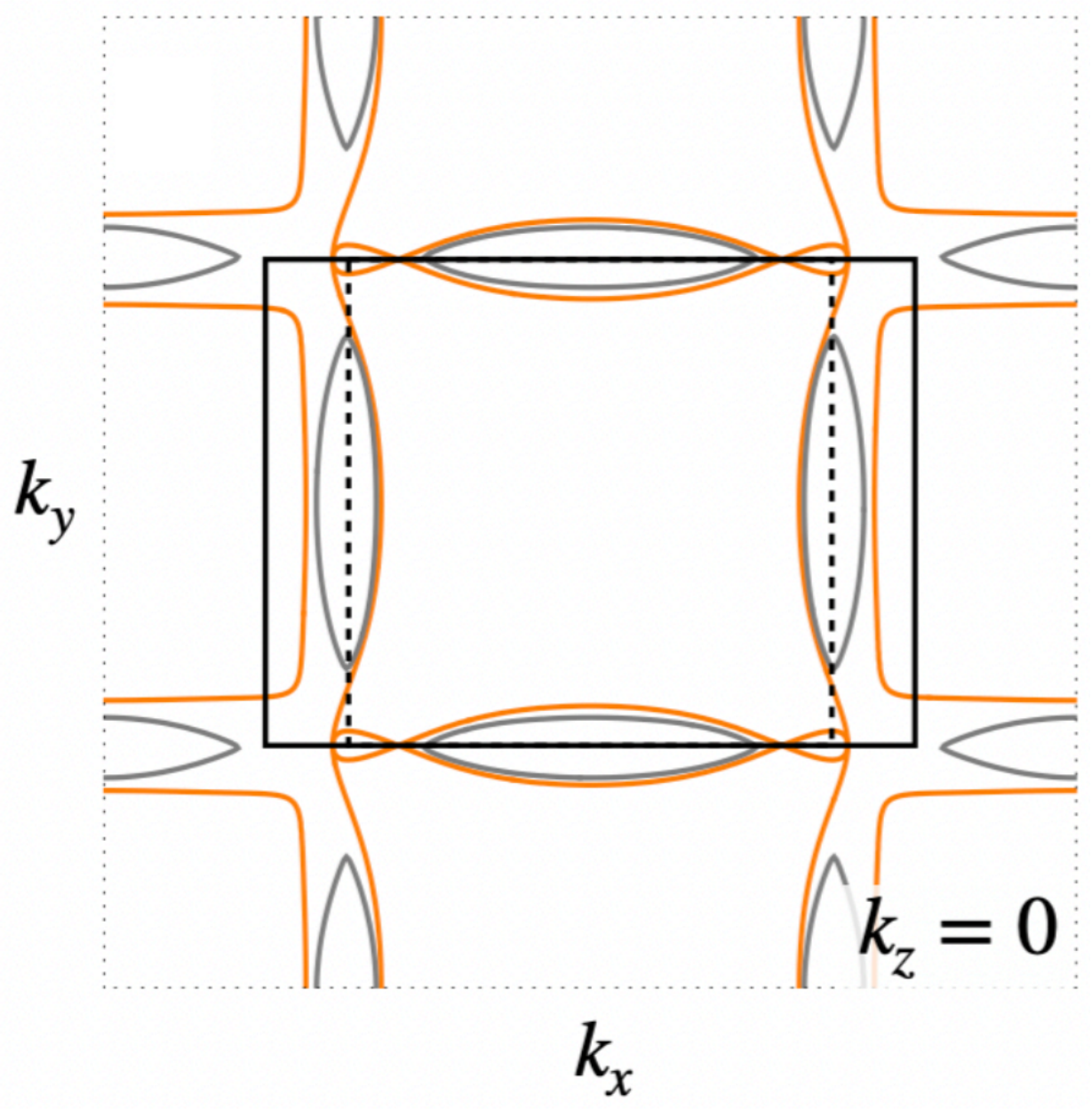}} \\
(d) & \phantom{xx} & (e) & \phantom{xx} & (f) \\
\includegraphics[width=0.25\linewidth]{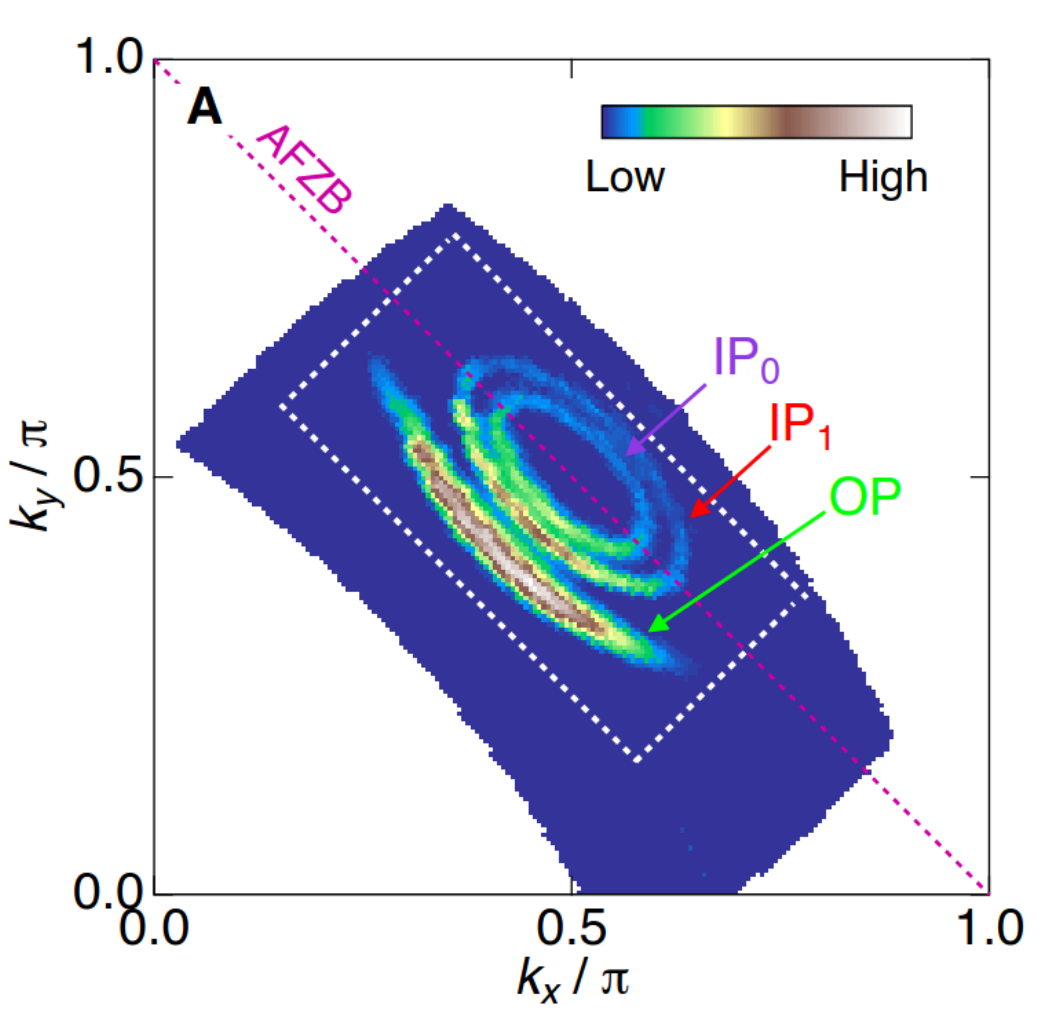} & \phantom{x} &
\raisebox{3.5mm}{\includegraphics[width=0.25\linewidth]{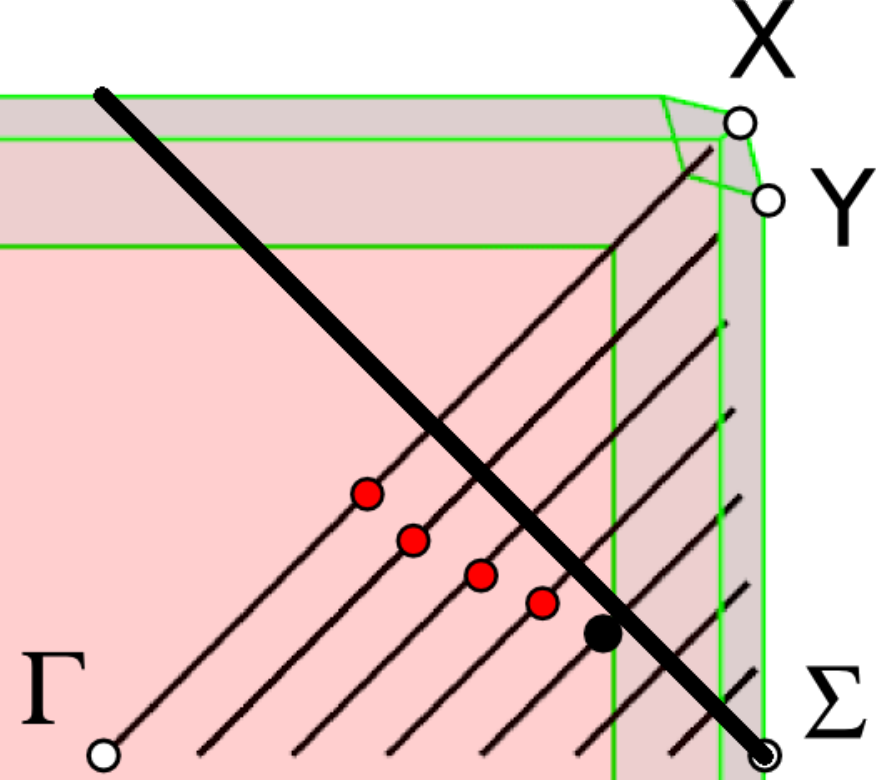}} & \phantom{x} &
\raisebox{1.5mm}{\includegraphics[width=0.26\linewidth]{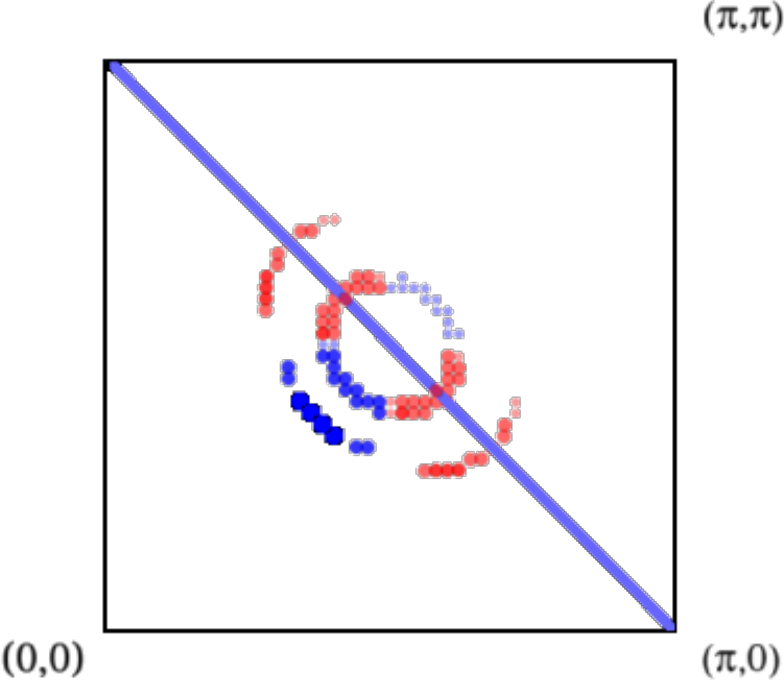}}
\end{tabular}
\caption{Top row: Various Fermi-surface reconstructions proposed in the literature. (a) Electronic, AF~\cite{Fang22}. (b) Electronic, CDW~\cite{Tabis21}. (c) Structural, LTO~\cite{Beck25}. (d) Measured ARPES response from a five-layer cuprate with different doping levels in the layers~\cite{Kunisada20}. The optimally doped plane gives rise to an arc that does not touch the antidiagonal, while the underdoped planes give rise to asymmetric pockets enclosing it. (e) One-body DFT+U calculation with dopant disorder~\cite{Lazic15}. Red dots: Fermi surface crossings not reaching the antidiagonal (black line). Black dot: Gap onset. (f) Open Fermi surface: Same as in Fig.~\ref{fignonint}f. Pocket: 5\% doping with parameter evolution, such that the plaquette with $\varepsilon_d=\varepsilon_{\mathrm{loc}}/2\to \varepsilon_d=\varepsilon_{\mathrm{loc}}$.
\label{figarcs}}
\end{figure}

The main effect of the projection in Fig.~\ref{fignonint}f is a realistic softening of the schematic cartoon in Fig.~\ref{fignonint}e. However, all the principal features are carried over by points of high weight ($>30\%$). There are in addition ``symmetry shadows,'' mirrored across the antidiagonal with much lower probability, not visible in the figure which has a cutoff below $30$\% weight. If the cutoff is set at close to $50$\%, the signature feature of an arc stopping at the antidiagonal of the zone is always preserved, but, depending on the parametrization, as low as $10$\% can be enough. Experimentally, sometimes the arc does not even touch the antidiagonal~\cite{Kunisada20}, so the Fermi surface is not reconstructed, which was also captured in a more elaborate one-body calculation that included dopant disorder~\cite{Lazic15}, as shown in Fig.~\ref{figarcs}.

The abrupt vertical shifts of the various colored leaves in Fig.~\ref{fignonint} are reminiscent of the ``waterfalls'' in ARPES, so we conjecture that the latter are also manifestations of large local chemical scales in the band picture, thus of similar provenance as the arcs, but far from the Fermi level. Such drastic effects are possible when the chemical scales related to local disorder are much larger than the physical scales related to conductivity, but much smaller than the binding scales that created the lattice, so lattice disorder can easily appear, affecting the metallicity without chemical change. That parameter regime is characteristic of octahedral tilts in cuprates, and of tetrahedral deformations in pnictides, but is not general among ionic metals. For example, delafossites crystallize spontaneously in so well-ordered lattices that studying disorder requires introducing it intentionally~\cite{Sunko19,Usui19}. We attribute this difference to the apical oxygen being strongly bound in the delafossites, while it is weakly bound or even interstitial~\cite{Tsukada05,Adachi13} in the cuprates. In murunskite, perfect crystal order persists in spite of occupational and orbital disorder~\cite{Tolj21,Tolj25}, as we shall elaborate below.

Because the arc is a merely a projection effect of the change of basis, it is equivalent to the original (scBZ) Fermi surfaces of the supercell model with localized holes. In other words, the model remains the same simple one-body model, no matter which of two equivalent ways of looking at it, Fig.~\ref{fignonint}d or Fig.~\ref{fignonint}f, one prefers. The scBZ is not observed experimentally because the plaquettes with localized CuO$_4$ ``molecules'' are randomly distributed, which is the standard interpretation for such projection effects in disordered alloys, where, similarly, only the pcBZ is measured in ARPES~\cite{Popescu12}.

A number of FSR scenarios proposed in the literature, some~\cite{Fang22,Tabis21,Beck25} shown in Fig.~\ref{figarcs}a--c, when compared with reconstructions indicated experimentally in specific regimes~\cite{Comin14,Tabis21,Fang22}, do not by themselves explain arc-like Fermi surfaces, in which regions of the Fermi surface are already gapped. When reconstructing the full (ungapped) Fermi surface, it is not obvious how to avoid generating multiple pockets that are not observed, aside from the mirrored arcs. Namely, by definition, an FSR creates a new, smaller zone, implying that the Fermi surface viewed in the old zone should be symmetric with respect to the boundaries of the new zone, which seems to be the case with CDW reconstruction~\cite{Comin14,Tabis14,Tabis17} but does not seem to be with AF reconstruction, where it is common to observe arcs that do not even touch the boundary~\cite{Kunisada20}. Moreover, in the cuprates, even when a closed pocket is observed~\cite{Kunisada20}, its intensity is asymmetric across the would-be zone boundary. Usually, this issue is addressed theoretically by claiming that one is on the high-temperature side of the purported reconstruction, but with already developed long-range fluctuations corresponding to the low-temperature phase~\cite{Yang06}. Such an approach gives the required asymmetry, but implies strong correlations among the conducting electrons, mediated by soft bosons related to the impending transition. Matrix-element effects are a more practical alternative conjecture, but it is excluded when one observes an arc together with pockets in the same sample~\cite{Kunisada20}.

The simple local-disorder approach with unfolding into the pcBZ can in principle reproduce the whole range of observations, as shown in Fig.~\ref{figarcs}e--f. Notably, even when the arc closes on itself to create the pocket, the intensities on the two sides of the antidiagonal are not equal. Indeed, to make them equal would require perfectly consistent modulation of all parameters, which is clearly not realistic in an ionic setting. The separate blue and red segments allow for situations in which the blue segment appears as an arc that does not touch the antidiagonal of the zone. Further manipulation of the $s$-orbital energy and the overlap $t_{ps}$ gives rise to pcBZ Fermi surfaces with long straight segments, like rounded squares centered on the $(\pi,\pi)$ point, which can provide the CDW nesting wave vector. Such multi-band simulations display a combination of rigidity and flexibility, which reflects how the particular chemical situation determines the physical content. Conversely, one-band models with more than nearest-neighbor hoppings are efficient in similar modelling simply because Fourier series can easily fit smooth curves.

While the above modelling of local Coulomb effects through partial reparametrization of the supercell plaquettes was purely schematic, the salient kinematic effect is robust. It only depends on \emph{four} bands in the scBZ playing the role of \emph{one} in the pcBZ, so that a gap between the upper and lower two in the former is at half-filling in the latter. Once such a gap is obtained by any convenient variation of parameters, simple charge-conservation counting ensures that the projected arc in the pcBZ accounts for $p$ mobile charges, if the total hole filling in the PC band is $1+p$. Tight-binding reparametrizations, required for a realistic description of Fermi-surface evolution, would be much more involved, however, the earlier DFT+U calculation~\cite{Lazic15} is proof that they are possible, because it implemented the same physics realistically in large supercells, with dopant effects calculated self-consistently. In that more complete approach, certain artefacts of the schematic tight-binding model do not appear, in particular, the symmetry-shadows, suppressed by a cutoff above, are unobservable. An additional step would be to average over the local disorder explicitly, whence the scBZ would cease to be observable even theoretically~\cite{Berlijn12}.

\section{From cuprates to pnictides.}

\subsection{Quantum kinematics and dynamics in Hund's rule.}

Hund's rule describes the order in which atomic orbitals are filled in the gas phase. It states that (a) every orbital in a sublevel is singly occupied before any orbital is doubly occupied, and (b) the electrons in singly occupied orbitals maximize the total spin. This rule is remarkably without exception across the periodic table. Unfortunately, the first attempt to explain it, by Slater in 1929~\cite{Slater29}, was shown to be incorrect only in 1965, and then by quantum chemists~\cite{Davidson65}, so it persisted in the perception of theoretical physicists even in the 1990s. 

Slater's explanation was plausible at face value. Two electrons in the same orbital have to be of opposite spins by the Pauli principle, i.e.\ in the antisymmetric spin state. Because their mutual Coulomb repulsion is maximal in the symmetric orbital state, it is optimal for them to be in an antisymmetric orbital state instead, which requires the spin state to be symmetric, i.e., of maximal spin.

The correct explanation is quite different. Valence electrons in the antisymmetric orbital state do not screen each other from the net positive charge of the closed-orbital core ion. The unscreened Coulomb attraction between them and the core shrinks the atom, resulting in an energy gain. Hund's rule is a many-body kinematic effect selected by aggregate core-valence attraction, not a two-body dynamic effect selected by constraint to a single orbital~\cite{Yamanaka05}.

In brief, the size of the atom is a variational parameter. Textbook expositions~\cite{Mattis65} of Slater's argument all follow his hidden assumption that the one-body orbitals comprising the two-electron wave function have the same Bohr radius in the symmetric and antisymmetric configurations. Relaxing this condition, one can even obtain Hund's rule in helium in a reduced calculation, in which only the $s$-wave (isotropic) components of the competing orbital wave functions are retained~\cite{Sajeev08}.

The last result shows that intraorbital repulsion is immaterial to the effect, even if it (misleadingly) gives the same result. Namely, if it were significant, then the \emph{angular} correlations of the two electrons in He would matter, because the effect of repulsion is to keep the two electrons on the opposite sides of the nucleus. These are neglected in the calculation reduced to $s$-waves~\cite{Sajeev08}, yet Hund's rule is still obtained.

\subsection{Mechanism of Hund's rule in ionic functional materials.}

In the solid state, the role of atomic size is played by the lattice constant. In ionic materials, its value is set by the overlaps between the valence orbitals of the metal cations and ligand anions. The key distinction to be made is the role of the ligands in the physical functionality. As always, this functionality must conform to the underlying chemistry, which controls the larger energy scales. The question is to what extent the two are independent.

In cuprates, the oxygen ligands both bridge the copper sites, and are responsible for the metallization of the doped holes. Hence, their physical and chemical roles cannot be separated. The lattice constant is adjusted by the copper-oxygen charge transfer, which is a central-field effect in the language of Hund's rule, while the conductivity occurs via O $2p$ holes. On the other hand, importantly, physics and chemistry are separated from the point of view of the coppers, where the $3d$ orbitals that account for the binding are localized, while metallic conduction proceeds dominantly by the empty $s$ orbital through its overlap with O $2p_{x,y}$ orbitals, as discussed above. The net effect is that the strong Coulomb interactions are played out in the insulating sector, while the conducting sector is an uncorrelated Fermi liquid.

In the following, we argue that in the pnictides, different subsets of Fe $3d$ orbitals are responsible for metallicity and binding. The $e_g$ orbitals bind to the ligands in a tetrahedral coordination, while the $t_{2g}$ orbitals overlap directly to provide conductivity. The ligands determine the binding, but not the functionality~\cite{Sunko20a}.

\subsection{Binding and correlations in the pnictides.}

In iron pnictides several bands cross the Fermi level. Due to hybridization of the various Fe $3d$ orbitals, one expects the local-orbital character of the bands to evolve with the wave-vector. A recent detailed assignment of the orbital symmetries at the Fermi energy in LiFeAs attributed all five observed Fermi surfaces to $t_{2g}$ symmetry~\cite{Fink19}. Thus, we are observing a division of roles among the local orbitals, such that the low-energy physical property (conductivity) is realized by the $t_{2g}$ orbitals, while the high-energy chemical property (binding) is relegated to the $e_g$ orbitals. This interpretation is corroborated by an earlier calculation in BaFeAs~\cite{Fink09}, which shows $t_{2g}$ bands dispersing through the Fermi level, the $d_{z^2}$ band deep below it, and the $d_{x^2-y^2}$ band starting at similarly high energy, but dispersing toward the $\Gamma$ point where it gives rise to a small Fermi surface. 

The orbital situation in the pnictides exceptionally vindicates Slater's original explanation of Hund's rule. Namely, the high-energy chemical binding by the $e_g$ orbitals means that the low-energy $t_{2g}$ orbitals do not set the lattice constant, so the physics in them plays out at an externally fixed atom size. Hence, the generally incorrect assumption, that the symmetric and antisymmetric orbital wave functions have the same effective Bohr radius, is specifically correct for the Fe $t_{2g}$ orbitals in the pnictides. The outcome is that angular correlations between the metallic valence electrons become important, which can be modelled by the multi-band Hubbard model with intraorbital repulsion. Thus, intra-orbital interactions appear in the metallic state, manifesting themselves as magnetic correlations up to second neighbors~\cite{Si16}.

While zeroth-order ARPES effects such as the arcs can be modelled without strong interactions among the remaining conducting electrons in the cuprates, the same is not true of the pnictides. In accord with the discussion above, the main qualitative effect is of the order of the residual interactions, $J\sim 0.2$~eV, not the bare Hubbard $U_d$. It is an isolated shift of a high-symmetry (van Hove) point in ARPES with respect to the best non-interacting fit. This shift could only be modelled by a combined DFT-DMFT calculation, confirming its origin in strong intraorbital repulsion, whose description originally motivated the development of the DMFT approach~\cite{Derondeau17}. Such a molecular-field approach is appropriate for the pnictides, because the AF fluctuations are fast, so the time-average of one site is equal to the space average of the sample, i.e., every site is representative of the material.

A more quantitative effect of the DMFT correction is a strong renormalization of the effective mass, which improves fits to the data significantly better for the electron pockets around the $M$ point than for the hole pockets around the $\Gamma$ point~\cite{Fink19}. There seems to be a separation of roles between the various orbitals, analogous to the one in cuprates, with the role of the localized hole possibly played by electrons whose mass is enhanced by AF correlations~\cite{Kumar23}. Indeed, some recent studies indicate that SC in the pnictides is orbital-selective, stressing the importance of these electron pockets in particular~\cite{Klemm24}.

\subsection{Dissipative features in the pnictide spectrum.}

In pnictides, it was argued early on that the smooth evolution with doping of the exponent $\alpha$ in the power-law temperature dependence $T^\alpha$ of resistivity is a signature of underlying quantum criticality. It changes from nearly $\alpha=2$ in the parent ordered compounds to $\alpha=1$ at the doping level with the highest T$_c$, and back to $\alpha=2$ at higher dopings. Such reasoning is also invoked whenever linear-like resistivity appears in the phase diagram above the maximum superconducting temperature (for example in cuprates). However, at least in iron pnictides, this scenario raises difficulties: it is unclear how to interpret the coefficient that changes smoothly from $2$ to $1$ in the ordered phase and then reverts smoothly to $2$ in the disordered phase~\cite{Gooch09,Jiang09,Dai15}. Moreover, conduction in a multiband system is never straightforward. In cuprates, this quantum-critical type of scenario is not consistent with experimental observations listed above (see also \cite{Barisic22}).

In an attempt to understand the multiband contributions, optical conductivity in 122 compounds was analyzed~\cite{Barisic10,Wu10}. Although the optical response remains quite featureless up to rather high energies, the first and simplest approach identified essentially two components. One of them is entirely temperature independent (incoherent) and can be modelled by a very broad Drude term, while the other captures the full temperature dependence of the system. At high temperatures it exhibits Drude-like behavior with an optical scattering rate proportional to $T^2$. Upon lowering the temperature and entering the ordered state, it either becomes partially gapped in the underdoped regime or develops superconductivity. 

What is particularly appealing about this decomposition is that two types of bands indeed exist in these systems: one flat, which should become incoherent once $kT$ exceeds the bandwidth, and others that are rather broad. Moreover, ARPES reveals two distinct superconducting gaps, again suggesting the presence of two different electronic subsystems. Finally, by extracting the temperature dependence of the resistivity of the narrow subsystem—simply by subtracting the temperature-independent contribution to the conductivity from the overall resistivity, which shows a power-law behavior with $\alpha<2$---a pure Fermi-liquid-like $T^2$ dependence was obtained in the temperature range from T$_c$ up to $300$~K~\cite{Barisic10,Wu10,Kumar23}. This establishes consistency between the optical conductivity data and resistivity.   

In the 111 compound, which is an intrinsic superconductor, ARPES does not reveal a flat band at the Fermi level. Instead, several bands cross it, and some of them form very shallow pockets~\cite{Fink19}. Therefore, one would expect that at low temperatures the resistivity exhibits no dissipative response. However, as the temperature increases and $k_BT$ exceeds the Fermi energy of a particular pocket, incoherence of that pocket should emerge. Indeed, similar to the cuprates in the pseudogap regime~\cite{Barisic22}, the resistivity in 111 compounds exhibits a quadratic temperature dependence below 150 K~\cite{Rullier-Albenque12}, the magnetoresistance obeys Kohler’s scaling~\cite{Rullier-Albenque12}, and the optical scattering rate follows Fermi-liquid scaling~\cite{Tytarenko15}. At higher temperatures, these behaviors deviate in a manner reminiscent of the ruthenates, where it has been argued~\cite{Kumar23} that the temperature marking the breakdown of low-temperature Fermi-liquid behavior corresponds to the Fermi energy of the pocket, as expected.

\section{Superconductivity in cuprates.}

In both cuprates and pnictides, the Fermi-liquid quasiparticles are the ones that become superconducting, most likely for different reasons, that is, through different mechanisms of Cooper pairing. Since cuprates exhibit the highest maximal transition temperature, they remain particularly intriguing. Here, we give a brief overview of a unique confluence of electronic features leading to high-T$_c$ SC that are at the same time universal across the cuprates. These are: (1) A localized hole interacting with a Fermi liquid of universal mobility; (2) degeneracy of planar O orbitals; (3) orbital overlaps and symmetries which facilitate 2D conductivity and SC.

\begin{figure}[t]
    \centering
\begin{tabular}{cc}
    \includegraphics[width=0.55\linewidth]{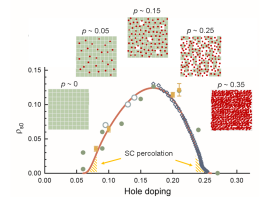} &
    \includegraphics[width=0.35\linewidth]{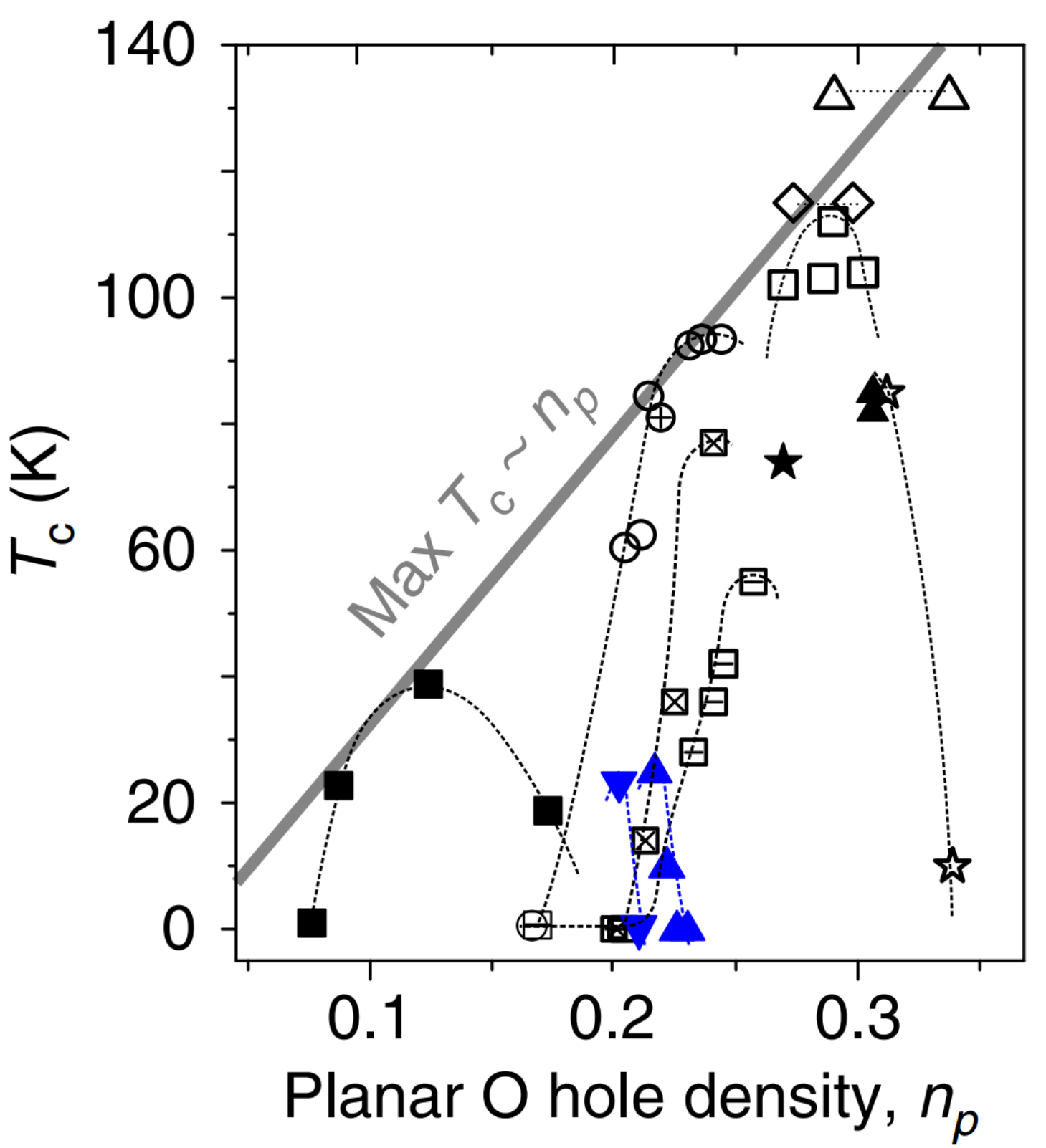}
\\ (a) & (b)
\end{tabular}
    \caption{(a) Evolution of the carrier density with doping. On the underdoped side, there are many localized holes (green squares) but few free carriers (red dots). On the overdoped side, it is the opposite. The symbols are measurements in LSCO. The full curve is Eq.~(\ref{rhoS}). The shaded areas are affected by percolative corrections. From Ref.~\cite{Pelc19}. (b) Evolution of T$_c$ with charge density on planar O sites $n_O$, called $2n_p$ in Ref.~\cite{Rybicki16}. From Ref.~\cite{Rybicki16}.}
    \label{figrhoS}
\end{figure}

\subsection{Localized hole as the SC glue.}

One can observe directly on the overdoped side of the phase diagram that SC vanishes concomitantly with the disappearance of the localized hole. The straightforward conjecture was that the localized hole is essential for the gluing mechanism of the itinerant charges. This led to the proposition that the superfluid density is nothing but the probability of itinerant charges to find a scattering center across which to pair (with percolative corrections at low superfluid densities~\cite{Pelc19,Barisic22}): 
\begin{equation}
\rho_S = n_{\mathrm{eff}} \cdot (O_S n_{\mathrm{loc}}),
\label{rhoS}
\end{equation}
where $n_{\mathrm{eff}}$ (and consequently $n_{\mathrm{loc}}$) are determined directly from the normal state as discussed above, while $O_S$ is a compound-dependent constant. Keeping in mind that the scattering rate and effective mass (Fermi velocity) are universal, the phenomenology of all key aspects of cuprates is fully captured with Eqs.~(\ref{chargebalance}) and~(\ref{rhoS}). Notably, the localized charge $n_{\mathrm{loc}}$ is responsible for all the ``strangeness'' of these compounds, including the pseudogap phenomenon and the superconducting glue. This simple approach captures the evolution of superfluid density across the superconducting dome without any fitting parameters, as shown in Fig.~\ref{figrhoS}~\cite{Pelc19}.

Finally, it is often argued in the literature that \emph{incoherent} carriers (sometimes called Planckian) are the ones that become superconducting. Optical conductivity provides a rather direct resolution of this question, as well as showing clearly the essentially localized nature of the non-FL ones~\cite{Kumar23}. Across the entire phase diagram, the spectral weight associated with the delta-peak at $\omega = 0$ originates from the coherent, FL-like carriers. In the underdoped regime, where the density of localized charge equals one, the corresponding spectral weight gives rise to Homes’ law. On the overdoped side, although the missing spectral weight scales with $n_{\mathrm{loc}}$ (sometimes referred to as $n_{\mathrm{incoh}}$, the incoherent component), it is again transferred from the itinerant part of the spectrum.

\subsection{Lattice symmetries and the SC mechanism.}

It was recognized quite early that the degeneracy of planar oxygens ($2p_x$ and $2p_y$) plays an important role in the SC mechanism. Based on analyses of the low-temperature tetragonal (LTT)–low-temperature orthorhombic (LTO)–high-temperature tetragonal (HTT) transitions in LSCO, it was concluded that Cu--O charge transfer stabilizes the LTO phase, still permitting O--O charge fluctuations. Static charge transfer between the two oxygen sites stabilizes the LTT phase, splitting the energies of the O $2p_x$ and $2p_y$ sites, in contrast to the HTT and LTO phases, where the two remain degenerate. Once the LTT phase is stabilized, the $2p_x$--$2p_y$ fluctuations are suppressed, while SC is (almost) completely lost, revealing the importance of these fluctuations for the SC~\cite{Barisic90}. However, it remained unclear why a mechanism involving these fluctuations would disappear on the overdoped side of the phase diagram.

A follow-up study~\cite{Barisic93,Tutis96} clarified the role of the large Hubbard repulsion $U_d$ in the Cu $3d$ orbital. It strongly suppresses the intracell $3d$--$2p$ fluctuations, while having no effect on the $2p_x$--$2p_y$ fluctuations. An even earlier study pointed out that the Hubbard $U_d$ would become ineffective as soon as the $3d$ orbital were covalently hybridized with the $2p_{x,y}$ orbitals, because that entails a significant spreading-out of the charge~\cite{Friedel88a}.

From the present point of view, it is the localized hole ionically bound to the Cu $3d$ orbital, with a spatial spread over the O $2p$ orbitals, that exhibits the behaviors described above, including providing the glue for superconductivity. How it spreads depends on its local structural environment (LTT, LTO, or HTT), where the superconducting mechanism requires unhindered $2p_x$–$2p_y$ fluctuations. This environment is manifested as bond-angle disorder, which is present in all the materials, but its actual realization (stripes, nematicity, complete randomness) is material-dependent, as is T$_c$ itself. The delocalization of the hole on the overdoped side corresponds to the evolution of the Cu–-O bond from ionic to covalent.

Yet another reason why oxygen degeneracy is essential for high T$_c$ is that Cooper pairing involves scattering between time-reversed pairs with zero total center-of-mass momentum~\cite{Mahan90}. This condition cannot be realized uniformly within a single unit cell if the oxygen sites are not degenerate.

\begin{figure}[t]
    \centering
\begin{tabular}{cc}
    \includegraphics[width=0.45\linewidth]{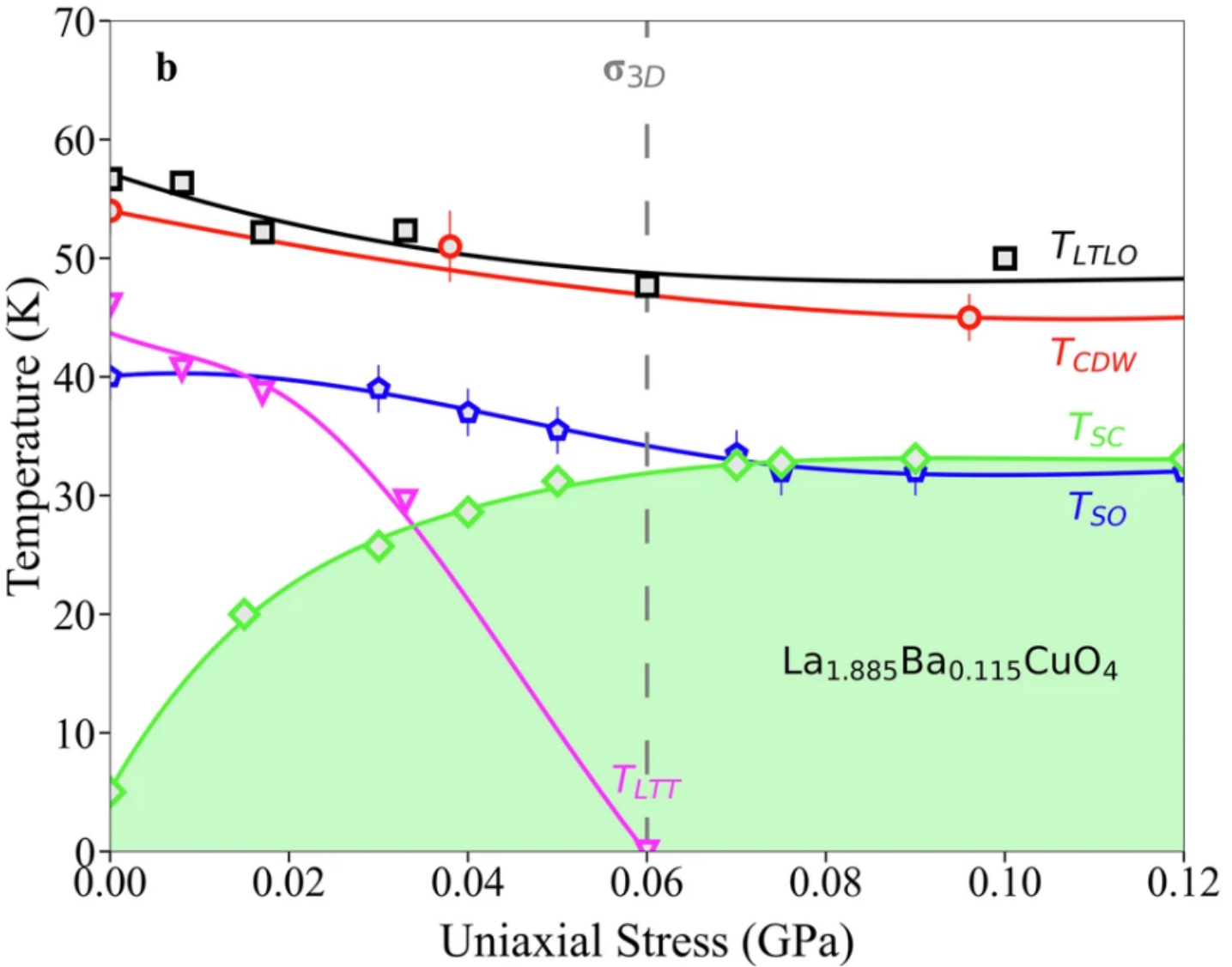}
&   \includegraphics[width=0.45\linewidth]{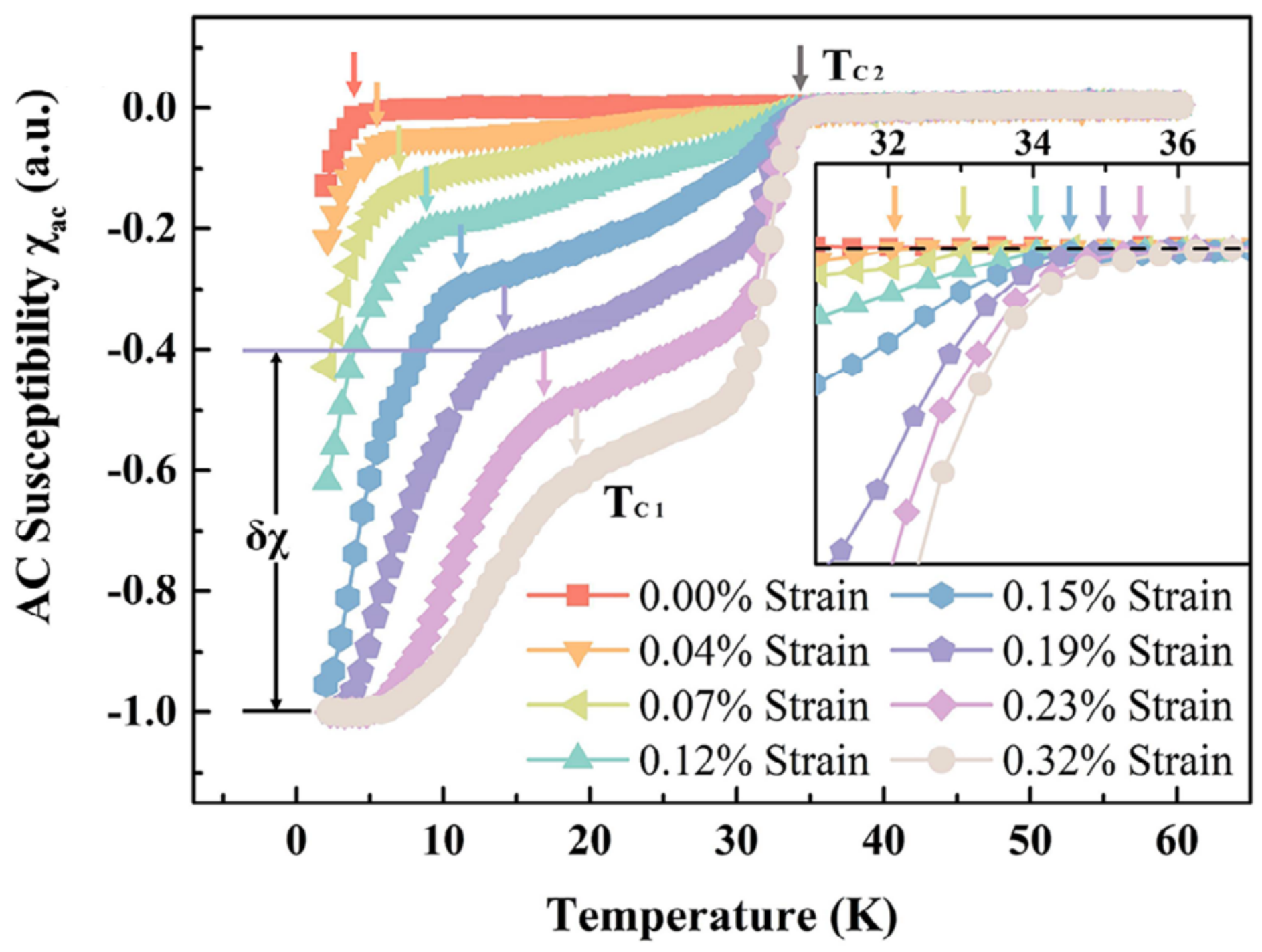}
\\
(a)  &  (b)
\end{tabular}
    \caption{(a) Recovery of SC ($T_{SC}$) under uniaxial strain which suppresses the LTT transition ($T_{LTT}$) in La$_{2-x}$Ba$_x$CuO$_4$ at doping $x=0.115$. All the other collective modes (charge density wave CDW, spin order SO and low-temperature less-orthorhombic order LTLO) are insensitive to SC. From Ref.~\cite{Thomarat24}.
    (b) Observation of continuous recovery of SC T$_c$ in La$_{2-x}$Ba$_x$CuO$_4$ at doping $x=0.125$ as the LTT tilt is suppressed. Note that there are two SC structural phases simultaneously present in the sample, of which only the LTT one is affected by the strain. From Ref.~\cite{Gao25}.}
    \label{figlttstrain}
\end{figure}

The detrimental effect of lifting the degeneracy of the two oxygen sites in the planar CuO$_2$ unit cell on superconductivity can be inferred from numerous experimental probes over decades, e.g., \cite{Axe89,Barisic90,Bianconi96,Lawler10,Pelc15,Massee20,Vinograd21,OMahony22,Thomarat24,Gao25}. A particularly elegant demonstration was recently achieved using uniaxial strain to remove the low-temperature tetragonal (LTT) tilt of O octahedra in La$_{2-x}$Ba$_x$CuO$_4$ at doping $x=0.115$, which suppresses the SC T$_c$ by breaking the oxygen degeneracy~\cite{Thomarat24}. It is clear from Fig.~\ref{figlttstrain}a that three other collective modes in the same material are insensitive to SC, while LTT is anticorrelated with it.

Similarly, it was found that the same compound separates structurally into stripes in which an LTT tilt is found, and those in which the O sites remained degenerate. Strain reduced the number of LTT stripes \emph{and} increased their SC T$_c$, while the T$_c$ of the O-degenerate phase remained unchanged, as can be seen in Fig.~\ref{figlttstrain}b~\cite{Gao25}. This observation resolves decades of controversy regarding stripes in cuprates, in particular with respect to alternative interpretations in terms of electronic, as opposed to structural, phase separation.

An immediate consequence of the above discussion is that, in compounds with higher T$_c$, the \emph{localized} hole also spreads onto the neighboring oxygen sites, which gives physical meaning to the coefficient $O_s$ in Eq.~(\ref{rhoS}). A probe that directly measures the average charge distribution between the Cu and O sites is $^{17}$O nuclear quadrupole resonance (NQR)~\cite{Rybicki16}:
\begin{equation}
n_{\mathrm{Cu}} + n_{\mathrm{O}}=1+p=n_{\mathrm{eff}} + n_{\mathrm{loc}},
\end{equation}
cf.\ Eq.~(\ref{chargebalance}). Because each of the four terms is known individually, one can infer how the localized hole is distributed between Cu and O sites on the underdoped side, where one can assume that the Cu $3d$ orbital is fully localized:
\begin{equation}
n_{\mathrm{Cu}}=n_{\mathrm{loc}} - \Omega,\quad
n_{\mathrm{O}}=n_{\mathrm{eff}} + \Omega.
\end{equation}
It was found that $n_O$ tracks T$_c$, as depicted in Fig.~\ref{figrhoS}b. However, more can be said. We observe that the SC domes in that figure, referring to different compounds, are rigidly shifted without much change in width. We know that the leftmost (lowest T$_c$) point in each dome corresponds to the doping $p$ in the underdoped regime, where $n_{\mathrm{eff}}=p$. Therefore, the rigid shift of the whole domes corresponds to increasing $\Omega$, the average density of the localized hole on the oxygens. We conclude that the gray line in the figure means that $\Omega$ tracks T$_c$, i.e., the more the \emph{localized} hole is shared between Cu and O, the higher the T$_c$. In this way, the origin of the material-dependent tuning parameter $O_S$ in Eq.~(\ref{rhoS}) is fully uncovered.

\subsection{Role of the relevant orbitals.}

SC in the the $d$-wave channel is not affected by the Coulomb repulsion between the O $2p_x$ and $2p_y$ orbitals, because the relevant matrix element vanishes by symmetry~\cite{Valkov16}. Furthermore, the carriers on the BZ diagonal $k_x=k_y$ are orthogonal to the $s$ orbital in Hilbert space, so the $s$--$p$ Fermi liquid, which we argue becomes superconducting, does not exist there. Both symmetry constraints allow $d$-wave SC by the $s$--$p$ carriers in 2D. More specifically, the effective $s$ orbital comprises the planar Cu $4s$ orbital and the out-of-plane apical O $2p_z$ and Cu $d_{z^2}$ orbitals. It was found~\cite{Pavarini01} that the SC T$_c$ was higher when the $s$ orbital was nearly pure $4s$, i.e., uncoupled from the third dimension, indicating that SC in the cuprates is natively 2D, as already mentioned above. Further along the same lines, pure $d$-wave SC is observed by surface probes, while bulk probes find a small $s$-wave component~\cite{Mueller97}. We conjecture that this component comes from pairs tunneling between planes via the $s$ orbital. The same confluence of orbital symmetries explains the very small residual resistivity when SC is removed by a magnetic field. Namely, both the $s$--$2p$ and $2p_x$--$2p_y$ channels are not affected by the disorder induced by the localized hole on the $3d_{x^2-z^2}$ orbital, as long as the $2p_{x,y}$ orbitals remain degenerate. The observation of Matthiessen's rule in this residual resistivity upon introducing point defects by irradiation (in the first place by displacing O atoms from their normal lattice sites) is direct evidence that the carriers are indeed a Fermi liquid~\cite{Rullier-Albenque03,Rullier-Albenque08,Solovjov25}.

SC in pnictides is generally assumed to be more ``normal,'' i.e., BCS-like, than in the cuprates. The above observations moderate this statement in two ways. First, pairs scattering via a localized hole are necessarily Cooper pairs, since pairs bound by the localized hole could not move, because the localized hole is not moving, so they are not BEC-like. The BCS-like mechanism is not phononic in either case. Second, orbital selectivity has been inferred in the pnictides as well~\cite{Klemm24}, so the pnictide and cuprate scenarios do not appear so different after all, especially if one were to show that the magnons created in the electron pockets were the glue in pnictide SC, analogously to the localized hole in the cuprates.

\section{Murunskite.}

\begin{figure}[t]
    \centering
    \includegraphics[width=\linewidth]{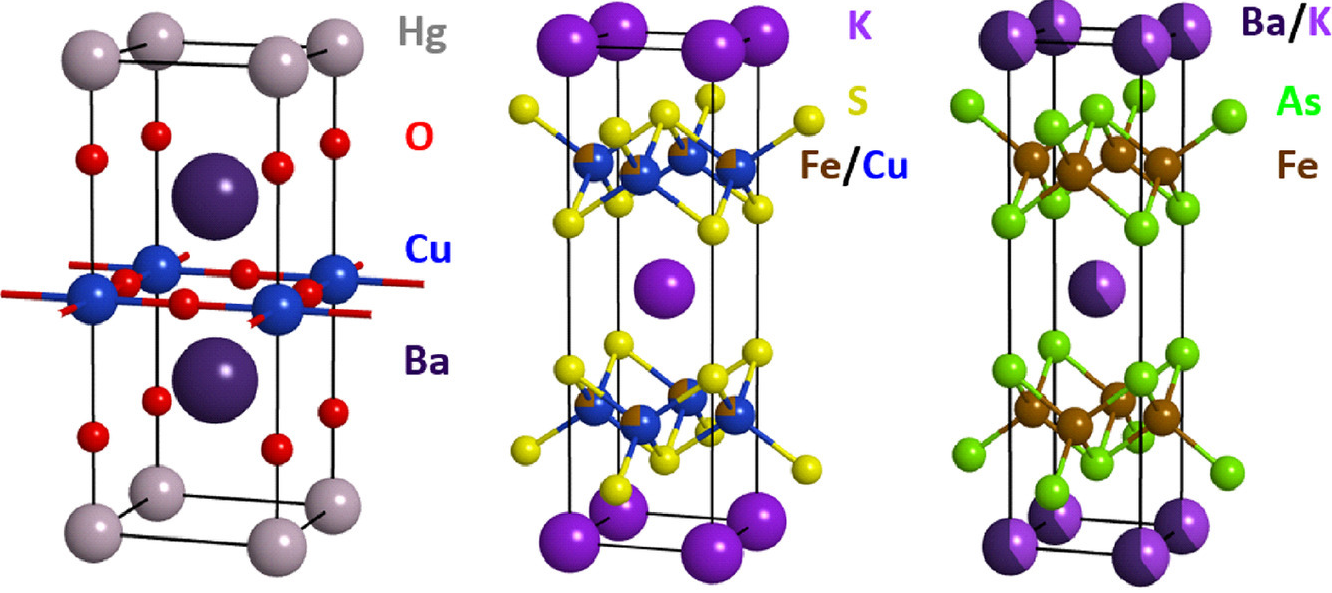}
    \caption{Murunskite (middle) between a mercury cuprate (left) and arsenic ferropnicitde (right). Transition metals: Cu (blue), Fe (brown). Spacers: Ba (dark purple), Hg (gray), K (purple).
Ligands: O (red), S (yellow), As (green). The Fe atoms in murunskite are in the Cu positions. From Ref.~\cite{Tolj21}.}
    \label{figmurunskite}
\end{figure}

Murunskite (K$_2$Cu$_3$FeS$_4$) is a scarce sulfosalt mineral, originally found in the Murunskii massif in Siberia in the form of small ($<0.001$~mm$^3$) incrustations~\cite{Dobrovolskaya82}, and more recently synthesized in large ($>20$~mm$^3$) single crystals~\cite{Tolj21}. In Fig.~\ref{figmurunskite}, we see that it interpolates between the cuprates and pnictides in several interesting ways. The metal cation position is occupied by Cu and Fe in a 3:1 ratio, distributed at random. The ligands are S, neighboring to O above, and P and As to the left, in the periodic table. Finally, it is isostructural to the pnictides, yet semiconducting like most undoped cuprates.

Murunskite is a test case for the translation of insights gained above to related transition-metal compounds. The ligands are active in cuprate and passive in pnictide functionality. What is sulfur doing in murunskite? Its orbitals are partially open, and it dominates the density of states near the top of the filled (valence) band~\cite{Tolj21}. That makes it more similar to oxygen in the cuprates than to arsenic in the pnictides. However, the functionality is magnetic, rather than superconducting~\cite{Tolj25}. Pnictides are indeed magnetic, but this magnetism is associated with metallic Fe orbitals, while murunskite is an insulator, so our analogies appear to be breaking down at first sight.

On the contrary, the analogies hold, but only because murunskite magnetism is not ordinary~\cite{Tolj25}. Local magnetic moments in murunskite are due to the Fe atoms, as one would expect, but these are distributed at random in $1/4$ of the metal cation positions, the other $3/4$ being taken up by non-magnetic Cu in the closed-shell Cu$^+$ ($3d^{10}$) configuration. Nevertheless, murunskite exhibits nearly commensurate magnetic order at 97 K, even though the Fe atoms also come in two different environments and oxidation states, Fe$^{2+}$ and Fe$^{3+}$, also distributed at random. While real-space disorder is usually a secondary consideration, in murunskite it plays a pivotal role and must be accounted for from the outset.

\begin{figure}[t]
    \centering
        \begin{tabular}{cc}
        (a) & (b) \\
        \multicolumn{2}{l}{\hskip -2mm\includegraphics[height=0.32\linewidth]{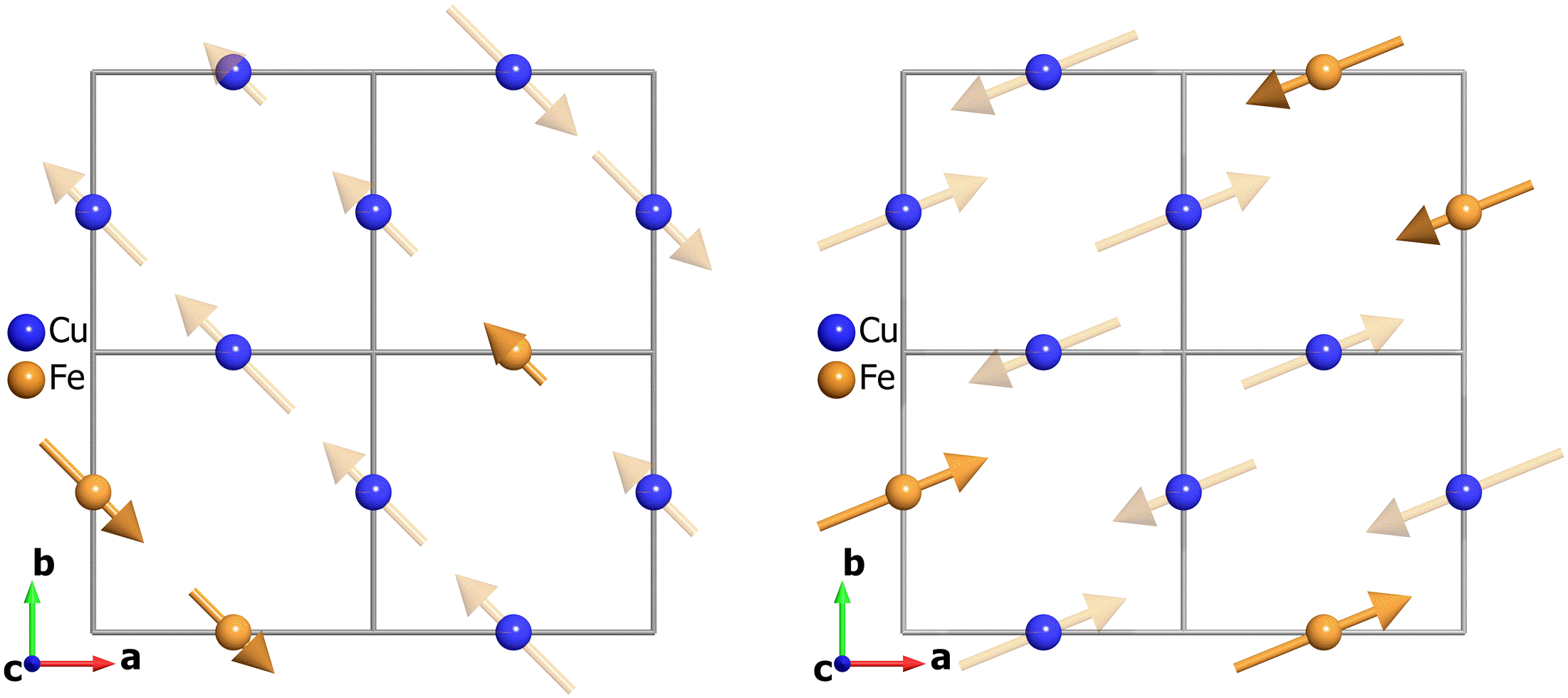}}
        \\
        (c) & (d)
        \\
        \includegraphics[height=0.30\linewidth]{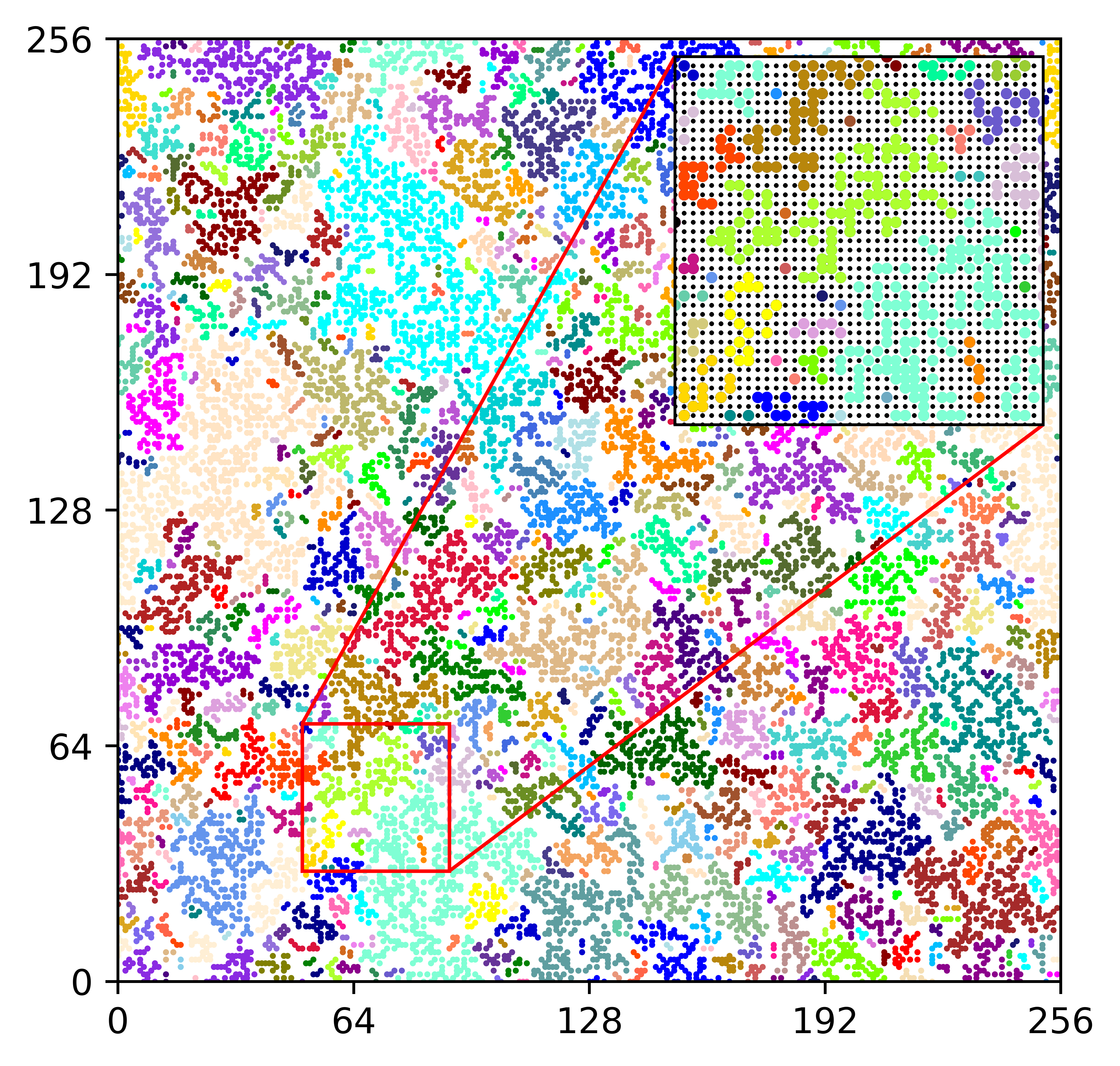}
        & 
        \hskip -6mm\raisebox{-3mm}{\includegraphics[height=0.32\linewidth]{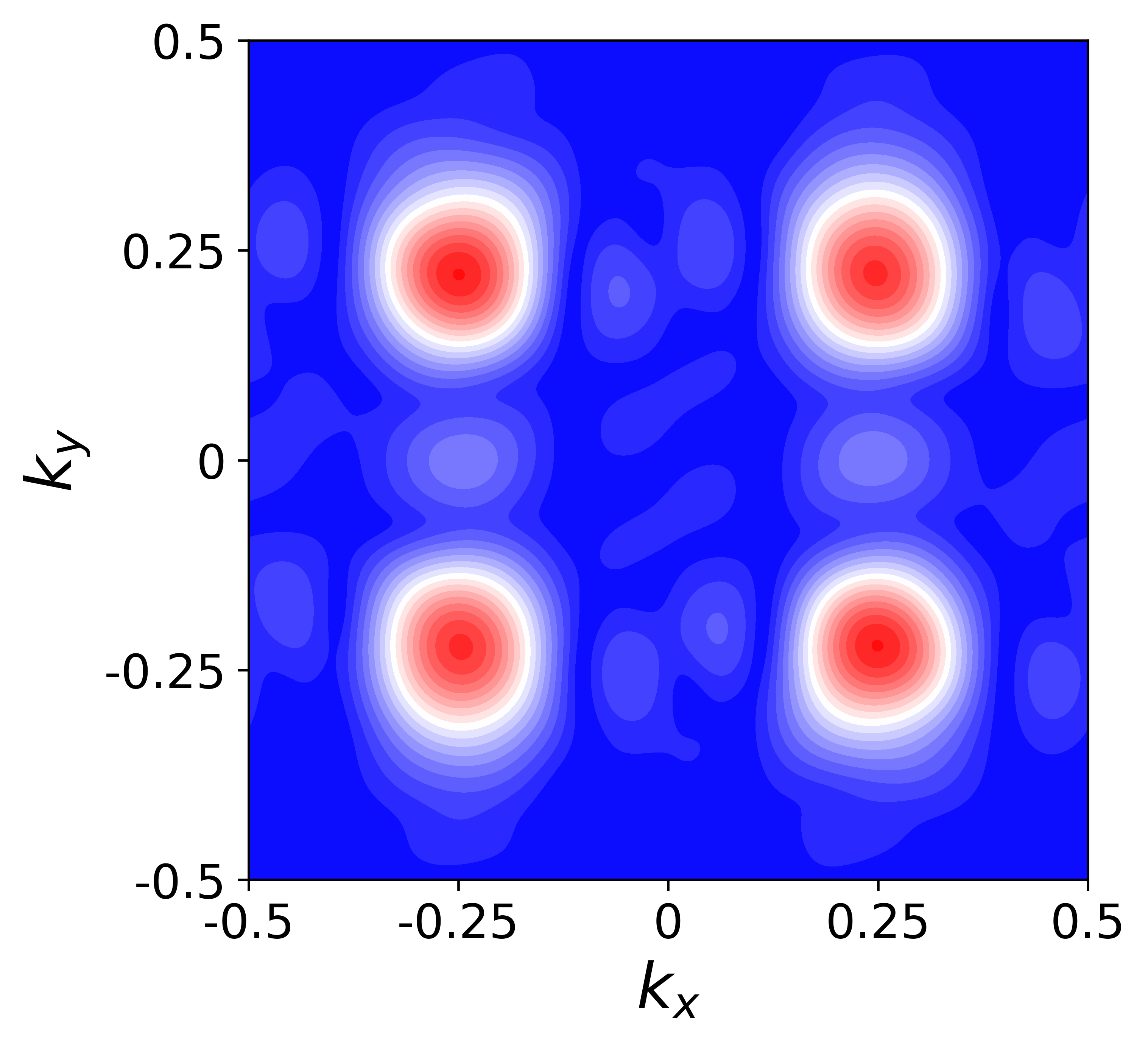}}
        \end{tabular}
    \caption{(a--b) Real-space image of Rietveld refinements of murunskite neutron scattering data, with two peaks at nearly commensurate quarter-zone vectors $k_1$ and $k_2$. The non-magnetic Cu ions are shown in blue. Faded arrows indicate that the inferred local magnetic moments are absent. From Ref.~\cite{Tolj25}. (a) $k_1= (0.266(3), 0.266(3), 0)$. (b) $k_2= (0.24(2), 0.76(2), 0) = (0.24(2), 1 - 0.24(2), 0)$. (c) Simulated AF-like clusters, each in a different color. The magnified inset shows the Cu ions as black dots. (d) Fourier transform of the image in (c). From Ref.~\cite{Tolj25}. Units of $2\pi/a$.}
    \label{figmagnetic}
\end{figure}

The conundrum and its proposed resolution are shown in Fig.~\ref{figmagnetic}. In the top row, a Rietveld refinement of neutron scattering data is given for the point crystal structure, assuming a local moment at equivalent (Wickoff) metal cation positions. The data are resolved by two nearly-quarter-zone wave vectors, $k_1= (0.266(3), 0.266(3), 0)$ and $k_2= (0.24(2), 0.76(2), 0) = (0.24(2), 1 - 0.24(2), 0)$. The problem is that $3/4$ of Wickoff sites are non-magnetic Cu, so there are no local moments there. It turns out that Fe ions, which have a magnetic moment, couple ferromagnetically into dimers as nearest neighbors, and these dimers couple antiferromagnetically via next-nearest interactions. The resulting $++--$ (quarter-zone AF) configurations form tree-like percolating clusters that are interrupted every time they encounter a Cu wall more than one atom thick, shown in the simulated configuration and its Fourer transform in the bottom row. The quarter-zone scattering signal of individual clusters is amplified macroscopically because the monocrystal lattice is nearly perfect, despite the occupational disorder. 

The analogies with cuprates are extensive. First, local disorder has a qualitative effect on the spectroscopy, such that one cannot interpret the measurements without taking it into account first. Because of it, Fermi arcs appear in cuprates, and quarter-zone AF in murunskite, instead of the closed Fermi surfaces and half-zone AF in the pnictides, where every site is occupied by Fe ions. At the same time, the scattering data itself appears quite orderly. The generic reason for that is that a lot of disorder in real space can be absorbed into peak widths in inverse space. Thus, the disordered-alloy paradigm can be applied both to cuprates, where it appears in the charge channel, and to murunskite, where it is in the spin channel.

Second is the role of ligands in creating the emergent order. In cuprates, the open O orbitals metallize and give rise to SC. XPS measurements in murunskite show that S exists in at least two oxidation states that compensate the local environments, comprising two different oxidation states Fe$^{3+}$ and Fe$^{2+}$ and non-magnetic Cu$^+$, without any disorder appearing in the atomic positions. This electronic plasticity enables the amplification of the quarter-zone AF, as described above. At the same time, it appears to play a role in mediating (at least) the second neighbor magnetic interactions, although that remains to be proven. Such a dual role of the oxygen ligands in the cuprates is already established. A combined analysis of NMR, optical, and transport measurements shows that the O $2p$ orbitals simultaneously partake in ionic bonding with the Cu $3d$ orbital, and metallic covalent bonding with the Cu $4s$ orbital~\cite{Barisic22}.

\section{Discussion.}

Bulk materials have been sought for their mechanical properties since antiquity, extended in the 19th century by the conductivity and ferromagnetism of elemental metals. Research in ionically or covalently bound functional materials, with its emphasis on high-T$_c$ SC and exotic magnetism, is in this sense against nature. The reason is that electrons in such materials are localized by default, while quantum electronic functionality requires coherent motion across at least a few atomic positions.

In this review, we have emphasized the conditions which facilitate local coherence. The primary considerations are kinematic: lattice topology and orbital symmetries. Functionality of ionically bound materials is essentially determined by the ligand anions, whose role varies from case to case. Such a ``chemical'' point of view begs the question, can simple physical models be generally useful. Current trends such as the materials genome initiative~\cite{dePablo19} and artificial intelligence (AI)~\cite{Papadimitriou24} seem to suggest little practical confidence in them. We are not so pessimistic.

Extensive experience has yielded the important insight that the Hall mobility is \emph{quantitatively} the same in all cuprates, despite the bewildering variety of compounds and non-universal properties in the phase diagram, such as CDW's or Lifshitz transitions. It motivates the search for \emph{chemical invariants} that could explain such universality. Our prime candidate is the large $s$--$2p$ overlap, which has been identified as responsible for the large effective $2p_x$--$2p_y$ hopping. The details of that original identification are important for our narrative.

The Cu $4s$ orbital does not stand out in an unconstrained DFT calculation, because it is far from the Fermi level and empty. The importance of the large $4s$--$2p$ overlap $t_{ps}\sim 2$~eV was realized only with realistic constraints, namely, excluding unphysical extremely delocalized filled orbitals from the optimization. In such a calculation, the bulk condensation energy is primarily due to nearest-neighbor interactions. Succinctly, a major physical player was uncovered by the computational technique only when the optimization was restrained in a chemically reasonable manner by human oversight~\cite{Andersen01}. In our picture, the second-order hopping $2p_x$--$4s$--$2p_y$ via the overlap $t_{ps}$ is primarily responsible for the universal Fermi-liquid properties of the cuprates, because the Cu $4s$ orbital is always large and empty, thus, it is a chemical invariant of the cuprate class of materials.

Armed with this insight, we can assert that the Cu $3d$ orbital is localized, i.e., not conducting, in the whole underdoped part of the phase diagram, and begins to delocalize only upon overdoping. This delocalization corresponds to a local (first-order-like) change in the nature of the chemical bond from ionic to covalent. Consequently, the transition is a crossover, as observed experimentally, rather than a true phase transition. We are then in a position to translate our reasoning to the pnictides, and to murunskite. Surprisingly, the analogies run opposite to expectations: Magnetic insulating murunskite is more like the SC cuprates than are the SC pnictides.

The pnictides are in the opposite physical regime than the cuprates, namely (orbital/band selective) Mott-Hubbard rather than charge-transfer. Every site is representative of the sample, and the electronic sector is essentially uncoupled from the lattice. The ensuing nearly-AF metal emerges for a subtle fundamental reason: Local orbital states cannot relax strong correlations by varying their radius, because they are comprised of overlapping $t_{2g}$ orbitals that are all equally correlated, while their radius is fixed externally by the functionally passive $e_g$ orbitals. The same is not true of the cuprates, where the correlated Cu $3d$ orbitals are bridged by the uncorrelated O $2p$ orbitals, so that relaxation of strong correlations is possible by charge transfer, causing the separation of the electronic sector into a disordered strongly correlated non-conducting $3d$--$2p$ sector, and an uncorrelated ordered, indeed SC, $s$--$2p$ sector.

Murunskite is structurally like pnictides, but electronically like cuprates, because the ligands are active in the functionality. It shows a similar interconnection of short-range disorder in real space and long-range order in inverse space. Both can be described as disordered alloys in real space, but functionalized by self-doping instead of alloying. That the preceding sentence is even meaningful, let alone true, is in our opinion a significant benefit of the translational approach to functional materials, reviewed here. In such an approach, the use of language is opposite to that of AI. Instead of making the most probable statements retrospectively, one makes singular statements prospectively. Such is the statement that the Hall mobility in cuprates is universal. Without a perspective to illuminate it, it is just a fun fact, like many others. Similarly, the parallel between murunskite and cuprates is in the manner of putting oneself in a new situation and inventing new words for it. Simple physical models, like our models of the Fermi arcs and quarter-zone AF in murunskite, arise naturally in meaningful contexts as an expression of human understanding.

In both cuprates and murunskite, the ligands compensate disorder that was imposed externally, by large local lattice deformations induced by doping in the first, and random atomic positions in the second case. In both cases, they overcompensate by enabling an emergent electronic order, SC in the cuprates and quarter-zone AF in murunskite.

Last but not least, the emergent order is local in both cases, witnessed by the very high second critical field in the cuprates, which implies a very short coherence length of Cooper pairs, and by the cluster-like magnetic structure observed in our simulations in murunskite. Such locality opens up the possibility of creating small (nano-size) samples which retain the essential physical functionality.

\begin{table}[t]
    \centering
    \begin{tabular}{l|ccc}
    Property & Cuprates & Pnictides & Murunskite  \\ \hline
    Cation coordination & octahedral & tetrahedral & tetrahedral  \\ 
    Open ligand orbitals & yes & no & yes  \\
    Short-range coherence & yes & no & yes  \\
    Self-doping & yes & yes & yes  \\ 
    Static AF & yes & yes & yes  \\ 
    Insulating parent & yes & no & yes  \\ 
    Doping possible & yes & yes & ?  \\ 
    Fermi liquid subsystem & yes & yes & ?  \\ 
    Second electronic subsystem & gapped & incoherent & ?  \\ 
    Superconductivity & yes & yes & ?  \\ 
    \end{tabular}
    \caption{Comparison of principal features of cuprates, pnictides, and murunskite.}
    \label{tabcompare}
\end{table}

In pnictides, because their structure is fixed by ionic ligands that are electronically passive, the metallic state nematicity is probably electronically driven~\cite{Li2017,Ishida20}. As discussed above, the net effect is that the conductivity is well described by a multiband Hubbard model in which the metal cation orbitals overlap directly, like in elemental metals, but with a strong residual interaction among the carriers. The price is that the interactions bring a van Hove singularity of the dispersion in the vicinity of the Fermi energy, thus creating a dissipative reservoir of static carriers, accessible even at low temperature. Such a reservoir is visible in the optical response, in contrast to the clearly gapped mid-infrared feature in the cuprates, associated with the localized $3d$ hole that is shared with the neighboring oxygens in an ionic ``CuO$_4$ molecule''~\cite{Kumar23}. We compare the principal features of the three classes of compounds in Table~\ref{tabcompare}.

Conversely, the superconducting nickelates are structurally like cuprates, but electronically like pnictides. They are in the Mott-Hubbard regime, with closed ligand orbitals and metallicity derived from doping the Ni $3d$ orbitals, which can be modelled by the Hubbard model with on-site repulsion~\cite{Wang24,Sakakibara24}. They corroborate the insight that ligand orbitals fix the atomic size for the metallic orbitals when they are at much higher energy than the latter.

We believe that we have only scratched the surface of what is possible with the translational approach so far. Murunskite, the first compound we synthesized in a conscious application of this approach, has yielded a new and unexpected form of magnetic order~\cite{Tolj25}. Important implications for ruthenates have also been uncovered~\cite{Kumar23}. Methodologically, one is motivated to focus on the role of ligands, and on finding chemical invariants, in all ionically bound functional materials. Extending this attitude further afield, one notices that AF order in covalent organic crystals is found across distances exceeding any known mechanism~\cite{Liu20,Lapidus21}. Even more intriguingly, naturally evolved components of photosynthetic light-harvesting complexes are capable of coherent conduction at room temperature across protein complexes with a mass of $40$~kDa at the low end~\cite{Wen09,Sarovar10,Klymchenko25}. Fundamentally, coherence over more than nearest neighbors is likely to shape investigations in quantum functionality for many years to come.

\section*{Acknowledgements.}

DKS gratefully acknowledges conversations with E. Tutiš. The work at the University of Zagreb was supported by the Croatian Science Foundation under Project No. IP-2022-10-3382 and the project CeNIKS co-financed by the Croatian Government and the European Union through the European Regional Development Fund-Competitiveness and Cohesion Operational Program (Grant No. KK.01.1.1.02.0013). The work at TU Wien was supported by the Austrian Science Fund (FWF) [10.55776/F86; 10.55776/P35945].

\section*{References.}
\bibliographystyle{unsrt}
\bibliography{cuprates_pnictides_sulfosalts.bib}

@article{Bednorz86,
  author =        {Bednorz, J. G. and M\"uller, K. A.},
  journal =       {Z. Physik B - Condensed Matter},
  number =        {2},
  pages =         {189},
  title =         {Possible high-{T}$_c$ superconductivity in the
                   {Ba--La--Cu--O} system},
  volume =        {64},
  year =          {1986},
  doi =           {10.1007/BF01303701},
}

@article{Anderson87,
  author =        {P.~W.~Anderson},
  journal =       {Science},
  pages =         {1196--8},
  title =         {The resonating valence bond state in {La$_2$CuO$_4$}
                   and superconductivity},
  volume =        {235},
  year =          {1987},
  doi =           {},
}

@article{Keimer15,
  author =        {Keimer, B. and Kivelson, S. A. and Norman, M. R. and
                   Uchida, S. and Zaanen, J.},
  journal =       {Nature},
  month =         {Feb},
  number =        {7538},
  pages =         {179-186},
  title =         {From quantum matter to high-temperature
                   superconductivity in copper oxides},
  volume =        {518},
  year =          {2015},
  doi =           {10.1038/nature14165},
  issn =          {1476-4687},
}

@article{Phillips22,
  author =        {Philip W. Phillips and Nigel E. Hussey and
                   Peter Abbamonte},
  journal =       {Science},
  number =        {6602},
  pages =         {eabh4273},
  title =         {Stranger than metals},
  volume =        {377},
  year =          {2022},
  doi =           {10.1126/science.abh4273},
}

@article{Zaanen85,
  author =        {Zaanen, J. and Sawatzky, G. A. and Allen, J. W.},
  journal =       {Phys. Rev. Lett.},
  month =         {Jul},
  pages =         {418--421},
  publisher =     {American Physical Society},
  title =         {Band gaps and electronic structure of
                   transition-metal compounds},
  volume =        {55},
  year =          {1985},
  doi =           {10.1103/PhysRevLett.55.418},
}

@article{Barisic87,
  author =        {S Bari{\v{s}}i{\'{c}} and I Batisti{\'{c}} and
                   J Friedel},
  journal =       {Europhysics Letters ({EPL})},
  month =         {jun},
  number =        {11},
  pages =         {1231--1236},
  publisher =     {{IOP} Publishing},
  title =         {Electron-Phonon Model for High-{T}$_c$ Layered-Metal
                   Oxides},
  volume =        {3},
  year =          {1987},
  doi =           {10.1209/0295-5075/3/11/013},
}

@article{Emery87,
  author =        {V.~J.~Emery},
  journal =       {Phys. Rev. Lett.},
  pages =         {2794--7},
  title =         {Theory of high-{T}$_c$ superconductivity in oxides},
  volume =        {58},
  year =          {1987},
}

@inproceedings{Mazumdar89,
  author =        {S. Mazumdar},
  booktitle =     {Interacting electrons in reduced dimensions},
  editor =        {Dionys Baeriswyl and David K. Campbell},
  pages =         {315--329},
  publisher =     {Plenum Press, New York},
  title =         {A unified theoretical approach to superconductors
                   with strong {Coulomb} correlations: the organics,
                   {LiTi$_2$O$_4$}, electron- and hole-doped copper
                   oxides and doped {BaBiO$_3$}},
  year =          {1989},
}

@article{Barlingay90,
  author =        {Barlingay, C. and
                   Garc\'{\i}a-V\'azquez, Valent\'{\i}n and
                   Falco, Charles M. and Mazumdar, S. and Risbud, S. H.},
  journal =       {Phys. Rev. B},
  month =         {Mar},
  pages =         {4797--4800},
  publisher =     {American Physical Society},
  title =         {Effects of zinc substitution on the electron
                   superconductor
                   {Nd$_{1.85}$Ce$_{0.15}$CuO$_{4-\delta}$}},
  volume =        {41},
  year =          {1990},
  doi =           {10.1103/PhysRevB.41.4797},
}

@article{Barisic90,
  author =        {S.~Bari\v{s}i\'{c} and J.~Zelenko},
  journal =       {Solid State Communications},
  note =          {},
  number =        {5},
  pages =         {367--370},
  title =         {Electron mechanism for the structural phase
                   transitions in {La$_{2-x}$Ba$_x$CuO$_4$}},
  volume =        {74},
  year =          {1990},
  doi =           {10.1016/0038-1098(90)90504-5},
}

@article{Eskes93,
  author =        {Eskes, Henk and Jefferson, John H.},
  journal =       {Phys. Rev. B},
  month =         {Oct},
  pages =         {9788--9798},
  publisher =     {American Physical Society},
  title =         {Superexchange in the cuprates},
  volume =        {48},
  year =          {1993},
  doi =           {10.1103/PhysRevB.48.9788},
}

@article{Barriquand94,
  author =        {Barriquand, F. and Sawatzky, G. A.},
  journal =       {Phys. Rev. B},
  month =         {Dec},
  pages =         {16649--16667},
  publisher =     {American Physical Society},
  title =         {Superexchange, hole-hole interactions, and oxygen
                   spin dynamics in high-${\mathrm{T}}_{\mathrm{c}}$
                   superconductors},
  volume =        {50},
  year =          {1994},
  doi =           {10.1103/PhysRevB.50.16649},
}

@article{Barisic22,
  author =        {Barišić, N. and Sunko, D. K.},
  journal =       {Journal of Superconductivity and Novel Magnetism},
  month =         jul,
  number =        {7},
  pages =         {1781--1799},
  title =         {High-{T}$_c$ {Cuprates}: a Story of Two Electronic
                   Subsystems.},
  volume =        {35},
  year =          {2022},
  doi =           {10.1007/s10948-022-06183-y},
  language =      {en},
}

@article{Barisic15,
  author =        {Barišić, N. and Chan, M. K. and Veit, M. J. and
                   Dorow, C. J. and Ge, Y. and Tang, Y. and Tabis, W. and
                   Yu, G. and Zhao, X. and Greven, M.},
  journal =       {arXiv},
  month =         jul,
  title =         {Evidence for a universal {Fermi}-liquid scattering
                   rate throughout the phase diagram of the copper-oxide
                   superconductors},
  year =          {2015},
  doi =           {10.48550/arXiv.1507.07885},
  url =           {http://arxiv.org/abs/1507.07885},
}

@article{Barisic19,
  author =        {Barišić, N and Chan, M K and Veit, M J and
                   Dorow, C J and Ge, Y and Li, Y and Tabis, W and
                   Tang, Y and Yu, G and Zhao, X and Greven, M},
  journal =       {New Journal of Physics},
  month =         {nov},
  number =        {11},
  pages =         {113007},
  publisher =     {IOP Publishing},
  title =         {Evidence for a universal {F}ermi-liquid scattering
                   rate throughout the phase diagram of the copper-oxide
                   superconductors},
  volume =        {21},
  year =          {2019},
  doi =           {10.1088/1367-2630/ab4d0f},
}

@article{Pelc19,
  author =        {Pelc, D. and Pop{\v c}evi{\'c}, P. and Po{\v z}ek, M. and
                   Greven, M. and Bari{\v s}i{\'c}, N.},
  journal =       {Science Advances},
  number =        {1},
  publisher =     {American Association for the Advancement of Science},
  title =         {Unusual behavior of cuprates explained by
                   heterogeneous charge localization},
  volume =        {5},
  year =          {2019},
  doi =           {10.1126/sciadv.aau4538},
}

@article{Badoux16,
  author =        {Badoux, S. and Tabis, W. and Lalibert{\'e}, F. and
                   Grissonnanche, G. and Vignolle, B. and Vignolles, D. and
                   B{\'e}ard, J. and Bonn, D. A. and Hardy, W. N. and
                   Liang, R. and Doiron-Leyraud, N. and Taillefer, Louis and
                   Proust, Cyril},
  journal =       {Nature},
  month =         {Mar},
  number =        {7593},
  pages =         {210-214},
  title =         {Change of carrier density at the pseudogap critical
                   point of a cuprate superconductor},
  volume =        {531},
  year =          {2016},
  doi =           {10.1038/nature16983},
  issn =          {1476-4687},
}

@article{Putzke21,
  author =        {Putzke, Carsten and Benhabib, Siham and
                   Tabis, Wojciech and Ayres, Jake and Wang, Zhaosheng and
                   Malone, Liam and Licciardello, Salvatore and
                   Lu, Jianming and Kondo, Takeshi and
                   Takeuchi, Tsunehiro and Hussey, Nigel E. and
                   Cooper, John R. and Carrington, Antony},
  journal =       {Nature Physics},
  month =         {Jul},
  number =        {7},
  pages =         {826-831},
  title =         {Reduced Hall carrier density in the overdoped strange
                   metal regime of cuprate superconductors},
  volume =        {17},
  year =          {2021},
  doi =           {10.1038/s41567-021-01197-0},
  issn =          {1745-2481},
}

@article{Nicholls25,
  author =        {Nicholls, R. and Hinlopen, R. D. H. and Ayres, J. and
                   Kotte, T. and F\"orster, T. and Park, J. and
                   Sourd, J. and Carrington, A. and Hussey, N. E.},
  journal =       {Phys. Rev. B},
  month =         {Aug},
  pages =         {064511},
  publisher =     {American Physical Society},
  title =         {Doping dependence of the low-temperature planar
                   carrier density in overdoped
                   YBa$_2$Cu$_3$O$_{7-\delta}$},
  volume =        {112},
  year =          {2025},
  doi =           {10.1103/6kk9-xq3z},
}

@article{Li16,
  author =        {Li, Yangmu and Tabis, W. and Yu, G. and
                     Bari\ifmmode \check{s}\else \v{s}\fi{}i\ifmmode
  \acute{c}\else \'{c}\fi{}, N. and Greven, M.},
  journal =       {Phys. Rev. Lett.},
  month =         {Nov},
  pages =         {197001},
  publisher =     {American Physical Society},
  title =         {Hidden Fermi-liquid Charge Transport in the
                   Antiferromagnetic Phase of the Electron-Doped Cuprate
                   Superconductors},
  volume =        {117},
  year =          {2016},
  doi =           {10.1103/PhysRevLett.117.197001},
}

@article{Pelc18,
  author =        {Pelc, Damjan and Vu{\v{c}}kovi{\'{c}}, Marija and
                   Grbi{\'{c}}, Mihael S. and Po{\v{z}}ek, Miroslav and
                   Yu, Guichuan and Sasagawa, Takao and Greven, Martin and
                   Bari{\v{s}}i{\'{c}}, Neven},
  journal =       {Nature Communications},
  month =         {Oct},
  number =        {1},
  pages =         {4327},
  title =         {Emergence of superconductivity in the cuprates via a
                   universal percolation process},
  volume =        {9},
  year =          {2018},
  doi =           {10.1038/s41467-018-06707-y},
  issn =          {2041-1723},
}

@article{Popcevic18,
  author =        {Pop{\v{c}}evi{\'{c}}, Petar and Pelc, Damjan and
                   Tang, Yang and Velebit, Kristijan and
                   Anderson, Zachary and Nagarajan, Vikram and
                   Yu, Guichuan and Po{\v{z}}ek, Miroslav and
                   Bari{\v{s}}i{\'{c}}, Neven and Greven, Martin},
  journal =       {npj Quantum Materials},
  month =         {Sep},
  number =        {1},
  pages =         {42},
  title =         {Percolative nature of the direct-current
                   paraconductivity in cuprate superconductors},
  volume =        {3},
  year =          {2018},
  doi =           {10.1038/s41535-018-0115-2},
  issn =          {2397-4648},
}

@article{Phillips03,
  author =        {J C Phillips and A Saxena and A R Bishop},
  journal =       {Reports on Progress in Physics},
  month =         {nov},
  number =        {12},
  pages =         {2111--2182},
  publisher =     {{IOP} Publishing},
  title =         {Pseudogaps, dopants, and strong disorder in cuprate
                   high-temperature superconductors},
  volume =        {66},
  year =          {2003},
  doi =           {10.1088/0034-4885/66/12/r02},
}

@article{Kresin06,
  author =        {Vladimir Z. Kresin and Yurii N. Ovchinnikov and
                   Stuart A. Wolf},
  journal =       {Physics Reports},
  number =        {5},
  pages =         {231-259},
  title =         {Inhomogeneous superconductivity and the
                   “pseudogap” state of novel superconductors},
  volume =        {431},
  year =          {2006},
  doi =           {https://doi.org/10.1016/j.physrep.2006.05.006},
  issn =          {0370-1573},
  url =           {https://www.sciencedirect.com/science/article/pii/
                  S0370157306001633},
}

@article{Fratini10,
  author =        {Fratini, Michela and Poccia, Nicola and
                   Ricci, Alessandro and Campi, Gaetano and
                   Burghammer, Manfred and Aeppli, Gabriel and
                   Bianconi, Antonio},
  journal =       {Nature},
  month =         aug,
  number =        {7308},
  pages =         {841--844},
  title =         {Scale-free structural organization of oxygen
                   interstitials in La$_2$CuO$_{4+y}$},
  volume =        {466},
  year =          {2010},
  doi =           {10.1038/nature09260},
  issn =          {1476-4687},
  url =           {https://doi.org/10.1038/nature09260},
}

@article{Norman98,
  author =        {Norman, M. R. and Ding, H. and Randeria, M. and
                   Campuzano, J. C. and Yokoya, T. and Takeuchi, T. and
                   Takahashi, T. and Mochiku, T. and Kadowaki, K. and
                   Guptasarma, P. and Hinks, D. G.},
  journal =       {Nature},
  month =         {Mar},
  number =        {6672},
  pages =         {157-160},
  title =         {Destruction of the Fermi surface in underdoped
                   high-Tc superconductors},
  volume =        {392},
  year =          {1998},
  doi =           {10.1038/32366},
  issn =          {1476-4687},
}

@article{Lee07-1,
  author =        {Lee, W. S. and Vishik, I. M. and Tanaka, K. and
                   Lu, D. H. and Sasagawa, T. and Nagaosa, N. and
                   Devereaux, T. P. and Hussain, Z. and Shen, Z.-X.},
  journal =       {Nature},
  month =         {Nov},
  number =        {7166},
  pages =         {81--84},
  publisher =     {Nature Publishing Group},
  title =         {Abrupt onset of a second energy gap at the
                   superconducting transition of underdoped Bi2212},
  volume =        {450},
  year =          {2007},
  doi =           {10.1038/nature06219},
}

@article{Kunisada20,
  author =        {So Kunisada and Shunsuke Isono and Yoshimitsu Kohama and
                   Shiro Sakai and Cédric Bareille and
                   Shunsuke Sakuragi and Ryo Noguchi and Kifu Kurokawa and
                   Kenta Kuroda and Yukiaki Ishida and Shintaro Adachi and
                   Ryotaro Sekine and Timur K. Kim and Cephise Cacho and
                   Shik Shin and Takami Tohyama and Kazuyasu Tokiwa and
                   Takeshi Kondo},
  journal =       {Science},
  number =        {6505},
  pages =         {833-838},
  title =         {Observation of small Fermi pockets protected by clean
                   CuO<sub>2</sub> sheets of a high-<i>T</i><sub>c</sub>
                   superconductor},
  volume =        {369},
  year =          {2020},
  doi =           {10.1126/science.aay7311},
}

@article{Bozin05,
  author =        {Bo\ifmmode \check{z}\else \v{z}\fi{}in, E. S. and
                   Billinge, S. J. L.},
  journal =       {Phys. Rev. B},
  month =         {Nov},
  pages =         {174427},
  publisher =     {American Physical Society},
  title =         {Nominal doping and partition of doped holes between
                   planar and apical orbitals in
                   {La$_{2-x}$Sr$_x$CuO$_4$}},
  volume =        {72},
  year =          {2005},
  doi =           {10.1103/PhysRevB.72.174427},
}

@article{Pelc15,
  author =        {D Pelc and M Po\v{z}ek and V Despoja and D K Sunko},
  journal =       {New Journal of Physics},
  number =        {8},
  pages =         {083033},
  title =         {Mechanism of metallization and superconductivity
                   suppression in
                   {YBa$_2($Cu$_{0.97}$Zn$_{0.03})_3$O$_{6.92}$}
                   revealed by {$^{67}$Zn} {NQR}},
  volume =        {17},
  year =          {2015},
  doi =           {10.1088/1367-2630/17/8/083033},
}

@article{Mazumdar18,
  author =        {Mazumdar, Sumit},
  journal =       {Phys. Rev. B},
  pages =         {205153},
  publisher =     {American Physical Society},
  title =         {Valence transition model of the pseudogap, charge
                   order, and super\-conductivity in electron-doped and
                   hole-doped copper oxides},
  volume =        {98},
  year =          {2018},
  doi =           {10.1103/PhysRevB.98.205153},
}

@article{Lee06,
  author =        {Lee, Kee Hag and Hoffmann, Roald},
  journal =       {The Journal of Physical Chemistry A},
  number =        {2},
  pages =         {609-617},
  title =         {{Oxygen Interstitials in Superconducting La2CuO4:
                   Their Valence State and Role}},
  volume =        {110},
  year =          {2006},
  doi =           {10.1021/jp053154f},
}

@article{Berlijn12,
  author =        {Berlijn, Tom and Lin, Chia-Hui and Garber, William and
                   Ku, Wei},
  journal =       {Phys. Rev. Lett.},
  month =         {May},
  pages =         {207003},
  publisher =     {American Physical Society},
  title =         {Do Transition-Metal Substitutions Dope Carriers in
                   Iron-Based Superconductors?},
  volume =        {108},
  year =          {2012},
  doi =           {10.1103/PhysRevLett.108.207003},
}

@article{Salluzzo08,
  author =        {Salluzzo, M. and Ghiringhelli, G. and Cezar, J. C. and
                   Brookes, N. B. and De Luca, G. M. and Fracassi, F. and
                   Vaglio, R.},
  journal =       {Phys. Rev. Lett.},
  month =         {Feb},
  pages =         {056810},
  publisher =     {American Physical Society},
  title =         {Indirect Electric Field Doping of the
                   ${\mathrm{CuO}}_{2}$ Planes of the Cuprate
                       ${\mathrm{NdBa}}_{2}{\mathrm{Cu}}_{3}{\mathrm{O}}_{7}$
  Superconductor},
  volume =        {100},
  year =          {2008},
  doi =           {10.1103/PhysRevLett.100.056810},
}

@article{Tolj21,
  author =        {Davor Tolj and Trpimir Iv\v si\'c and
                   Ivica \v{Z}ivkovi\'c and Konstantin Semeniuk and
                   Edoardo Martino and Ana Akrap and Priyanka Reddy and
                   Benjamin Klebel-Knobloch and Ivor Lon\v cari\'c and
                   L\'aszl\'o Forr\'o and Neven Bari\v si\'c and
                   Henrik M. Ronnow and Denis K. Sunko},
  journal =       {Applied Materials Today},
  pages =         {101096},
  title =         {Synthesis of murunskite single crystals: A bridge
                   between cuprates and pnictides},
  volume =        {24},
  year =          {2021},
  doi =           {https://doi.org/10.1016/j.apmt.2021.101096},
  issn =          {2352-9407},
  url =           {https://www.sciencedirect.com/science/article/pii/
                  S235294072100161X},
}

@article{Montanaro24,
  author =        {Montanaro, Angela and Rigoni, Enrico Maria and
                   Giusti, Francesca and Barba, Luisa and
                   Chita, Giuseppe and Glerean, Filippo and
                   Jarc, Giacomo and Mathengattil, Shahla Y. and
                   Boschini, Fabio and Eisaki, Hiroshi and
                   Greven, Martin and Damascelli, Andrea and
                   Giannetti, Claudio and Mihailovic, Dragan and
                   Kabanov, Viktor and Fausti, Daniele},
  journal =       {Phys. Rev. B},
  month =         {Sep},
  pages =         {125102},
  publisher =     {American Physical Society},
  title =         {Dynamics of nonthermal states in optimally doped
  {$\mathrm{B}{\mathrm{i}}_{2}\mathrm{S}{\mathrm{r}}_{2}\mathrm{C}{\mathrm{a}}_{0.92}{\mathrm{Y}}_{0.08}\mathrm{C}{\mathrm{u}}_{2}{\mathrm{O}}_{8+\ensuremath{\delta}}$}
  revealed by midinfrared three-pulse spectroscopy},
  volume =        {110},
  year =          {2024},
  doi =           {10.1103/PhysRevB.110.125102},
}

@article{Martinez91,
  author =        {J. C. Martinez and O. Laborde and J. J. Préjean and
                   C. Chappert and J. P. Renard and J. Karpinski and
                   E. Kaldis},
  journal =       {Europhysics Letters},
  month =         {apr},
  number =        {7},
  pages =         {693},
  title =         {Study of the Magneto-Resistive Transition of a
                   {Y$_2$Ba$_4$Cu$_8$O$_{16}$} Single Crystal},
  volume =        {14},
  year =          {1991},
  doi =           {10.1209/0295-5075/14/7/015},
}

@article{Fink19,
  author =        {Fink, J. and Nayak, J. and Rienks, E. D. L. and
                   Bannies, J. and Wurmehl, S. and Aswartham, S. and
                   Morozov, I. and Kappenberger, R. and ElGhazali, M. A. and
                   Craco, L. and Rosner, H. and Felser, C. and
                   B\"uchner, B.},
  journal =       {Phys. Rev. B},
  month =         {Jun},
  pages =         {245156},
  publisher =     {American Physical Society},
  title =         {Evidence of hot and cold spots on the Fermi surface
                   of LiFeAs},
  volume =        {99},
  year =          {2019},
  doi =           {10.1103/PhysRevB.99.245156},
}

@article{Wang24,
  author =        {Wang, Bai Yang and Lee, Kyuho and Goodge, Berit H.},
  journal =       {Annual Review of Condensed Matter Physics},
  number =        {Volume 15, 2024},
  pages =         {305-324},
  publisher =     {Annual Reviews},
  type =          {Journal Article},
  title =         {Experimental Progress in Superconducting Nickelates},
  volume =        {15},
  year =          {2024},
  doi =
  {https://doi.org/10.1146/annurev-conmatphys-032922-093307},
  issn =          {1947-5462},
}

@article{Tsukada05,
  author =        {A. Tsukada and Y. Krockenberger and M. Noda and
                   H. Yamamoto and D. Manske and L. Alff and M. Naito},
  journal =       {Solid State Communications},
  note =          {},
  number =        {7},
  pages =         {427--431},
  title =         {New class of {T'}-structure cuprate superconductors},
  volume =        {133},
  year =          {2005},
  doi =           {10.1016/j.ssc.2004.12.011},
  issn =          {0038-1098},
}

@article{Adachi13,
  author =        {Adachi, Tadashi and Mori, Yosuke and Takahashi, Akira and
                   Kato, Masatsune and Nishizaki, Terukazu and
                   Sasaki, Takahiko and Kobayashi, Norio and
                   Koike, Yoji},
  journal =       {Journal of the Physical Society of Japan},
  number =        {6},
  pages =         {063713},
  title =         {Evolution of the Electronic State through the
                   Reduction Annealing in Electron-Doped
                   Pr1.3-xLa0.7CexCuO4+\u03b4 (x=0.10) Single Crystals:
                   Antiferromagnetism, Kondo Effect, and
                   Superconductivity},
  volume =        {82},
  year =          {2013},
  doi =           {10.7566/JPSJ.82.063713},
}

@article{NBarisic13,
  author =        {{Bari\v si\'c}, Neven and Chan, Mun K. and Li, Yuan and
                   Yu, Guichuan and Zhao, Xudong and Dressel, Martin and
                   Smontara, Ana and Greven, Martin},
  journal =       {PNAS},
  number =        {30},
  pages =         {12235-12240},
  title =         {Universal sheet resistance and revised phase diagram
                   of the cuprate high-temperature superconductors},
  volume =        {110},
  year =          {2013},
  doi =           {10.1073/pnas.1301989110},
}

@article{Zhou03,
  author =        {Zhou, X. J. and Yoshida, T. and Lanzara, A. and
                   Bogdanov, P. V. and Kellar, S. A. and Shen, K. M. and
                   Yang, W. L. and Ronning, F. and Sasagawa, T. and
                   Kakeshita, T. and Noda, T. and Eisaki, H. and
                   Uchida, S. and Lin, C. T. and Zhou, F. and
                   Xiong, J. W. and Ti, W. X. and Zhao, Z. X. and
                   Fujimori, A. and Hussain, Z. and Shen, Z.-X.},
  journal =       {Nature},
  month =         {May},
  number =        {6938},
  pages =         {398-398},
  title =         {Universal nodal Fermi velocity},
  volume =        {423},
  year =          {2003},
  doi =           {10.1038/423398a},
  issn =          {1476-4687},
}

@inproceedings{Andersen01,
  author =        {O.~K.~Andersen and E.~Pavarini and I.~Dasgupta and
                   T.~Saha-Dasgupta and O.~Jepsen},
  booktitle =     {{Proceedings of the 4th Asian Workshop on
                   First-Principles Electronic Structure Calculations}},
  publisher =     {Taipei, Taiwan},
  title =         {Bandstructures of hole-doped cuprates; correlation
                   with {T}$_{c\,\mathrm{max}}$},
  year =          {2001},
}

@article{Pavarini01,
  author =        {Pavarini, E. and Dasgupta, I. and Saha-Dasgupta, T. and
                   Jepsen, O. and Andersen, O. K.},
  journal =       {Phys. Rev. Lett.},
  month =         {Jul},
  pages =         {047003},
  publisher =     {American Physical Society},
  title =         {Band-Structure Trend in Hole-Doped Cuprates and
                   Correlation with
                   {${\mathit{T}}_{\mathit{c}\mathrm{max}}$}},
  volume =        {87},
  year =          {2001},
  doi =           {10.1103/PhysRevLett.87.047003},
}

@article{Qimiao90,
  author =        {{Qimiao Si} and {Ju H.~Kim} and {Jian Ping Lu} and
                   K.~Levin},
  journal =       {Phys. Rev. B},
  pages =         {1033--6},
  title =         {Phenomenological description of the copper oxides as
                   almost localized {Fermi} liquids},
  volume =        {42},
  year =          {1990},
}

@article{Qimiao93,
  author =        {{Qimiao Si} and {Yuyao Zha} and K.~Levin and
                   J.~P.~Lu},
  journal =       {Phys. Rev. B},
  pages =         {9055--76},
  title =         {Comparison of spin dynamics in
                   {YBa$_2$Cu$_3$O$_{7-\delta}$} and
                   {La$_{2-x}$Sr$_x$CuO$_4$}: Effects of {Fermi} surface
                   geometry},
  volume =        {47},
  year =          {1993},
}

@article{Mrkonjic03,
  author =        {I.~Mrkonji\'{c} and S.~Bari\v{s}i\'{c}},
  journal =       {Eur. Phys. J. B},
  pages =         {69--84},
  title =         {Singular behavior of the {E}mery model with {O}-{O}
                   hopping for high-{T}$_c$ superconductors},
  volume =        {34},
  year =          {2003},
}

@article{Sunko07,
  author =        {Sunko, D. K. and
                     Bari\ifmmode \check{s}\else \v{s}\fi{}i\ifmmode
  \acute{c}\else \'{c}\fi{}, S.},
  journal =       {Phys. Rev. B},
  month =         {Feb},
  pages =         {060506(R)},
  publisher =     {American Physical Society},
  title =         {Electronic pseudogap of optimally doped
                   high-temperature
  {${\mathrm{Nd}}_{2-x}{\mathrm{Ce}}_{x}\mathrm{Cu}{\mathrm{O}}_{4}$}
  superconductors},
  volume =        {75},
  year =          {2007},
  doi =           {10.1103/PhysRevB.75.060506},
}

@article{Hashimoto08,
  author =        {Hashimoto, M. and Yoshida, T. and Yagi, H. and
                   Takizawa, M. and Fujimori, A. and Kubota, M. and
                   Ono, K. and Tanaka, K. and Lu, D. H. and Shen, Z.-X. and
                   Ono, S. and Ando, Yoichi},
  journal =       {Phys. Rev. B},
  month =         {Mar},
  pages =         {094516},
  publisher =     {American Physical Society},
  title =         {Doping evolution of the electronic structure in the
                   single-layer cuprate
                   {Bi$_{2}$Sr$_{2-x}$La$_{x}$CuO$_{6+\delta}$}:
                   {C}omparison with other single-layer cuprates},
  volume =        {77},
  year =          {2008},
  doi =           {10.1103/PhysRevB.77.094516},
}

@article{Lazic15,
  author =        {P.~{Lazi\'c} and D.~K.~Sunko},
  journal =       {EPL (Europhysics Letters)},
  number =        {3},
  pages =         {37011},
  title =         {Fermi arcs and pseudogap emerging from dimensional
                   crossover at the {Fermi} surface in
                   {La$_{2-x}$Sr$_x$CuO$_4$}},
  volume =        {112},
  year =          {2015},
  doi =           {10.1209/0295-5075/112/37011},
}

@article{Hohenberg67,
  author =        {Hohenberg, P. C.},
  journal =       {Phys. Rev.},
  month =         {Jun},
  pages =         {383--386},
  publisher =     {American Physical Society},
  title =         {Existence of Long-Range Order in One and Two
                   Dimensions},
  volume =        {158},
  year =          {1967},
  doi =           {10.1103/PhysRev.158.383},
}

@article{Mermin66,
  author =        {Mermin, N. D. and Wagner, H.},
  journal =       {Phys. Rev. Lett.},
  month =         {Nov},
  pages =         {1133--1136},
  publisher =     {American Physical Society},
  title =         {{Absence of Ferromagnetism or Antiferromagnetism in
                   One- or Two-Dimensional Isotropic Heisenberg Models}},
  volume =        {17},
  year =          {1966},
  doi =           {10.1103/PhysRevLett.17.1133},
}

@article{Palle21,
  author =        {Grgur Palle and D K Sunko},
  journal =       {Journal of Physics A: Mathematical and Theoretical},
  month =         {jul},
  number =        {31},
  pages =         {315001},
  publisher =     {{IOP} Publishing},
  title =         {Physical limitations of the
                   {Hohenberg--Mermin--Wagner} theorem},
  volume =        {54},
  year =          {2021},
  doi =           {10.1088/1751-8121/ac0a9d},
}

@article{Anderson59,
  author =        {P.~W.~Anderson},
  journal =       {Phys. Rev.},
  pages =         {2--13},
  title =         {New approach to the theory of superexchange
                   interactions},
  volume =        {115},
  year =          {1959},
}

@article{Onose01,
  author =        {Y.~Onose and Y.~Taguchi and K.~Ishizaka and
                   Y.~Tokura},
  journal =       {Phys. Rev. Lett.},
  pages =         {217001},
  title =         {Doping Dependence of Pseudogap and Related Charge
                   Dynamics in {Nd$_{2-x}$Ce$_x$CuO$_4$}},
  volume =        {87},
  year =          {2001},
}

@article{Onose04,
  author =        {Onose, Y. and Taguchi, Y. and Ishizaka, K. and
                   Tokura, Y.},
  journal =       {Phys. Rev. B},
  month =         {Jan},
  pages =         {024504},
  publisher =     {American Physical Society},
  title =         {Charge dynamics in underdoped
  ${\mathrm{Nd}}_{2\ensuremath{-}x}{\mathrm{Ce}}_{x}{\mathrm{CuO}}_{4}:$
  Pseudogap and related phenomena},
  volume =        {69},
  year =          {2004},
  doi =           {10.1103/PhysRevB.69.024504},
}

@article{Hirsch89,
  author =        {Hirsch, J. E. and Marsiglio, F.},
  journal =       {Phys. Rev. B},
  month =         {Jun},
  pages =         {11515--11525},
  publisher =     {American Physical Society},
  title =         {Superconducting state in an oxygen hole metal},
  volume =        {39},
  year =          {1989},
  doi =           {10.1103/PhysRevB.39.11515},
  url =           {https://link.aps.org/doi/10.1103/PhysRevB.39.11515},
}

@article{Li19,
  author =        {Yangmu Li and W. Tabis and Y. Tang and G. Yu and
                   J. Jaroszynski and N. Bari\v{s}i\'{c} and M. Greven},
  journal =       {Science Advances},
  number =        {2},
  pages =         {eaap7349},
  title =         {Hole pocket-driven superconductivity and its
                   universal features in the electron-doped cuprates},
  volume =        {5},
  year =          {2019},
  doi =           {10.1126/sciadv.aap7349},
}

@incollection{Fournier98,
  address =       {Boston, MA},
  author =        {Fournier, P. and Maiser, E. and Greene, R. L.},
  booktitle =     {The Gap Symmetry and Fluctuations in High-Tc
                   Superconductors},
  editor =        {Bok, Julien and Deutscher, Guy and Pavuna, Davor and
                   Wolf, Stuart A.},
  pages =         {145--158},
  publisher =     {Springer US},
  title =         {Current Research Issues for the Electron-Doped
                   Cuprates},
  year =          {1998},
  doi =           {10.1007/0-306-47081-0_9},
  isbn =          {978-0-306-47081-3},
}

@phdthesis{Li17,
  author =        {Yangmu Li},
  note =          {https://hdl.handle.net/11299/193423, pp. 97--100},
  school =        {University of Minnesota},
  title =         {Neutron Scattering, Muon Spin Rotation/Relaxation,
                   and Charge Transport Study of the Electron-Doped
                   Cuprate Superconductors},
  year =          {2017},
  url =           {https://hdl.handle.net/11299/193423},
}

@article{Jiang21,
  author =        {Shengtao Jiang and Douglas J. Scalapino and
                   Steven R. White},
  journal =       {Proceedings of the National Academy of Sciences},
  number =        {44},
  pages =         {e2109978118},
  title =         {Ground-state phase diagram of the {$t$--$t'$--$J$}
                   model},
  volume =        {118},
  year =          {2021},
  doi =           {10.1073/pnas.2109978118},
}

@article{Ando04,
  author =        {Ando, Yoichi and Kurita, Y. and Komiya, Seiki and
                   Ono, S. and Segawa, Kouji},
  journal =       {Phys. Rev. Lett.},
  month =         {May},
  number =        {19},
  pages =         {197001},
  publisher =     {American Physical Society},
  title =         {Evolution of the {Hall} Coefficient and the Peculiar
                   Electronic Structure of the Cuprate Superconductors},
  volume =        {92},
  year =          {2004},
  doi =           {10.1103/PhysRevLett.92.197001},
}

@article{Chan14,
  author =        {Chan, M. K. and Veit, M. J. and Dorow, C. J. and
                   Ge, Y. and Li, Y. and Tabis, W. and Tang, Y. and
                   Zhao, X. and
                     Bari\ifmmode \check{s}\else \v{s}\fi{}i\ifmmode
  \acute{c}\else \'{c}\fi{}, N. and Greven, M.},
  journal =       {Phys. Rev. Lett.},
  month =         {Oct},
  pages =         {177005},
  publisher =     {American Physical Society},
  title =         {In-Plane Magnetoresistance Obeys Kohler's Rule in the
                   Pseudogap Phase of Cuprate Superconductors},
  volume =        {113},
  year =          {2014},
  doi =           {10.1103/PhysRevLett.113.177005},
}

@article{Mirzaei12,
  author =        {Mirzaei, Seyed Iman and Stricker, Damien and
                   Hancock, Jason N. and Berthod, Christophe and
                   Georges, Antoine and van Heumen, Erik and
                   Chan, Mun K. and Zhao, Xudong and Li, Yuan and
                   Greven, Martin and Bari\v{s}i\'{c}, Neven and
                   van der Marel, Dirk},
  journal =       {PNAS},
  number =        {15},
  pages =         {5774-5778},
  title =         {Spectroscopic evidence for Fermi liquid-like energy
                   and temperature dependence of the relaxation rate in
                   the pseudogap phase of the cuprates},
  volume =        {110},
  year =          {2013},
  doi =           {10.1073/pnas.1218846110},
}

@article{Kumar23,
  author =        {Kumar, C. M. N. and Akrap, A. and Homes, C. C. and
                   Martino, E. and Klebel-Knobloch, B. and Tabis, W. and
                   Bari\v{s}i\'{c}, O. S. and Sunko, D. K. and
                   Bari\v{s}i\'{c} N.},
  journal =       {Phys. Rev. B},
  month =         {Apr},
  pages =         {144515},
  publisher =     {American Physical Society},
  title =         {Characterization of two electronic subsystems in
                   cuprates through optical conductivity},
  volume =        {107},
  year =          {2023},
  doi =           {10.1103/PhysRevB.107.144515},
}

@article{Chien91,
  author =        {Chien, T. R. and Wang, Z. Z. and Ong, N. P.},
  journal =       {Phys. Rev. Lett.},
  month =         {Oct},
  pages =         {2088--2091},
  publisher =     {American Physical Society},
  title =         {Effect of {Zn} impurities on the normal-state Hall
                   angle in single-crystal {${\mathrm{YBa}}_{2}
                   {\mathrm{Cu}}_{3\mathrm{\ensuremath{-}}\mathit{x}}
                   {\mathrm{Zn}}_{\mathit{x}}
  {\mathrm{O}}_{7\mathrm{\ensuremath{-}}\mathrm{\ensuremath{\delta}}}$}},
  volume =        {67},
  year =          {1991},
  doi =           {10.1103/PhysRevLett.67.2088},
}

@misc{NBarisic15,
  author =        {N. Bari\v{s}i\'{c} and M. K. Chan and M. J. Veit and
                   C. J. Dorow and Y. Ge and Y. Tang and W. Tabis and
                   G. Yu and X. Zhao and M. Greven},
  note =          {{arXiv}:1507.07885v1},
  title =         {Evidence for a universal Fermi-liquid scattering rate
                   throughout the phase diagram of the copper-oxide
                   superconductors},
  year =          {2015},
}

@article{Klebel23,
  author =        {Klebel-Knobloch, B. and Tabi{\'{s}}, W. and
                   Gala, M. A. and Bari{\v{s}}i{\'{c}}, O. S. and
                   Sunko, D. K. and Bari{\v{s}}i{\'{c}}, N.},
  journal =       {Scientific Reports},
  month =         {Aug},
  number =        {1},
  pages =         {13562},
  title =         {Transport properties and doping evolution of the
                   Fermi surface in cuprates},
  volume =        {13},
  year =          {2023},
  doi =           {10.1038/s41598-023-39813-z},
  issn =          {2045-2322},
}

@article{Zhong22,
  author =        {Yong Zhong and Zhuoyu Chen and Su-Di Chen and
                   Ke-Jun Xu and Makoto Hashimoto and Yu He and
                   Shin-ichi Uchida and Donghui Lu and Sung-Kwan Mo and
                   Zhi-Xun Shen},
  journal =       {Proceedings of the National Academy of Sciences},
  number =        {32},
  pages =         {e2204630119},
  title =         {Differentiated roles of Lifshitz transition on
                   thermodynamics and superconductivity in
  La<sub>2-</sub><i><sub>x</sub></i>Sr<i><sub>x</sub></i>CuO<sub>4</sub>},
  volume =        {119},
  year =          {2022},
  doi =           {10.1073/pnas.2204630119},
  url =           {https://www.pnas.org/doi/abs/10.1073/pnas.2204630119},
}

@article{Gorkov06,
  author =        {Gor'kov, Lev P. and Teitel'baum, Gregory B.},
  journal =       {Phys. Rev. Lett.},
  month =         {Dec},
  pages =         {247003},
  publisher =     {American Physical Society},
  title =         {Interplay of Externally Doped and Thermally Activated
                   Holes in
                       ${\mathrm{La}}_{2-x}{\mathrm{Sr}}_{x}{\mathrm{CuO}}_{4}$
  and Their Impact on the Pseudogap Crossover},
  volume =        {97},
  year =          {2006},
  doi =           {10.1103/PhysRevLett.97.247003},
}

@article{Rullier-Albenque08,
  author =        {F. Rullier-Albenque and H. Alloul and F. Balakirev and
                   C. Proust},
  journal =       {{EPL} (Europhysics Letters)},
  month =         {jan},
  number =        {3},
  pages =         {37008},
  publisher =     {{IOP} Publishing},
  title =         {Disorder, metal-insulator crossover and phase diagram
                   in high-{T}$_c$ cuprates},
  volume =        {81},
  year =          {2008},
  doi =           {10.1209/0295-5075/81/37008},
}

@article{NBarisic13a,
  author =        {Bari{\v{s}}i{\'{c}}, Neven and Badoux, Sven and
                   Chan, Mun K. and Dorow, Chelsey and Tabis, Wojciech and
                   Vignolle, Baptiste and Yu, Guichuan and
                   B{\'e}ard, J{\'e}r{\^o}me and Zhao, Xudong and
                   Proust, Cyril and Greven, Martin},
  journal =       {Nature Physics},
  month =         {Dec},
  number =        {12},
  pages =         {761-764},
  title =         {Universal quantum oscillations in the underdoped
                   cuprate superconductors},
  volume =        {9},
  year =          {2013},
  doi =           {10.1038/nphys2792},
  issn =          {1745-2481},
}

@article{Doiron-Leyraud07,
  author =        {Doiron-Leyraud, Nicolas and Proust, Cyril and
                   LeBoeuf, David and Levallois, Julien and
                   Bonnemaison, Jean-Baptiste and Liang, Ruixing and
                   Bonn, D. A. and Hardy, W. N. and Taillefer, Louis},
  journal =       {Nature},
  month =         {May},
  number =        {7144},
  pages =         {565-568},
  title =         {Quantum oscillations and the Fermi surface in an
                   underdoped high-Tc superconductor},
  volume =        {447},
  year =          {2007},
  doi =           {10.1038/nature05872},
  issn =          {1476-4687},
}

@article{Vuckovic25,
  author =          {Vu\ifmmode \check{c}\else \v{c}\fi{}kovi\ifmmode
  \acute{c}\else \'{c}\fi{}, Marija and Najev, Ana and Yu, Biqiong and
  Sasagawa, Takao and Bielinski, Nina and Bari\ifmmode \check{s}\else
  \v{s}\fi{}i\ifmmode \acute{c}\else \'{c}\fi{}, Neven and Greven, Martin and
  Pelc, Damjan and Po\ifmmode \check{z}\else \v{z}\fi{}ek, Miroslav},
  journal =       {Phys. Rev. B},
  month =         {May},
  pages =         {184509},
  publisher =     {American Physical Society},
  title =         {Cu NMR study of lightly doped
  ${\mathrm{La}}_{\mathit{2}\ensuremath{-}\mathit{x}}{\mathrm{Sr}}_{\mathit{x}}{\mathrm{CuO}}_{4}$},
  volume =        {111},
  year =          {2025},
  doi =           {10.1103/PhysRevB.111.184509},
  url =           {https://link.aps.org/doi/10.1103/PhysRevB.111.184509},
}

@article{Mirzaei13,
  author =        {Mirzaei, Seyed Iman and Stricker, Damien and
                   Hancock, Jason N. and Berthod, Christophe and
                   Georges, Antoine and van Heumen, Erik and
                   Chan, Mun K. and Zhao, Xudong and Li, Yuan and
                   Greven, Martin and Bari\v{s}i\'{c}, Neven and
                   van der Marel, Dirk},
  journal =       {PNAS},
  number =        {15},
  pages =         {5774-5778},
  title =         {Spectroscopic evidence for Fermi liquid-like energy
                   and temperature dependence of the relaxation rate in
                   the pseudogap phase of the cuprates},
  volume =        {110},
  year =          {2013},
  doi =           {10.1073/pnas.1218846110},
}

@article{Norman95,
  author =        {M.~R.~Norman and M.~Randeria and H.~Ding and
                   J.~C.~Campuzano},
  journal =       {Phys. Rev. B},
  pages =         {615--22},
  title =         {Phenomenological models for the gap anisotropy of
                   {BSCCO} as measured by angle-resolved photoemission
                   spectroscopy},
  volume =        {52},
  year =          {1995},
}

@article{Fang22,
  author =        {Fang, Yawen and Grissonnanche, Ga{\"e}l and
                   Legros, Ana{\"e}lle and Verret, Simon and
                   Lalibert{\'e}, Francis and Collignon, Cl{\'e}ment and
                   Ataei, Amirreza and Dion, Maxime and Zhou, Jianshi and
                   Graf, David and Lawler, Michael J. and
                   Goddard, Paul A. and Taillefer, Louis and
                   Ramshaw, B. J.},
  journal =       {Nature Physics},
  month =         {May},
  number =        {5},
  pages =         {558-564},
  title =         {Fermi surface transformation at the pseudogap
                   critical point of a cuprate superconductor},
  volume =        {18},
  year =          {2022},
  doi =           {10.1038/s41567-022-01514-1},
  issn =          {1745-2481},
}

@misc{Tabis21,
  author =        {W. Tabi\'{s} and P. Pop\v{c}evi\'{c} and
                   B. Klebel-Knobloch and I. Biało and C. M. N. Kumar and
                   B. Vignolle and M. Greven and N. Bari\v{s}i\'{c}},
  note =          {{arXiv}:2106.07457},
  title =         {Arc-to-pocket transition and quantitative
                   understanding of transport properties in cuprate
                   superconductors},
  year =          {2021},
}

@misc{Beck25,
  author =        {Sophie Beck and Aline Ramires},
  note =          {{arXiv}:2511.02508},
  title =         {Structure-property relation in the cuprates: a
                   possible explanation for the pseudogap},
  year =          {2025},
  url =           {https://arxiv.org/abs/2511.02508},
}

@article{Popescu12,
  author =        {Popescu, Voicu and Zunger, Alex},
  journal =       {Phys. Rev. B},
  month =         {Feb},
  pages =         {085201},
  publisher =     {American Physical Society},
  title =         {Extracting {$E$} versus $\vec{k}$ effective band
                   structure from supercell calculations on alloys and
                   impurities},
  volume =        {85},
  year =          {2012},
  doi =           {10.1103/PhysRevB.85.085201},
}

@article{Sunko19,
  author =        {Sunko, V. and McGuinness, P. H. and Chang, C. S. and
                   Zhakina, E. and Khim, S. and Dreyer, C. E. and
                   Konczykowski, M. and Borrmann, H. and Moll, P. J. W. and
                   K\"onig, M. and Muller, D. A. and Mackenzie, A. P.},
  journal =       {Phys. Rev. X},
  month =         {Apr},
  pages =         {021018},
  publisher =     {American Physical Society},
  title =         {Controlled Introduction of Defects to Delafossite
                   Metals by Electron Irradiation},
  volume =        {10},
  year =          {2020},
  doi =           {10.1103/PhysRevX.10.021018},
}

@article{Usui19,
  author =        {Usui, Hidetomo and Ochi, Masayuki and Kitamura, Sota and
                   Oka, Takashi and Ogura, Daisuke and Rosner, Helge and
                   Haverkort, Maurits W. and Sunko, Veronika and
                   King, Philip D. C. and Mackenzie, Andrew P. and
                   Kuroki, Kazuhiko},
  journal =       {Phys. Rev. Mater.},
  month =         {Apr},
  pages =         {045002},
  publisher =     {American Physical Society},
  title =         {Hidden kagome-lattice picture and origin of high
                   conductivity in delafossite {PtCoO}$_2$},
  volume =        {3},
  year =          {2019},
  doi =           {10.1103/PhysRevMaterials.3.045002},
}

@article{Tolj25,
  author =        {Tolj, Davor and Reddy, Priyanka and Živković, Ivica and
                   Akšamović, Luka and Soh, Jian Rui and
                   Kom\c{e}dera, Kamila and Biało, Izabela and
                   Chogondahalli Muniraju, Naveen Kumar and
                   Ivšić, Trpimir and Novak, Mario and Zaharko, Oksana and
                   Ritter, Clemens and LaGrange, Thomas and
                   Tabiś, Wojciech and Batistić, Ivo and
                   Forró, László and Rønnow, Henrik M. and
                   Sunko, Denis K. and Barišić, Neven},
  journal =       {Advanced Functional Materials},
  number =        {40},
  pages =         {2500099},
  title =         {High-Entropy Magnetism of Murunskite},
  volume =        {35},
  year =          {2025},
  doi =           {10.1002/adfm.202500099},
}

@article{Comin14,
  author =        {R. Comin and A. Frano and M. M. Yee and Y. Yoshida and
                   H. Eisaki and E. Schierle and E. Weschke and
                   R. Sutarto and F. He and A. Soumyanarayanan and
                   Yang He and M. Le Tacon and I. S. Elfimov and
                   Jennifer E. Hoffman and G. A. Sawatzky and B. Keimer and
                   A. Damascelli},
  journal =       {Science},
  number =        {6169},
  pages =         {390-392},
  title =         {Charge Order Driven by Fermi-Arc Instability in
  Bi<sub>2</sub>Sr<sub>2\&\#x2212;<i>x</i></sub>La<sub><i>x</i></sub>CuO<sub>6+\&\#x3b4;</sub>},
  volume =        {343},
  year =          {2014},
  doi =           {10.1126/science.1242996},
  url =           {https://www.science.org/doi/abs/10.1126/science.1242996},
}

@article{Tabis14,
  author =        {Tabis, W. and Li, Y. and Le Tacon, M. and
                   Braicovich, L. and Kreyssig, A. and Minola, M. and
                   Dellea, G. and Weschke, E. and Veit, M. J. and
                   Ramazanoglu, M. and Goldman, A. I. and Schmitt, T. and
                   Ghiringhelli, G. and Bari{\v{s}}i{\'c}, N. and
                   Chan, M. K. and Dorow, C. J. and Yu, G. and Zhao, X. and
                   Keimer, B. and Greven, M.},
  journal =       {Nature Communications},
  pages =         {5875},
  title =         {Charge order and its connection with Fermi-liquid
                   charge transport in a pristine high-$T_c$ cuprate},
  volume =        {5},
  year =          {2014},
  doi =           {10.1038/ncomms6875},
  issn =          {2041-1723},
  url =           {https://doi.org/10.1038/ncomms6875},
}

@article{Tabis17,
  author =        {Tabis, W. and Yu, B. and Bialo, I. and Bluschke, M. and
                   Kolodziej, T. and Kozlowski, A. and Blackburn, E. and
                   Sen, K. and Forgan, E. M. and Zimmermann, M. v. and
                   Tang, Y. and Weschke, E. and Vignolle, B. and
                   Hepting, M. and Gretarsson, H. and Sutarto, R. and
                   He, F. and Le Tacon, M. and
                     Bari\ifmmode \check{s}\else \v{s}\fi{}i\ifmmode
  \acute{c}\else \'{c}\fi{}, N. and Yu, G. and Greven, M.},
  journal =       {Phys. Rev. B},
  month =         {Oct},
  pages =         {134510},
  publisher =     {American Physical Society},
  title =         {Synchrotron x-ray scattering study of
                   charge-density-wave order in
  ${\mathrm{HgBa}}_{2}{\mathrm{CuO}}_{4+\ensuremath{\delta}}$},
  volume =        {96},
  year =          {2017},
  doi =           {10.1103/PhysRevB.96.134510},
}

@article{Yang06,
  author =        {Yang, Kai-Yu and Rice, T. M. and Zhang, Fu-Chun},
  journal =       {Phys. Rev. B},
  month =         {May},
  pages =         {174501},
  publisher =     {American Physical Society},
  title =         {Phenomenological theory of the pseudogap state},
  volume =        {73},
  year =          {2006},
  doi =           {10.1103/PhysRevB.73.174501},
}

@article{Slater29,
  author =        {Slater, J. C.},
  journal =       {Phys. Rev.},
  month =         {Nov},
  pages =         {1293--1322},
  publisher =     {American Physical Society},
  title =         {The Theory of Complex Spectra},
  volume =        {34},
  year =          {1929},
  doi =           {10.1103/PhysRev.34.1293},
}

@article{Davidson65,
  author =        {Davidson,Ernest R.},
  journal =       {The Journal of Chemical Physics},
  number =        {12},
  pages =         {4199--4200},
  title =         {Single-Configuration Calculations on Excited States
                   of Helium. II},
  volume =        {42},
  year =          {1965},
  doi =           {10.1063/1.1695919},
}

@article{Yamanaka05,
  author =        {Yamanaka, S. and Koizumi, K. and Kitagawa, Y. and
                   Kawakami, T. and Okumura, M. and Yamaguchi, K.},
  journal =       {International Journal of Quantum Chemistry},
  number =        {6},
  pages =         {687--700},
  publisher =     {Wiley Subscription Services, Inc., A Wiley Company},
  title =         {Chemical bonding, less screening, and {Hund's} rule
                   revisited},
  volume =        {105},
  year =          {2005},
  doi =           {10.1002/qua.20784},
  issn =          {1097-461X},
}

@book{Mattis65,
  address =       {New York},
  author =        {D.~C.~Mattis},
  publisher =     {Harper and Row},
  title =         {The Theory of Magnetism},
  year =          {1965},
}

@article{Sajeev08,
  author =        {Sajeev,Y. and Sindelka,M. and Moiseyev,N.},
  journal =       {The Journal of Chemical Physics},
  number =        {6},
  pages =         {061101},
  title =         {Hund's multiplicity rule: From atoms to quantum dots},
  volume =        {128},
  year =          {2008},
  doi =           {10.1063/1.2837456},
}

@article{Sunko20a,
  author =        {Sunko, D. K.},
  journal =       {Journal of Superconductivity and Novel Magnetism},
  month =         {Jan},
  number =        {1},
  pages =         {27-33},
  title =         {High-Temperature Superconductors as Ionic Metals},
  volume =        {33},
  year =          {2020},
  doi =           {10.1007/s10948-019-05280-9},
  issn =          {1557-1947},
}

@article{Fink09,
  author =        {Fink, J. and Thirupathaiah, S. and Ovsyannikov, R. and
                   D\"urr, H. A. and Follath, R. and Huang, Y. and
                   de Jong, S. and Golden, M. S. and Zhang, Yu-Zhong and
                   Jeschke, H. O. and Valent\'{\i}, R. and Felser, C. and
                   Dastjani Farahani, S. and Rotter, M. and
                   Johrendt, D.},
  journal =       {Phys. Rev. B},
  month =         {Apr},
  pages =         {155118},
  publisher =     {American Physical Society},
  title =         {Electronic structure studies of ${ ext{BaFe}}_{2}{
                   ext{As}}_{2}$ by angle-resolved photoemission
                   spectroscopy},
  volume =        {79},
  year =          {2009},
  doi =           {10.1103/PhysRevB.79.155118},
}

@article{Si16,
  author =        {Si, Qimiao and Yu, Rong and Abrahams, Elihu},
  journal =       {Nature Reviews Materials},
  pages =         {16017},
  publisher =     {Macmillan Publishers Limited},
  title =         {High-temperature superconductivity in iron pnictides
                   and chalcogenides},
  volume =        {1},
  year =          {2016},
  doi =           {10.1038/natrevmats.2016.17},
}

@article{Derondeau17,
  author =        {Derondeau, Gerald and Bisti, Federico and
                   Kobayashi, Masaki and Braun, J\"urgen and
                   Ebert, Hubert and Rogalev, Victor A. and Shi, Ming and
                   Schmitt, Thorsten and Ma, Junzhang and Ding, Hong and
                   Strocov, Vladimir N. and Min\'ar, J\'an},
  journal =       {Scientific Reports},
  number =        {1},
  pages =         {8787},
  title =         {Fermi surface and effective masses in photoemission
                   response of the {(Ba$_{1-x}$K$_x$)Fe$_2$As$_2$}
                   superconductor},
  volume =        {7},
  year =          {2017},
  doi =           {10.1038/s41598-017-09480-y},
}

@article{Klemm24,
  author =        {Mason L. Klemm and Shirin Mozaffari and Rui Zhang and
                   Brian W. Casas and Alexei E. Koshelev and Ming Yi and
                   Luis Balicas and Pengcheng Dai},
  journal =       {Cell Reports Physical Science},
  number =        {2},
  pages =         {101816},
  title =         {Nematic superconductivity from selective orbital
                   pairing in iron pnictide single crystals},
  volume =        {5},
  year =          {2024},
  doi =           {https://doi.org/10.1016/j.xcrp.2024.101816},
  issn =          {2666-3864},
}

@article{Gooch09,
  author =        {Gooch, Melissa and Lv, Bing and Lorenz, Bernd and
                   Guloy, Arnold M. and Chu, Ching-Wu},
  journal =       {Phys. Rev. B},
  month =         {Mar},
  pages =         {104504},
  publisher =     {American Physical Society},
  title =         {Evidence of quantum criticality in the phase diagram
                   of K$_x$Sr$_{1-x}$Fe$_2$As$_2$ from measurements of
                   transport and thermoelectricity},
  volume =        {79},
  year =          {2009},
  doi =           {10.1103/PhysRevB.79.104504},
}

@article{Jiang09,
  author =        {Jiang, Shuai and Xing, Hui and Xuan, Guofang and
                   Wang, Cao and Ren, Zhi and Feng, Chunmu and
                   Dai, Jianhui and Xu, Zhu’an and Cao, Guanghan},
  journal =       {Journal of Physics: Condensed Matter},
  month =         {aug},
  number =        {38},
  pages =         {382203},
  publisher =     {},
  title =         {Superconductivity up to 30 K in the vicinity of the
                   quantum critical point in
                   {BaFe$_2$(As$_{1-x}$P$_x$)$_2$}},
  volume =        {21},
  year =          {2009},
  doi =           {10.1088/0953-8984/21/38/382203},
  url =           {https://doi.org/10.1088/0953-8984/21/38/382203},
}

@article{Dai15,
  author =        {Dai, Y. M. and Miao, H. and Xing, L. Y. and
                   Wang, X. C. and Wang, P. S. and Xiao, H. and Qian, T. and
                   Richard, P. and Qiu, X. G. and Yu, W. and Jin, C. Q. and
                   Wang, Z. and Johnson, P. D. and Homes, C. C. and
                   Ding, H.},
  journal =       {Phys. Rev. X},
  month =         {Sep},
  pages =         {031035},
  publisher =     {American Physical Society},
  title =         {Spin-Fluctuation-Induced Non-Fermi-Liquid Behavior
                   with Suppressed Superconductivity in
                   {LiFe$_{1-x}$Co$_x$As}},
  volume =        {5},
  year =          {2015},
  doi =           {10.1103/PhysRevX.5.031035},
  url =           {https://link.aps.org/doi/10.1103/PhysRevX.5.031035},
}

@article{Barisic10,
  author =        {Bari\v{s}i\'{c}, N. and Wu, D. and Dressel, M. and
                   Li, L. J. and Cao, G. H. and Xu, Z. A.},
  journal =       {Phys. Rev. B},
  month =         {Aug},
  pages =         {054518},
  publisher =     {American Physical Society},
  title =         {Electrodynamics of electron-doped iron pnictide
                   superconductors: Normal-state properties},
  volume =        {82},
  year =          {2010},
  doi =           {10.1103/PhysRevB.82.054518},
}

@article{Wu10,
  author =        {Wu, D. and
                     Bari\ifmmode \check{s}\else \v{s}\fi{}i\ifmmode
  \acute{c}\else \'{c}\fi{}, N. and Kallina, P. and Faridian, A. and Gorshunov,
  B. and Drichko, N. and Li, L. J. and Lin, X. and Cao, G. H. and Xu, Z. A. and
  Wang, N. L. and Dressel, M.},
  journal =       {Phys. Rev. B},
  month =         {Mar},
  pages =         {100512},
  publisher =     {American Physical Society},
  title =         {Optical investigations of the normal and
                   superconducting states reveal two electronic
                   subsystems in iron pnictides},
  volume =        {81},
  year =          {2010},
  doi =           {10.1103/PhysRevB.81.100512},
}

@article{Rullier-Albenque12,
  author =        {Rullier-Albenque, F. and Colson, D. and Forget, A. and
                   Alloul, H.},
  journal =       {Phys. Rev. Lett.},
  month =         {Nov},
  pages =         {187005},
  publisher =     {American Physical Society},
  title =         {Multiorbital Effects on the Transport and the
                   Superconducting Fluctuations in LiFeAs},
  volume =        {109},
  year =          {2012},
  doi =           {10.1103/PhysRevLett.109.187005},
  url =           {https://link.aps.org/doi/10.1103/PhysRevLett.109.187005},
}

@article{Tytarenko15,
  author =        {Tytarenko, Alona and Huang, Yingkai and
                   de Visser, Anne and Johnston, Steve and
                   van Heumen, Erik},
  journal =       {Scientific Reports},
  month =         {Jul},
  number =        {1},
  pages =         {12421},
  title =         {Direct observation of a Fermi liquid-like normal
                   state in an iron-pnictide superconductor},
  volume =        {5},
  year =          {2015},
  doi =           {10.1038/srep12421},
  issn =          {2045-2322},
}

@article{Rybicki16,
  author =        {Rybicki, Damian and Jurkutat, Michael and
                   Reichardt, Steven and Kapusta, Czes{\l}aw and
                   Haase, J{\"u}rgen},
  journal =       {Nature Communications},
  month =         {May},
  number =        {1},
  pages =         {11413},
  title =         {Perspective on the phase diagram of cuprate
                   high-temperature superconductors},
  volume =        {7},
  year =          {2016},
  doi =           {10.1038/ncomms11413},
  issn =          {2041-1723},
}

@article{Barisic93,
  author =        {S. Barišić and E. Tutiš},
  journal =       {Solid State Communications},
  number =        {6},
  pages =         {557-561},
  title =         {Effect of strong electron correlations on the
                   electron-phonon coupling in high {T}$_c$
                   superconductors},
  volume =        {87},
  year =          {1993},
  doi =           {10.1016/0038-1098(93)90596-F},
  issn =          {0038-1098},
}

@inproceedings{Tutis96,
  address =       {Berlin, Heidelberg},
  author =        {Tuti{\v{s}}, E. and Nik{\v{s}}i{\'{c}}, H. and
                   Bari{\v{s}}i{\'{c}}, S.},
  booktitle =     {From Quantum Mechanics to Technology},
  editor =        {Petru, Zygmunt and Przystawa, Jerzy and
                   Rapcewicz, Krzysztof},
  pages =         {161--175},
  publisher =     {Springer Berlin Heidelberg},
  title =         {Charge dynamics in cuprate superconductors},
  year =          {1996},
  doi =           {10.1007/BFb0106021},
  isbn =          {978-3-540-70724-0},
}

@article{Friedel88a,
  author =        {Jacques Friedel},
  journal =       {J. Phys. France},
  pages =         {1091--1095},
  title =         {On the non local character of {Wannier} functions in
                   hybrid bands},
  volume =        {49},
  year =          {1988},
  doi =           {10.1051/jphys:019880049070109100},
}

@book{Mahan90,
  address =       {New York},
  author =        {G.~D.~Mahan},
  publisher =     {Plenum},
  title =         {Many-Particle Physics, second edition},
  year =          {1990},
}

@article{Thomarat24,
  author =        {Thomarat, Laure and Elson, Frank and
                   Nocerino, Elisabetta and Das, Debarchan and
                   Ivashko, Oleh and Bartkowiak, Marek and
                   M{\aa}nsson, Martin and Sassa, Yasmine and
                   Adachi, Tadashi and Zimmermann, Martin v. and
                   Luetkens, Hubertus and Chang, Johan and
                   Janoschek, Marc and Guguchia, Zurab and
                   Simutis, Gediminas},
  journal =       {Communications Physics},
  month =         {Aug},
  number =        {1},
  pages =         {271},
  title =         {Tuning of charge order by uniaxial stress in a
                   cuprate superconductor},
  volume =        {7},
  year =          {2024},
  doi =           {10.1038/s42005-024-01760-0},
  issn =          {2399-3650},
}

@article{Gao25,
  author =        {Gao, Baizhi and Nikbin, Ehsan and Johnstone, Graham and
                   Shi, Ze and Heath, Christopher and
                   Appathurai, Narayan and Moreno, Beatriz Diaz and
                   Rahemtulla, Al and Gu, G. D. and Tranquada, John M. and
                   Howe, Jane Y. and Kim, Young-June},
  journal =       {Advanced Materials},
  note =          {Online Version of Record before inclusion in an
                   issue.},
  number =        {},
  pages =         {e09308},
  title =         {Structural Phase Separation and Enhanced
                   Superconductivity in La1.875Ba0.125CuO4 Under
                   Uniaxial Strain},
  volume =        {},
  year =          {2025},
  doi =           {https://doi.org/10.1002/adma.202509308},
}

@article{Axe89,
  author =        {Axe, J. D. and Moudden, A. H. and Hohlwein, D. and
                   Cox, D. E. and Mohanty, K. M. and Moodenbaugh, A. R. and
                   Xu, Youwen},
  journal =       {Phys. Rev. Lett.},
  month =         {Jun},
  pages =         {2751--2754},
  publisher =     {American Physical Society},
  title =         {Structural phase transformations and
                   superconductivity in {La$_{2-x}$Ba$_x$CuO$_4$}},
  volume =        {62},
  year =          {1989},
  doi =           {10.1103/PhysRevLett.62.2751},
}

@article{Bianconi96,
  author =        {Bianconi, A. and Saini, N. L. and Lanzara, A. and
                   Missori, M. and Rossetti, T. and Oyanagi, H. and
                   Yamaguchi, H. and Oka, K. and Ito, T.},
  journal =       {Phys. Rev. Lett.},
  month =         {Apr},
  pages =         {3412--3415},
  publisher =     {American Physical Society},
  title =         {Determination of the Local Lattice Distortions in the
                   Cu${\mathrm{O}}_{2}$ Plane of
  L${\mathrm{a}}_{1.85}$S${\mathrm{r}}_{0.15}$Cu${\mathrm{O}}_{4}$},
  volume =        {76},
  year =          {1996},
  doi =           {10.1103/PhysRevLett.76.3412},
}

@article{Lawler10,
  author =        {Lawler, M. J. and Fujita, K. and Lee, Jhinhwan and
                   Schmidt, A. R. and Kohsaka, Y. and Kim, Chung Koo and
                   Eisaki, H. and Uchida, S. and Davis, J. C. and
                   Sethna, J. P. and Kim, Eun-Ah},
  journal =       {Nature},
  month =         {Jul},
  number =        {7304},
  pages =         {347-351},
  title =         {Intra-unit-cell electronic nematicity of the high-Tc
                   copper-oxide pseudogap states},
  volume =        {466},
  year =          {2010},
  doi =           {10.1038/nature09169},
  issn =          {1476-4687},
}

@article{Massee20,
  author =        {F. Massee and Y. K. Huang and M. Aprili},
  journal =       {Science},
  number =        {6473},
  pages =         {68-71},
  title =         {Atomic manipulation of the gap in
  Bi<sub>2</sub>Sr<sub>2</sub>CaCu<sub>2</sub>O<sub>8+x</sub>},
  volume =        {367},
  year =          {2020},
  doi =           {10.1126/science.aaw7964},
}

@article{Vinograd21,
  author =        {Vinograd, Igor and Zhou, Rui and Hirata, Michihiro and
                   Wu, Tao and Mayaffre, Hadrien and Kr{\"a}mer, Steffen and
                   Liang, Ruixing and Hardy, W. N. and Bonn, D. A. and
                   Julien, Marc-Henri},
  journal =       {Nature Communications},
  month =         {Jun},
  number =        {1},
  pages =         {3274},
  title =         {Locally commensurate charge-density wave with
                   three-unit-cell periodicity in YBa2Cu3Oy},
  volume =        {12},
  year =          {2021},
  doi =           {10.1038/s41467-021-23140-w},
  issn =          {2041-1723},
}

@article{OMahony22,
  author =        {Shane M. O’Mahony and Wangping Ren and
                   Weijiong Chen and Yi Xue Chong and Xiaolong Liu and
                   H. Eisaki and S. Uchida and M. H. Hamidian and
                   J. C. Séamus Davis},
  journal =       {Proceedings of the National Academy of Sciences},
  number =        {37},
  pages =         {e2207449119},
  title =         {On the electron pairing mechanism of copper-oxide
                   high temperature superconductivity},
  volume =        {119},
  year =          {2022},
  doi =           {10.1073/pnas.2207449119},
}

@article{Valkov16,
  author =        {V. V. Val'kov and D. M. Dzebisashvili and
                   M. M. Korovushkin and A. F. Barabanov},
  journal =       {JETP Lett.},
  pages =         {385},
  title =         {Stability of the superconducting $d_{x^2-y^2}$ phase
                   in high-{T}$_c$ superconductors with respect to the
                   intersite {Coulomb} repulsion of holes at oxygen},
  volume =        {103},
  year =          {2016},
  doi =           {10.1134/S0021364016060114},
}

@inproceedings{Mueller97,
  address =       {Dordrecht},
  author =        {M{\"u}ller, K. Alex and Keller, Hugo},
  booktitle =     {High-Tc Superconductivity 1996: Ten Years after the
                   Discovery},
  editor =        {Kaldis, E. and Liarokapis, E. and M{\"u}ller, K. A.},
  pages =         {7--29},
  publisher =     {Springer Netherlands},
  title =         {'s' and `d' Wave Symmetry Components in
                   High-Temperature Cuprate Superconductors},
  year =          {1997},
  doi =           {10.1007/978-94-011-5554-0\_2},
  isbn =          {978-94-011-5554-0},
}

@article{Rullier-Albenque03,
  author =        {Rullier-Albenque, F. and Alloul, H. and Tourbot, R.},
  journal =       {Phys. Rev. Lett.},
  month =         {Jul},
  pages =         {047001},
  publisher =     {American Physical Society},
  title =         {Influence of Pair Breaking and Phase Fluctuations on
                   Disordered High ${T}_{c}$ Cuprate Superconductors},
  volume =        {91},
  year =          {2003},
  doi =           {10.1103/PhysRevLett.91.047001},
  url =           {https://link.aps.org/doi/10.1103/PhysRevLett.91.047001},
}

@article{Solovjov25,
  author =        {Solovjov, A. L. and Rogacki, K. and Shytov, N. V. and
                   Petrenko, E. V. and Bludova, L. V. and Chroneos, A. and
                   Vovk, R. V.},
  journal =       {Phys. Rev. B},
  month =         {May},
  pages =         {174508},
  publisher =     {American Physical Society},
  title =         {Influence of strong electron irradiation on
                   fluctuation conductivity and pseudogap in
                   {YBa$_2$Cu$_3$O$_{7-\delta}$} single crystals},
  volume =        {111},
  year =          {2025},
  doi =           {10.1103/PhysRevB.111.174508},
  url =           {https://link.aps.org/doi/10.1103/PhysRevB.111.174508},
}

@article{Dobrovolskaya82,
  author =        {M. G. Dobrovol'skaya},
  journal =       {International Geology Review},
  number =        {9},
  pages =         {1109--1114},
  publisher =     {GSA Website},
  title =         {Murunskite, {K$_2$Cu$_3$FeS$_4$}, a new sulfide of
                   potassium, copper, and iron},
  volume =        {24},
  year =          {1982},
  doi =           {10.1080/00206818209451049},
}

@article{dePablo19,
  author =        {de Pablo, Juan J. and Jackson, Nicholas E. and
                   Webb, Michael A. and Chen, Long-Qing and
                   Moore, Joel E. and Morgan, Dane and Jacobs, Ryan and
                   Pollock, Tresa and Schlom, Darrell G. and
                   Toberer, Eric S. and Analytis, James and
                   Dabo, Ismaila and DeLongchamp, Dean M. and
                   Fiete, Gregory A. and Grason, Gregory M. and
                   Hautier, Geoffroy and Mo, Yifei and Rajan, Krishna and
                   Reed, Evan J. and Rodriguez, Efrain and
                   Stevanovic, Vladan and Suntivich, Jin and
                   Thornton, Katsuyo and Zhao, Ji-Cheng},
  journal =       {npj Computational Materials},
  month =         {Apr},
  number =        {1},
  pages =         {41},
  title =         {New frontiers for the materials genome initiative},
  volume =        {5},
  year =          {2019},
  doi =           {10.1038/s41524-019-0173-4},
  issn =          {2057-3960},
}

@article{Papadimitriou24,
  author =        {Ioannis Papadimitriou and Ilias Gialampoukidis and
                   Stefanos Vrochidis and Ioannis Kompatsiaris},
  journal =       {Computational Materials Science},
  pages =         {112793},
  title =         {{AI} methods in materials design, discovery and
                   manufacturing: A review},
  volume =        {235},
  year =          {2024},
  doi =           {https://doi.org/10.1016/j.commatsci.2024.112793},
  issn =          {0927-0256},
  url =           {https://www.sciencedirect.com/science/article/pii/
                  S0927025624000144},
}

@article{Li2017,
  author =        {Li, Jun and Pereira, Paulo J. and Yuan, Jie and
                   Lv, Yang-Yang and Jiang, Mei-Ping and Lu, Dachuan and
                   Lin, Zi-Quan and Liu, Yong-Jie and Wang, Jun-Feng and
                   Li, Liang and Ke, Xiaoxing and Van Tendeloo, Gustaaf and
                   Li, Meng-Yue and Feng, Hai-Luke and Hatano, Takeshi and
                   Wang, Hua-Bing and Wu, Pei-Heng and Yamaura, Kazunari and
                   Takayama-Muromachi, Eiji and Vanacken, Johan and
                   Chibotaru, Liviu F. and Moshchalkov, Victor V.},
  journal =       {Nature Communications},
  month =         {Dec},
  number =        {1},
  pages =         {1880},
  title =         {Nematic superconducting state in iron pnictide
                   superconductors},
  volume =        {8},
  year =          {2017},
  doi =           {10.1038/s41467-017-02016-y},
  issn =          {2041-1723},
}

@article{Ishida20,
  author =        {Kousuke Ishida and Masaya Tsujii and Suguru Hosoi and
                   Yuta Mizukami and Shigeyuki Ishida and Akira Iyo and
                   Hiroshi Eisaki and Thomas Wolf and Kai Grube and
                   Hilbert v. Löhneysen and Rafael M. Fernandes and
                   Takasada Shibauchi},
  journal =       {Proceedings of the National Academy of Sciences},
  number =        {12},
  pages =         {6424-6429},
  title =         {Novel electronic nematicity in heavily hole-doped
                   iron pnictide superconductors},
  volume =        {117},
  year =          {2020},
  doi =           {10.1073/pnas.1909172117},
}

@article{Sakakibara24,
  author =        {Sakakibara, Hirofumi and Kitamine, Naoya and
                   Ochi, Masayuki and Kuroki, Kazuhiko},
  journal =       {Phys. Rev. Lett.},
  month =         {Mar},
  pages =         {106002},
  publisher =     {American Physical Society},
  title =         {Possible High ${T}_{c}$ Superconductivity in
                   ${\mathrm{La}}_{3}{\mathrm{Ni}}_{2}{\mathrm{O}}_{7}$
                   under High Pressure through Manifestation of a Nearly
                   Half-Filled Bilayer Hubbard Model},
  volume =        {132},
  year =          {2024},
  doi =           {10.1103/PhysRevLett.132.106002},
  url =           {https://link.aps.org/doi/10.1103/PhysRevLett.132.106002},
}

@article{Liu20,
  author =        {Liu, Haoliang and Malissa, Hans and Stolley, Ryan M. and
                   Singh, Jaspal and Groesbeck, Matthew and Popli, Henna and
                   Kavand, Marzieh and Chong, Su Kong and
                   Deshpande, Vikram V. and Miller, Joel S. and
                   Boehme, Christoph and Vardeny, Z. Valy},
  journal =       {Advanced Materials},
  number =        {39},
  pages =         {2002663},
  title =         {Spin Wave Excitation, Detection, and Utilization in
                   the Organic-Based Magnet, {V(TCNE) (TCNE =
                   Tetracyanoethylene)}},
  volume =        {32},
  year =          {2020},
  doi =           {https://doi.org/10.1002/adma.202002663},
}

@article{Lapidus21,
  author =        {Lapidus, Saul H. and Stephens, Peter W. and
                   Fumanal, Maria and Ribas-Ariño, Jordi and
                   Novoa, Juan J. and DaSilva, Jack G. and
                   Rheingold, Arnold L. and Miller, Joel S.},
  journal =       {Dalton Trans.},
  pages =         {11228-11242},
  publisher =     {The Royal Society of Chemistry},
  title =         {Low temperature structures and magnetic interactions
                   in the organic-based ferromagnetic and metamagnetic
                   polymorphs of decamethylferrocenium
                   7{,}7{,}8{,}8-tetracyano-p-quinodimethanide{,}
                   [{FeCp}*$_2$]$^{\cdot +}$[{TCNQ}]$^{\cdot -}$},
  volume =        {50},
  year =          {2021},
  doi =           {10.1039/D1DT02106K},
}

@article{Wen09,
  author =        {Jianzhong Wen and Hao Zhang and Michael L. Gross and
                   Robert E. Blankenship},
  journal =       {Proceedings of the National Academy of Sciences},
  number =        {15},
  pages =         {6134-6139},
  title =         {Membrane orientation of the {FMO} antenna protein
                   from \emph{Chlorobaculum tepidum} as determined by
                   mass spectrometry-based footprinting},
  volume =        {106},
  year =          {2009},
  doi =           {10.1073/pnas.0901691106},
}

@article{Sarovar10,
  author =        {Sarovar, Mohan and Ishizaki, Akihito and
                   Fleming, Graham R. and Whaley, K. Birgitta},
  journal =       {Nature Physics},
  month =         {Jun},
  number =        {6},
  pages =         {462-467},
  title =         {Quantum entanglement in photosynthetic
                   light-harvesting complexes},
  volume =        {6},
  year =          {2010},
  doi =           {10.1038/nphys1652},
  issn =          {1745-2481},
}

@article{Klymchenko25,
  author =        {Klymchenko, Andrey S. and Biswas, Deep Sekhar and
                   Didier, Pascal},
  journal =       {Advanced Materials},
  note =          {Online Version of Record before inclusion in an
                   issue.},
  number =        {},
  pages =         {e01237},
  title =         {Light-Harvesting Nanomaterials Based on Dyes for
                   Energy Transfer and Amplified Biosensing},
  volume =        {},
  year =          {2025},
  doi =           {https://doi.org/10.1002/adma.202501237},
}
\end{document}